\documentclass[10pt,letterpaper]{article}

\usepackage{amsmath}
\usepackage{amsthm}
\usepackage{amssymb}
\usepackage{graphicx}
\usepackage{dsfont}
\usepackage{enumerate}
\usepackage{mathabx}
\usepackage{mathdots}
\usepackage[margin=1.3in]{geometry}
\usepackage{color}
\usepackage{authblk}

\newtheorem{theorem}{Theorem}[section]

\newtheorem{algorithm}[theorem]{Algorithm}
\newtheorem{proposition}[theorem]{Proposition}

\DeclareMathOperator{\sech}{sech}
\DeclareMathOperator{\csch}{csch}
\DeclareMathOperator{\cn}{cn}
\DeclareMathOperator{\sn}{sn}
\DeclareMathOperator{\dn}{dn}
\DeclareMathOperator{\nc}{nc}
\DeclareMathOperator{\nd}{nd}
\DeclareMathOperator{\ns}{ns}
\DeclareMathOperator{\sac}{sc}
\DeclareMathOperator{\dc}{dc}
\DeclareMathOperator{\cs}{cs}
\DeclareMathOperator{\ds}{ds}
\DeclareMathOperator{\sd}{sd}
\DeclareMathOperator{\cd}{cd}
\DeclareMathOperator{\sign}{sign}
\DeclareMathOperator{\diag}{diag}

\title{Classification of the orthogonal separable webs for the Hamilton-Jacobi and Laplace-Beltrami equations on 3-dimensional Hyperbolic and de Sitter spaces}
\author{Carlos Valero\footnote{Department of Applied Mathematics, University
  of Waterloo, Waterloo, Ontario, N2L~3G1, Canada,
  email:\ cjvalero@uwaterloo.ca} \ and Raymond G. McLenaghan\footnote{Department of Applied Mathematics, University
  of Waterloo, Waterloo, Ontario, N2L~3G1, Canada,
  email:\ rgmlena@uwaterloo.ca}}

\begin{document}

\maketitle

\begin{abstract}
    We review the theory of orthogonal separation of variables on pseudo-Riemannian manifolds of constant non-zero curvature via concircular tensors and warped products. We then apply this theory simultaneously to both the three-dimensional Hyperbolic and de Sitter spaces, obtaining an invariant classification of the thirty-four orthogonal separable webs on each space, modulo action of the respective isometry groups. The inequivalent coordinate charts adapted to each web are also determined and listed. The results obtained for Hyperbolic 3-space agree with those in the literature, while the results for de Sitter 3-space appear to be new.
\end{abstract}

\tableofcontents

\newpage

\section{Introduction}\label{sec1}

In this paper, we consider two partial differential equations on an $n$-dimensional pseudo-Riemannian manifold $(M,g)$ which appear frequently in mathematical physics. The first is the Hamilton-Jacobi (HJ) equation for the geodesics, which in coordinates $q$ takes the form
\begin{equation}\label{eq:Hamilton-Jacobi}
	\frac{1}{2}g^{ij}\frac{\partial W}{\partial q^i}\frac{\partial W}{\partial q^j}= E,
\end{equation}
where $E$ is a nonzero constant. The second is the Laplace-Beltrami (LB) equation 
\begin{equation}\label{eq:wave}
    \frac{1}{\sqrt{|\det{g}|}} \frac{\partial}{\partial q^i} \Big( \sqrt{|\det{g}|} g^{ij} \frac{\partial \varphi}{\partial q^j} \Big) +m^2\varphi= 0,
\end{equation}
where $m$ is a non-zero constant and $\det{g}$ is the determinant of the matrix representing $g$ in coordinates $q$. Note that in the Lorentzian case, the Laplace-Beltrami equation is often called the Klein-Gordon equation.

We say that equation \eqref{eq:Hamilton-Jacobi} is {\em additive separable} in a coordinate system $q$ if there exists a solution of the form
\begin{equation}\label{eq:sumansatz}
    W(q,c)=\sum_{i=1}^{n}W_i(q^i,c),
\end{equation}
where $c=(c_1,\ldots,c_n)$ denotes $n$ constants, such that $W(q,c)$ satisfies the {\em completeness relation}
\begin{equation}
    \det \Big(\frac{\partial^2W}{\partial c_i \partial q^j } \Big)\neq0.
\end{equation}
Similarly, we say that equation \eqref{eq:wave} is {\em product separable} in a coordinate system $q$ if there exists a solution of the form \cite{Benenti2002a}
\begin{equation}\label{eq:prodansatz}
    \varphi(q,c)=\prod_{i=1}^n\varphi_i(q^i,c),
\end{equation}
that depend on $2n$ parameters $c=(c_1,\ldots,c_{2n})$, satisfying the completeness relation \cite{Benenti2002a}
\begin{equation}
    \det \begin{pmatrix} \dfrac{\partial u_i}{\partial c} \\[9pt] \dfrac{\partial v_i}{\partial c} \end{pmatrix} \neq 0, \;\; u_i= \frac{\varphi_i'}{\varphi_i}, \;\; v_i=\frac{\varphi_i''}{\varphi_i}.
\end{equation}
Separation of variables (for either equation) in coordinates $q$ is said to be {\em orthogonal} if the coordinates $q$ are orthogonal, i.e. if $g_{ij} = 0$, for $i \neq j$. It is clear that if the HJ (resp. LB) equation is separable in coordinates $q$, then it is also separable in coordinates $q'$, where $q'$ is obtained from $q$ by an transformation whose Jacobian is diagonal. This motivates the introduction of the following geometric object in the study of separability \cite{Benenti1997}. An {\em orthogonal web} in $(M,g)$ is a set of $n$ mutually transversal and orthogonal foliations of dimension $n-1$. A coordinate system $q$ is said to be {\em adapted} to an orthogonal web if its leaves are locally given by the level sets of the coordinates, $q^i = c^i$. A {\em separable web} is an orthogonal web such that the HJ equation is separable in any coordinates adapted to the web \cite{Benenti2002a}. It can be shown that if the LB equation is separable in coordinates $q$, then the HJ equation is separable in the same coordinates \cite{Robertson1927}.  However, in a space of {\em constant curvature}, where the Riemann curvature tensor has the form
\begin{equation}
    R_{ijkl}=k(g_{ik}g_{jl}-g_{il}g_{jk}),
\end{equation}
for some constant $k$, the separability of the LB equation in coordinates $q$ is equivalent to the separability of the HJ equation in the same system \cite{Eisenhart1934}. Furthermore, the separable coordinates are necessarily orthogonal \cite{Kalnins1986b}. We say that two orthogonal webs in a pseudo-Riemannian manifold are {\em equivalent} if they can be mapped into one another by an isometry.

The purpose of this paper is to determine, and classify modulo isometry, the thirty-four orthogonal separable webs on 3-dimensional de Sitter space $\mathrm{dS}_3$, the Lorentzian space of constant positive curvature, via a new method based on concircular tensors. We find that this new approach allows us {\em en passant} to {\em simultaneously} determine and classify the thirty-four orthogonal separable webs on 3-dimensional hyperbolic space $\mathbb{H}^3$, the Riemannian space of constant negative curvature, in agreement with the results obtained by Olevsky \cite{Olevsky1950}. We therefore get a complete classification of separable webs on both $\mathrm{dS}_3$ and $\mathbb{H}^3$, ultimately due to the fact that each space can be isometrically embedded in 4-dimensional Minkowski space $\mathbb{E}^4_1$. It will become evident that the procedure presented here generalizes and allows one to obtain a complete classification of the separable webs on $\mathrm{dS}_n$ and $\mathbb{H}^n$ simultaneously. It is also obvious from the theory of concircular tensors that for all $n$, $\mathrm{dS}_n$ and $\mathbb{H}^n$ have the same number of orthogonal separable webs modulo isometries.

The approach used is this paper is based on the theory of concircular tensors and warped products developed by Rajaratnam \cite{Rajaratnam2014}, Rajaratnam and McLenaghan \cite{Rajaratnam2014b,Rajaratnam2014a} and Rajaratnam, McLenaghan and Valero \cite{Rajaratnam2016}, which is applicable to pseudo-Riemannian spaces of constant curvature. This theory is derived from Eisenhart's \cite{Eisenhart1934} characterization of orthogonal separability by means of valence-two Killing tensors which have simple eigenvalues and orthogonally integrable eigendirections, called {\em characteristic Killing tensors}. We recall that a {\em Killing tensor} is a symmetric tensor $K_{ij}$ which satisfies the equation \cite{Eisenhart1934}
\begin{equation}\label{eq:Killingtensor}
    \nabla_{(i}K_{jk)}=0,
\end{equation}
where $\nabla$ denotes the covariant derivative associated to the Levi-Civita connection of $g$.

The problem for $\mathbb{H}^3$ has been studied by other methods by different authors including Olevsky \cite{Olevsky1950}, Kalnins, Miller and Reid \cite{Kalnins1984}, Kalnins \cite{Kalnins1986b} and Adlam, McLenaghan and Smirnov \cite{Cochran2011}. For a review and comparison of these methods with those used in the present paper, see \cite{Rajaratnam2016}.  However, the results for $\mathrm{dS}_3$ appear to be new.

We shall denote the $n$-dimensional pseudo-Euclidean space with signature $\nu$ by $\mathbb{E}^n_\nu$. Hence, $n$-dimensional Euclidean and Minkowski space are denoted by $\mathbb{E}^n$ and $\mathbb{E}^n_1$ respectively. As $\mathbb{E}^n_\nu$ is a vector space, we make liberal use of the canonical isomorphism between $\mathbb{E}^n_\nu$ and $T_p\mathbb{E}^n_\nu$ to identify points and tangent vectors. Accordingly, we denote the metric in $\mathbb{E}^n_\nu$ by both $g$ and $\langle \cdot , \cdot \rangle$. For any nonzero real number $\kappa$, we define the hypersurface $\mathbb{E}^{n}_\nu(\kappa)$
$$\mathbb{E}^{n}_\nu(\kappa) := \{ p \in \mathbb{E}^{n}_\nu \ | \ \langle p, p \rangle = \kappa^{-1} \}$$
Note that $\mathbb{E}^{n}_\nu(\kappa)$ is an $(n-1)$-dimensional space of constant curvature, with signature $\nu$ if $\kappa > 0$, and signature $\nu - 1$ if $\kappa < 0$. Accordingly, we identify the sphere $\mathbb{S}^{n-1}$, hyperbolic space $\mathbb{H}^{n-1}$ and de Sitter space $\mathrm{dS}_{n-1}$ with any connected component of $\mathbb{E}^{n}(1)$, $\mathbb{E}^{n}_1(-1)$ and $\mathbb{E}^{n}_1(1)$, respectively. Also, if $N$ is an open submanifold of $\mathbb{E}^{n}_\nu$, we define $N(\kappa) := N \cap \mathbb{E}^{n}_\nu(\kappa)$.

Finally, this paper relies heavily on some elementary results regarding the classification of self-adjoint linear operators on $\mathbb{E}^n_\nu$. A review of the relevant material may be found in \cite{Rajaratnam2016}; for convenience, in appendix \ref{appA} we have summarized the main results for the case of $\mathbb{E}^n_1$, which is all that shall be needed in this paper (of course in $\mathbb{E}^n$ all self-adjoint operators are orthogonally diagonalizable). Any reader not familiar with this subject should look at the appendix before proceeding. The notation and definitions found therein will be used hereafter without comment.

\section{Concircular Tensors in $\mathbb{E}^n_\nu(\kappa)$}\label{sec2}

The introduction of concircular tensors greatly simplifies the determination and classification of separable webs in spaces of constant curvature. Let $(M,g)$ be a pseudo-Riemannian manifold. A {\em concircular 2-tensor} is defined to be a symmetric tensor of valence two which satisfies
\begin{equation}\label{eq:CTdef}
    \nabla_{k}L_{ij} = \alpha_{(i}g_{j)k}
\end{equation}
for some one-form $\alpha$. Since we will not need to work with higher-order tensors, by concircular tensor we will always mean a concircular 2-tensor. Furthermore, we will often denote concircular tensor by the acronym CT.

Let $L$ be a CT on an open subset $U$ of $M$. $L$ is called {\em orthogonal} (denoted by OCT) if it is pointwise diagonalizable. It is called {\em Benenti} if it has pointwise simple eigenfunctions, and if these eigenfunctions are also functionally independent, then $L$ is called {\em irreducible} (denoted by ICT). A CT which is not irreducible is called {\em reducible}. We will concern ourselves exclusively with OCTs, as these are the ones which induces separable webs.

In a pseudo-Euclidean space $\mathbb{E}^n_\nu$, one can show \cite{Crampin2003} that the general CT is given by
\begin{equation}\label{eq:CTflat}
    L = A + 2 w \odot r + m r \odot r
\end{equation}
where $A$ is a constant symmetric tensor, $w$ is a constant vector, $m \in \mathbb{R}$, $\odot$ is the symmetric tensor product, and $r$ is the radial vector field, given in pseudo-Cartesian coordinates $(x^i)$ by $r := x^i\partial_i$. Moreover, one can also show \cite{Crampin2003} that if $N$ is an umbilical submanifold\footnote{Recall that an umbilical submanifold is one where the second fundamental form is everywhere proportional to the metric. This includes the hypersurfaces $\mathbb{E}^n_\nu(\kappa)$.} of $M$ and $L$ is a CT on $M$, then the pullback of $L$ to $N$ is a CT on $N$. Using this observation, as well as a dimensional argument, one finds that the general CT on $\mathbb{E}^n_\nu(\kappa)$ of $\mathbb{E}^n_\nu$ is given by pulling back the general CT on $\mathbb{E}^n_\nu$ via the inclusion map $\iota$. Since $T_p\mathbb{E}^n_\nu(\kappa) = r^{\bot}$, using equation \eqref{eq:CTflat}, we find that
\begin{equation}\label{eq:CTsphere}
    \tilde{L} = \iota^*A
\end{equation}
gives the general CT in $\mathbb{E}^n_\nu(\kappa)$. This gives a bijective correspondence between CTs on $\mathbb{E}^n_\nu(\kappa)$ and constant symmetric tensors on $\mathbb{E}^n_\nu$. In equation \eqref{eq:CTsphere}, $A$ is called the {\em parameter tensor} associated with $\tilde{L}$. This correspondence allows the classification of CTs in $\mathbb{E}^n_\nu(\kappa)$ to be reformulated as a classification of self-adjoint operators in $\mathbb{E}^n_\nu$. For instance, the following proposition, which is proven in \cite{Rajaratnam2014a}, is useful for identifying reducible OCTs in spherical submanifolds of Euclidean and Minkowski spaces. We will use it frequently in section \ref{sec4}.

\begin{proposition}\label{reducible}
Let $\tilde{L} = \iota^*A$ be an OCT in $\mathbb{E}^n_\nu(\kappa)$ for $\nu \leq 1$. Then $\tilde{L}$ is reducible if and only if $A$ has a multi-dimensional real eigenspace.
\end{proposition}

We now define the following equivalence relation: two CTs $L$ and $L'$ on a pseudo-Riemannian manifold $(M,g)$ are {\em geometrically equivalent} if $\exists \, a \in \mathbb{R} \setminus \{0\}$, $b \in \mathbb{R}$, and an isometry $\Lambda$ such that
$$L' = a \Lambda_* L + bg$$
It can be shown \cite{Rajaratnam2014a} that if $L$ and $L'$ are CTs on a connected manifold with at least one of them non-constant, then their induced separable webs are related by isometry if and only if they are geometrically equivalent. For CTs in $\mathbb{E}^n_\nu(\kappa)$, we can translate this into a statement about their parameter tensors, which will also be used frequently in section \ref{sec4}:

\begin{proposition}\label{geo}
Let $L_1$ and $L_2$ be CTs in $\mathbb{E}^n_\nu$ with parameter tensors $A_1$ and $A_2$. Then $L_1 = \Lambda_*L_2$ for an isometry $\Lambda$ if and only if $A_1$ and $A_2$ have the same metric-Jordan canonical form. Thus, $L_1$ and $L_2$ are geometrically equivalent if and only if 
$$A_1 = a \tilde{A}_2 + bg$$
for $a \in \mathbb{R} \setminus \{0\}$, $b \in \mathbb{R}$, and where $\tilde{A}_2$ has the same metric-Jordan canonical form as $A_2$.
\end{proposition}

So, having reviewed the theory needed to classify CTs, we now record some technical results from \cite{Rajaratnam2014} which will be used in explicitly obtaining the transformation equations between pseudo-Cartesian coordinates in the ambient space, and the separable coordinates in $\mathbb{E}^n_\nu(\kappa)$ induced by an ICT. We shall only need the following special case: let $L$ be an ICT in $\mathbb{E}^n_\nu(\kappa)$, and suppose that its parameter tensor $A$, and the ambient metric $g$, take the following form in (lightcone)-Cartesian coordinates $(x^1,\dots,x^n)$:
\begin{equation}
    \label{ICTform}
    A = J_k(0)^T \oplus \mathrm{diag}(\lambda_{k+1}, \dots, \lambda_{n}) \ \ \ \ \ \ \ \ \ \ g = \epsilon_0 S_k \oplus \mathrm{diag}(\epsilon_{k+1}, \dots, \epsilon_{n})
\end{equation}
where $\epsilon_i = \pm 1$, and where $J_k(\lambda)$ and $S_k$ are the $k \times k$ Jordan and skew-normal matrices defined in appendix \ref{appA}. Moreover, let $p(\zeta)$ denote the characteristic polynomial of $L$, let $B(\zeta)$ denote the characteristic polynomial of $A$, and let $B_{U^{\bot}}(\zeta)$ be the characteristic polynomial of $A$ restricted to the subspace corresponding to $(x^{k+1}, \dots, x^n)$. We then have \cite{Rajaratnam2014} the following
\begin{equation}
\label{ICTa}
    \sum_{i=1}^{l+1}x^ix^{l+2-i} = \frac{1}{\kappa} \frac{\epsilon_0}{l!} \Big(\frac{d}{d\zeta}\Big)^l \Big( \frac{p(\zeta)}{B_{U^{\bot}}(\zeta)}\Big) \Big|_{\zeta=0} \ \ \ \ \ \ \ \ l = 0, \dots, k-1
\end{equation}
\begin{equation}
    \label{ICTb}
    (x^i)^2 = \frac{\epsilon_i}{\kappa} \frac{p(\lambda_i)}{B'(\lambda_i)} \ \ \ \ \ \ \ \ \ \ i = k+1, \dots, n
\end{equation}
If $(u^1,\dots,u^{n-1})$ are the eigenfunctions of $L$, then we may write 
\begin{equation}
    \label{p(z)}
    p(\zeta) = \prod_{i=1}^{n-1} (\zeta-u^i)
\end{equation}
which then yields the transformation equations between the ambient coordinates $(x^1,\dots,x^n)$ and the separable coordinates $(u^1,\dots,u^{n-1})$ on $\mathbb{E}^n_\nu(\kappa)$. The last result we quote from \cite{Rajaratnam2014} is the following: in the coordinates $(u^1,\dots,u^{n-1})$ induced by an ICT, the metric is diagonal and its diagonal elements are given by
\begin{equation}
    \label{ICTmetric}
    g_{ii} = -\frac{1}{4\kappa} \frac{\prod_{j \neq i}(u^i - u^j)}{\prod_{j=1}^{n}(u^i - \lambda_j)}
\end{equation}
for $i=1,\dots,n-1$, where $\lambda_1,\dots, \lambda_n$ are the eigenvalues of $A$.

\section{Warped Products in $\mathbb{E}^n_\nu(\kappa)$}\label{sec3}

We now discuss some fundamental ideas concerning warped products and their application to separable webs. Suppose we have, for each $i = 0,1,\dots,k$, a pseudo-Riemannian manifold $(M_i,g_i)$, and $k$ smooth positive functions $\rho_i : M_0 \rightarrow \mathbb{R}^+$ where $1 \leq i \leq k$. Then the {\em warped product} $M_0 \times_{\rho_1} M_1 \times_{\rho_2} \cdots \times_{\rho_k} M_k$ is defined to be the product manifold $M_0 \times \cdots \times M_k$ equipped with the pseudo-Riemannian metric given by
\begin{equation}
    \label{WPdef}
    g := \pi_0^*g_0 + \sum_{i=1}^{k} (\rho_i^2 \circ \pi_0) \pi_i^* g_i
\end{equation}
where $\pi_i : M_0 \times \cdots \times M_k \rightarrow M_i$ is the $i$-th projection. $M_0$ is called the geodesic factor, the $M_i$ for $i \geq 1$ are called the spherical factors, and the $\rho_i$ are called the warping functions. A map $\psi : M_0 \times_{\rho_1} \cdots \times_{\rho_k} M_k \rightarrow M$ which is a (local) isometry is called a (local) {\em warped product decomposition} of $M$. We will often use the terms warped product and warped product decomposition interchangeably.

Let $L$ be a CT in $M$ and let $\psi : N_0 \times_{\rho_1} \cdots \times_{\rho_n} N_k \rightarrow M$ be a warped product of $M$. Then we say that $\psi$ is adapted to $L$ if for each $i > 0$ and for all points $p \in N_i$, $\psi_*(T_pN_i)$ is an invariant subspace of $L$. In this case, one can show \cite{Rajaratnam2014a} that the restriction (via $\psi$) of $L$ to $N_0$ is a Benenti tensor, and therefore induces a separable web on $N_0$ which we may lift to $M$ using $\psi$. In particular, if the restriction of $L$ is an ICT, then its eigenfunctions give a set of separable coordinates on $N_0$. By choosing a separable web for each of the spherical factors and lifting them to $M$ via $\psi$, we hence obtain a separable web in $M$.

The goal of this section is to construct a warped product of $\mathbb{E}^n_\nu(\kappa)$ which is adapted to a given reducible CT. This will be done by first considering a related CT in the ambient space $\mathbb{E}^n_\nu$; constructing a warped product adapted to it; and then restricting the warped product back to $\mathbb{E}^n_\nu(\kappa)$. One can show that this restricted warped product is adapted to the original CT in $\mathbb{E}^n_\nu(\kappa)$. Most of what follows will be a very brief overview of results from \cite{Rajaratnam2014}.

A warped product of $\mathbb{E}^n_\nu$ is uniquely determined by the following {\em initial data}: a point $\bar{p} \in \mathbb{E}^n_\nu$; an orthogonal decomposition $T_{\bar{p}}\mathbb{E}^n_\nu = V_0 \obot \cdots \obot V_k$ of the tangent space at $\bar{p}$ into nontrivial (and therefore non-degenerate) subspaces, where $k > 0$; and $k$ vectors $a_1,\dots,a_k \in V_0$ which are pairwise orthogonal and linearly independent. It is of no loss of generality to assume that the warped product is in {\em canonical form}. This means that $\bar{p} \in V_0$ and $\langle \bar{p}, a_i \rangle = 1$ for each $i > 0$. We will now show how this data determines a warped product. 

One can show \cite{Rajaratnam2014} that each triple $(\bar{p}; V_i;a_i)$ for $i > 0$ uniquely determines a maximal connected and complete spherical submanifold $S_i$ of $\mathbb{E}^n_\nu$, and is such that $\bar{p} \in S_i$, $T_{\bar{p}}S_i = V_i$, and $S_i$ has mean curvature vector $-a_i$ at $\bar{p}$. $S_i$ is an open submanifold of $N_i$, where $N_i$ is called the {\em sphere determined by} $(\bar{p}; V_i;a_i)$ and takes one of the following forms

\begin{enumerate}[(i)]
    \item if $a_i = 0$, then $N_i = \bar{p} + V_i$ and is therefore pseudo-Euclidean
    \item if $a_i$ is non-null, then $N_i = c + \{ p \in \mathbb{R}a_i \obot V_i \ | \ \langle p, p \rangle = \langle a_i, a_i\rangle^{-1} \}$, where $c := \bar{p} - \langle a_i, a_i \rangle^{-1}a_i$. If $a_i$ is timelike (spacelike), then $N_i$ has constant negative (positive) curvature $\langle a_i, a_i \rangle$
    \item if $a_i$ is lightlike, then $N_i = \bar{p} + \{p - \frac{1}{2}\langle p, p \rangle a_i \ | \ p \in V_i \}$ and is isometric to a (parabolically-embedded) pseudo-Euclidean space
\end{enumerate}
In the first case, $N_i$ is simply a plane passing through $\bar{p}$. This is relevant for warped products in pseudo-Euclidean spaces, but won't be needed for constructing warped products in $\mathbb{E}^n_\nu(\kappa)$. We will henceforward assume that our warped product is {\em proper}, meaning $a_i \neq 0$ for all $i > 0$.

Continuing with our construction, we define $N_i$ to be the sphere determined by $(\bar{p}; V_i; a_i)$ for $i > 0$. For each $i > 0$, we define the function $\rho_i : V_0 \rightarrow \mathbb{R}$ by $\rho_i(p_0) = \langle p_0, a_i \rangle$. We then let $N_0$ be the open subset of $V_0$ where each $\rho_i$ is positive. This defines the warped product $N_0 \times_{\rho_1} \cdots \times_{\rho_k} N_k$ induced by the initial data $(\bar{p}; V_0 \obot \cdots \obot V_k ; a_1,\dots,a_k)$, as well as a warped product decomposition $\psi : N_0 \times_{\rho_1} \cdots \times_{\rho_k} N_k \rightarrow \mathbb{E}^n_\nu$. Rather than give the general expression for $\psi$ (which can be found in \cite{Rajaratnam2014}), we will describe it below for some simple but useful cases.

Let us write down a convenient expression for these warped products. We begin by considering warped products with only two factors, and later comment on how warped products with multiple factors can be constructed from these. First let $\psi : N_0 \times_{\rho_1} N_1 \rightarrow \mathbb{E}^n_\nu$ be a warped product determined by initial data $(\bar{p}; V_0 \obot V_1 ; a_1)$ where $a_1$ is non-null. Define $W_0 := V_0 \cap a_1^{\bot}$, and let $P_0 : \mathbb{E}^n_\nu \rightarrow W_0$ denote the orthogonal projection. Then $\psi$ takes the form
\begin{equation}
\label{non-nullWP}
    \psi(p_0,p_1) = P_0p_0 + \langle a_1, p_0 \rangle (p_1 - c)
\end{equation}
where $c := \bar{p} - \langle a_1 , a_1 \rangle^{-1}a_1$. Now consider the case where $a_1$ is lightlike. Then there is another lightlike vector $b \in V_0$ such that $\langle a_1, b \rangle = 1$. Here, we define $W_0 := V_0 \cap \mathrm{span}\{a_1,b\}^{\bot}$ and $W_1 := V_1$, and let $P_i : \mathbb{E}^n_\nu \rightarrow W_i$ for $i=0,1$ denote the orthogonal projection. Then $\psi$ takes the form
\begin{equation}
\label{nullWP}
    \psi(p_0,p_1) = P_0p_0 + (\langle b, p_0 \rangle - \frac{1}{2}\langle a_1, p_0 \rangle \langle P_1p_1, P_1p_1 \rangle)a_1 + \langle a_1, p_0 \rangle b + \langle a_1, p_0 \rangle P_1p_1
\end{equation}
A warped product with multiple factors can be constructed from two-factor warped products by inductively decomposing the geodesic factor in a compatible way (see \cite{Rajaratnam2014} for details). In this manner, one can show that the warped product $\psi : N_0 \times_{\rho_1} \cdots \times_{\rho_k} N_k \rightarrow \mathbb{E}^n_\nu$ determined by $(\bar{p}; V_0 \obot \cdots \obot V_k;a_1,\dots,a_k)$, where each $a_i$ is non-null, is given by
\begin{equation}
\label{multipleWP}
    \psi(p_0,\dots,p_k) = P_0p_0 + \sum_{i=1}^k \langle a_i, p_0 \rangle (p_i - c_i)
\end{equation}
where $c_i := \bar{p} - \langle a_i, a_i \rangle^{-1} a_i$ and $P_0$ is the orthogonal projection onto $V_0 \cap a_1^{\bot} \cap \cdots \cap a_k^{\bot}$. It also useful to know the images of these standard warped products. If $\psi$ is a warped product of the form \eqref{multipleWP}, and if $P_i : \mathbb{E}^n_\nu \rightarrow \mathbb{R}a_i \obot V_i$ are orthogonal projections, then we have \cite{Rajaratnam2014}
\begin{equation}
\label{nonnullimage}
    \mathrm{Im}(\psi) = \{p \in \mathbb{E}^n_\nu \ | \ \sign\, \langle P_i(p), P_i(p) \rangle = \sign\, \langle a_i, a_i \rangle, \ i=1,\dots,k \}
\end{equation}
If we require the $N_i$ to be connected, then for each $i > 0$ such that $N_i$ is disconnected, we impose the extra condition $\langle a_i, P_i(p) \rangle > 0$. If $\psi$ is a warped product of the form \eqref{nullWP}, then we have
\begin{equation}
\label{nullimage}
    \mathrm{Im}(\psi) = \{p \in \mathbb{E}^n_\nu \ | \ \langle a_1, p \rangle > 0 \}.
\end{equation}

For the remainder of this section, let us restrict ourselves to Euclidean and Minkowski spaces, i.e. $\nu \leq 1$. Consider a reducible CT $L$ in $\mathbb{E}^n_\nu(\kappa)$ with parameter tensor $A$. Since $\nu \leq 1$, we have that $L$ is reducible iff $A$ has a multi-dimensional eigenspace. Let us further consider the CT in $\mathbb{E}^n_\nu$ given by $L_c := A + r \odot r$. This CT is reducible, and hence, the algorithm given in chapter 9 of \cite{Rajaratnam2014} yields a (proper) warped product adapted to $L_c$. For completeness, we give the algorithm below, reiterating that we have assumed $\nu \leq 1$.

\begin{algorithm}
\label{WP}
    Let $\{E_i\}$ be the multidimensional eigenspaces of $A$. For each $i$, apply the following construction: 
    \begin{enumerate}[(i)]
        \item If $E_i$ is non-degenerate, choose a unit vector $a_i \in E_i$ and define $V_i := E_i \cap a_i^{\bot}$.
        \item If $E_i$ is a degenerate subspace, then there is a cycle $v_1,\dots,v_r$ of generalized eigenvectors of $A$, such that $v_r \in E_i$ is lightlike. Let $a_i := v_r$, and define $V_i := E_i \cap v_1^{\bot}$. Note that $V_i$ is non-degenerate, and in $\mathbb{E}^n_1$, $r \leq 3$.
    \end{enumerate}
    Define $V_0 := V_1^{\bot} \cap \cdots \cap V_k^{\bot}$, and let $\bar{p} \in \mathbb{E}^n_\nu$ be such that the warped product $\psi : N_0 \times_{\rho_1} \cdots \times_{\rho_k} N_k \rightarrow \mathbb{E}^n_\nu$ determined by initial data $(\bar{p}; V_0 \obot \cdots \obot V_k; a_1,\dots,a_k)$ is in canonical form. Then $\psi$ is a proper warped product adapted to $L_c = A + r \odot r$. 
\end{algorithm}
\noindent One can show \cite{Rajaratnam2014} that the restriction of $\psi$ to $N_0(\kappa) \times_{\rho_1} N_1 \times_{\rho_1} \cdots \times_{\rho_k} N_k$ yields a warped product of $\mathbb{E}^n_\nu(\kappa)$ which is adapted to $L$. Also, the restriction of $L$ to $N_0(\kappa)$ via $\psi$ is Benenti, and its parameter tensor is given by the restriction of $A$ to the subspace $V_0$ defined in algorithm \ref{WP}. 

This finally gives us a procedure for constructing separable webs from reducible CTs in $\mathbb{E}^n_\nu(\kappa)$. Given a reducible CT $L$ with parameter tensor $A$, we first use the above algorithm to construct a warped product of $\mathbb{E}^n_\nu$ which is adapted to $A + r \odot r$. We then restrict this warped product to $N_0(\kappa)$ to get a warped product which is adapted to $L$. Then, upon choosing a separable web (and corresponding separable coordinates) for each factor, we lift these to $\mathbb{E}^n_\nu(\kappa)$ via $\psi$, thus obtaining a separable web on $\mathbb{E}^n_\nu(\kappa)$. Note that this algorithm is recursive, in the sense that it requires knowledge of {\em all} separable webs in the lower-dimensional spaces which could appear as factors in a warped product decomposition of $\mathbb{E}^n_\nu(\kappa)$.

\section{Classification of Separable Webs in $\mathbb{H}^3$ and $\mathrm{dS}_3$}\label{sec4}

In this section, we finally apply the theory of concircular tensors reviewed above to determine and classify modulo isometry, the 34 separable webs in dS$_3$ and $\mathbb{H}^3$ simultaneously. We also work out, for each separable web, all the inequivalent coordinate charts adapted to the web (by inequivalent coordinate charts, we mean ones that cannot be mapped into one another by isometry).

Thus, for each web below, and for each corresponding coordinate chart, we give the transformation equations between the separable coordinates and pseudo-Cartesian coordinates in the ambient space, as well as the components of the metric in the separable coordinates. Note that we only give the coordinate transformations for a particular chart in each equivalence class. All other equivalent charts can then be obtained by isometry (often some combination of $x^i \mapsto -x^i$ and permutation of the spacelike coordinates $x,y,z$). 

As in \cite{Rajaratnam2016}, we will write the transformation equations in terms of transcendental functions, when possible. In many cases these will be the Jacobi elliptic functions, $\sn(u;a)$, $\cn(u;a)$, $\dn(u;a)$, etc. When dealing with these functions, $K(a)$ will always mean the complete elliptic integral of the first kind with parameter $a$, where $0 < a < 1$. For an overview of the Jacobi elliptic functions, see, for example, the book of Lawden \cite{Lawden2010}.

When constructing warped products of $\mathbb{E}^4_1$, we will require that our spherical factors be connected (see the remarks following equation \eqref{nonnullimage}). This is only relevant in the cases below where the spherical factors contain dS$_1$ or $\mathbb{H}^n$. Also, for consistency, we shall use the following lightcone coordinates ($\eta,\xi$) throughout this section,
$$\eta := t + x, \ \ \ \ \ \ \ \xi := \frac{1}{2}(x - t)$$
Note that $\langle \partial_\eta , \partial_\xi \rangle = 1$, and so these vectors form a skew-normal sequence for their span (see appendix $A$ for terminology and notation).

Lastly, according to the remarks at the end of section \ref{sec3}, we will need an exhaustive catalogue of the separable webs in the lower-dimensional spaces $\mathrm{dS}_2$ and $\mathbb{H}^2$. A list of the separable webs for $\mathrm{dS}_2$, their adapted coordinates, and their associated CTs can be found in \cite{Rajaratnam2016}, while the same information for $\mathbb{H}^2$ is tabulated in appendix \ref{appB} for convenience. \\

\vspace{5pt}

\noindent {\bf 4.1} \ $A = J_{-1}(1) \oplus J_1(0) \oplus J_1(0) \oplus J_1(0)$
\vspace{10pt}

\noindent If $A$ has a three-dimensional spacelike eigenspace, then the associated CT is reducible, and we may choose pseudo-Cartesian coordinates such that $A = \partial_t \odot \partial_t$. Then, upon choosing an eigenvector in the eigenspace of $A$, say $\partial_x$, algorithm \ref{WP} yields the warped product $\psi$ which decomposes the $A + r \odot r$ in $\mathbb{E}^4_1$. By equation \eqref{non-nullWP}, $\psi$ is given by
\begin{align*}
\psi &: N_0 \times_{\rho} \mathbb{S}^2 \rightarrow \mathbb{E}^4_1 \\
							&(t \partial_t + \tilde{x} \partial_x, p) \mapsto t \partial_t + \tilde{x}p
\end{align*}
where $N_0 = \{t\partial_t + \tilde{x} \partial_x \in \mathbb{E}^4_1 \ | \ \tilde{x} > 0 \}$ and $\rho(t\partial_t + \tilde{x} \partial_x) = \tilde{x}$. By equation \eqref{nonnullimage}, the image of $\psi$ is dense in $\mathbb{E}^4_1$. To obtain a warped product which decomposes the CT induced by $A$ on $\mathbb{H}^3$ or dS$_3$, we restrict $\psi$ to $N_0(-1)$ or $N_0(1)$ respectively. \\

\noindent {\em Restriction to $\mathbb{H}^3$} \\

\noindent Since $N_0$ is isometric to an open subset of $\mathbb{E}^2_1$, we have that $N_0(-1)$ is isometric to an open subset $\mathbb{H}^1$, and the restriction of $A$ induces the standard coordinate $u$ on $\mathbb{H}^1$. We then get two separable webs, corresponding to the two possible webs we may lift from $\mathbb{S}^2$. \\

\noindent H-1. {\em Spacelike rotational web I}
\begin{flalign*}
\left \{
\begin{aligned}
&ds^2 = du^2 + \sinh^2{u}\, (dv^2 + \sin^2{v}\, dw^2)\\
&t=\cosh{u}, \ \ \ \ x=\sinh{u}\, \cos{v}, \\
&y=\sinh{u}\, \sin{v}\, \sin{w}, \ \ \ \ z=\sinh{u}\, \sin{v}\, \cos{w} \\ 
&0 < u < \infty, \ \ 0 < v < \pi, \ \ 0 < w < 2\pi \ \ \ \ \ \ \ \ \ \ \ \  \ \ \ \ \ \ \ \ \ \ \ \ \ \ \  \ \ \ \ \ \ \ \ \ \ \ \ \ \ \ \ \ \  \ \ \ \ \ \ \ \ \ \ \ \ \ \ \  \ \ \ \ \ \ \ \ \ \ \ \ \ \ \ \ \ \ \ \ \ \ \ \ \ \ \ \ \ \ \ \ \ \ \ \ \ \ \ \ \ \ \ \ \ \ \ \ \ \ \ \ \ \ \ \ \ \ \ \ \ \ \ \ \ \ \ \ \ \ \ \ \ \ \ \ \ \ \ \ \ \ \ \ \ \ \ \ \ \ \ \ \ \ \  \ \ \ \ \ \ \ \ \ \ \ \ \ \ \  \ \ \ \ \ \ \ \ \ \ \ \ \ \ \ \ \ \ \ \ \ \ \ \ 
\end{aligned}
\right.
\end{flalign*}

\noindent H-2. {\em Hyperbolic-elliptic web I}
\begin{flalign*}
\left \{
\begin{aligned}
&ds^2 = du^2 + \sinh^2{u}\, (a^2 \cn^2(v;a) + b^2 \cn^2(w;b))(dv^2 + dw^2)\\
&t=\cosh{u}, \ \ \ \ x=\sinh{u}\, \sn(v;a) \dn(w;b), \\
&y=\sinh{u}\, \cn(v;a)\, \cn(w;b), \ \ \ \ z=\sinh{u}\, \dn(v;a)\, \sn(w;b) \\ 
&0 < u < \infty, \ \ 0 < v < 4K(a), \ \ -2K(b) < w < 2K(b), \ \ a^2 + b^2 = 1 \ \ \ \ \ \ \ \ \ \ \ \  \ \ \ \ \ \ \ \ \ \ \ \ \ \ \  \ \ \ \ \ \ \ \ \ \ \ \ \ \ \ \ \ \  \ \ \ \ \ \ \ \ \ \ \ \ \ \ \  \ \ \ \ \ \ \ \ \ \ \ \ \ \ \ \ \ \ \ \ \ \ \ \ \ \ \ \ \ \ \ \ \ \ \ \ \ \ \ \ \ \ \ \ \ \ \ \ \ \ \ \ \ \ \ \ \ \ \ \ \ \ \ \ \ \ \ \ \ \ \ \ \ \ \ \ \ \ \ \ \ \ \ \ \ \ \ \ \ \ \ \ \ \ \  \ \ \ \ \ \ \ \ \ \ \ \ \ \ \  \ \ \ \ \ \ \ \ \ \ \ \ \ \ \ \ \ \ \ \ \ \ \ \ \ \ \ \ \ \ \ \ \ \ \ \ \ \ \ \ \ \ \ \ \ \ \ \ \ \ \ \ \ \ \ \ \ \ \ \ \ \ \ \ \ \ \ \ \ \ \ \ \ \ \ \ \ \ \ \ \  
\end{aligned}
\right.
\end{flalign*}

\vspace{10pt}

\noindent {\em Restriction to }dS$_3$ \\

\noindent For the above warped product, $N_0(1)$ is isometric to dS$_1$, and the restriction of $A$ to the first factor induces the standard coordinate $u$ on dS$_1$. We again have two webs upon lifting the two from $\mathbb{S}^2$. \\

\noindent dS-1. {\em Spacelike rotational web I}
\begin{flalign*}
\left \{
\begin{aligned}
&ds^2 = -du^2 + \cosh^2{u}\, (dv^2 + \sin^2{v}\, dw^2)\\
&t=\sinh{u}, \ \ \ \ x=\cosh{u}\, \cos{v}, \\
&y=\cosh{u}\, \sin{v}\, \sin{w}, \ \ \ \ z=\cosh{u}\, \sin{v}\, \cos{w} \\ 
&-\infty < u < \infty, \ \ 0 < v < \pi, \ \ 0 < w < 2\pi \ \ \ \ \ \ \ \ \ \ \ \  \ \ \ \ \ \ \ \ \ \ \ \ \ \ \  \ \ \ \ \ \ \ \ \ \ \ \ \ \ \ \ \ \  \ \ \ \ \ \ \ \ \ \ \ \ \ \ \  \ \ \ \ \ \ \ \ \ \ \ \ \ \ \ \ \ \ \ \ \ \ \ \ \ \ \ \ \ \ \ \ \ \ \ \ \ \ \ \ \ \ \ \ \ \ \ \ \ \ \ \ \ \ \ \ \ \ \ \ \ \ \ \ \ \ \ \ \ \ \ \ \ \ \ \ \ \ \ \ \ \ \ \ \ \ \ \ \ \ \ \ \ \ \  \ \ \ \ \ \ \ \ \ \ \ \ \ \ \  \ \ \ \ \ \ \ \ \ \ \ \ \ \ \ \ \ \ \ \ \ \ \ \ \ \ \ \ \ \ \ \ \ \ \ \ \ \ \ \ \ \ \ \ \ \ \ \ \ \ \ \ \ \ \ \ \ \ \ \ \ \ \ \ \ \ \ \ \ \ \ \ \ \ \ \ \ \ \ \ \ 
\end{aligned}
\right.
\end{flalign*}

\vspace{10pt}

\noindent dS-2. {\em de Sitter-elliptic web I}
\begin{flalign*}
\left \{
\begin{aligned}
&ds^2 = -du^2 + \cosh^2{u}\ (a^2 \textrm{cn}^2(v;a) + b^2 \textrm{cn}^2(w;b))(dv^2 + dw^2)\\
&t=\sinh{u}, \ \ \ \ x=\sinh{u}\, \sn(v;a) \dn(w;b), \\
&y=\sinh{u}\, \cn(v;a)\, \cn(w;b), \ \ \ \ z=\sinh{u}\, \dn(v;a)\, \sn(w;b) \\ 
&-\infty < u < \infty, \ \ 0 < v < 4K(a), \ \ -2K(b) < w < 2K(b), \ \ a^2 + b^2 = 1 \ \ \ \ \ \ \ \ \ \ \ \  \ \ \ \ \ \ \ \ \ \ \ \ \ \ \  \ \ \ \ \ \ \ \ \ \ \ \ \ \ \ \ \ \  \ \ \ \ \ \ \ \ \ \ \ \ \ \ \  \ \ \ \ \ \ \ \ \ \ \ \ \ \ \ \ \ \ \ \ \ \ \ \ \ \ \ \ \ \ \ \ \ \ \ \ \ \ \ \ \ \ \ \ \ \ \ \ \ \ \ \ \ \ \ \ \ \ \ \ \ \ \ \ \ \ \ \ \ \ \ \ \ \ \ \ \ \ \ \ \ \ \ \ \ \ \ \ \ \ \ \ \ \ \  \ \ \ \ \ \ \ \ \ \ \ \ \ \ \  \ \ \ \ \ \ \ \ \ \ \ \ \ \ \ \ \ \ \ \ \ \ \ \ \ \ \ \ \ \ \ \ \ \ \ \ \ \ \ \ \ \ \ \ \ \ \ \ \ \ \ \ \ \ \ \ \ \ \ \ \ \ \ \ \ \ \ \ \ \ \ \ \ \ \ \ \ \ \ \ \ 
\end{aligned}
\right.
\end{flalign*}

\vspace{10pt}

\noindent {\bf 4.2} \ $A = J_{-1}(0) \oplus J_1(0) \oplus J_1(0) \oplus J_1(1)$
\vspace{10pt}

\noindent If $A$ has a three-dimensional Lorentzian eigenspace, then the associated CT is reducible, and we may choose pseudo-Cartesian coordinates such that $A = \partial_z \odot \partial_z$. Algorithm \ref{WP} gives two different warped products which decompose $A + r \odot r$, depending on whether we choose a timelike or spacelike unit vector in the eigenspace of $A$. This in turn will depend on whether the point of $\mathbb{E}^4_1$ through which we construct the warped product is timelike or spacelike.

Let us first choose a timelike eigenvector of $A$, say $-\partial_t$, in the construction of algorithm \ref{WP}. Then, by equation \eqref{non-nullWP} the resulting warped product $\psi_1$ which decomposes $A + r \odot r$ in $\mathbb{E}^4_1$ is
\begin{align*}
\psi_1 &: N_0 \times_{\rho} \mathbb{H}^2 \rightarrow \mathbb{E}^4_1 \\
							&(- \tilde{t} \partial_t + z \partial_z, p) \mapsto z \partial_z + \tilde{t}p
\end{align*}
where $N_0 = \{-\tilde{t} \partial_t + z \partial_z \in \mathbb{E}^4_1 \ | \ \tilde{t} > 0 \}$ and $\rho(-\tilde{t} \partial_t + z \partial_z) = \tilde{t}$. By equation \eqref{nonnullimage} and the remarks following it, the image of $\psi_1$ consists of all points $(t,x,y,z)$ such that $-t^2 + x^2 + y^2 < 0$ and $t > 0$. If instead we choose a spacelike eigenvector of $A$, say $\partial_x$, in algorithm \ref{WP}, we find that the resulting warped product $\psi_2$ which decomposes the $A + r \odot r$ in $\mathbb{E}^4_1$ is
\begin{align*}
\psi_2 &: N_0 \times_{\rho} \mathrm{dS}_2 \rightarrow \mathbb{E}^4_1 \\
							&(z \partial_z + \tilde{y} \partial_y, p) \mapsto z \partial_z + \tilde{y}p
\end{align*}
where $N_0 = \{z \partial_z + \tilde{y} \partial_y \in \mathbb{E}^4_1 \ | \ \tilde{y} > 0 \}$ and $\rho(z \partial_z + \tilde{y} \partial_y) = \tilde{y}$. By equation \eqref{nonnullimage}, the image of $\psi_2$ consists of all points $(t,x,y,z)$ such that $-t^2 + x^2 + y^2 > 0$. To obtain warped products which decompose the CT induced by $A$ on $\mathbb{H}^3$ or dS$_3$, we restrict $\psi_1$ and $\psi_2$ to $N_0(-1)$ or $N_0(1)$ respectively.  \\

\noindent {\em Restriction to $\mathbb{H}^3$} \\

\noindent First note that restricting $\psi_2$ to $N_0(-1)$ does not yield a warped product of $\mathbb{H}^3$. Now, for $\psi_1$, $N_0$ is isometric to an open subset of $\mathbb{E}^2_1$, and so $N_0(-1)$ is isometric to an open subset of $\mathbb{H}^1$. The restriction of $A$ induces the standard coordinate $u$ on $\mathbb{H}^1$. We then get nine separable webs, corresponding to the nine possible webs we may lift from $\mathbb{H}^2$. For a list of the nine webs and their adapted coordinates on $\mathbb{H}^2$, see appendix \ref{appB}. \\

\noindent H-3. {\em Hyperbolic-elliptic web II}
\begin{flalign*}
\left \{
\begin{aligned}
&ds^2 = du^2 + \cosh^2{u}\, (a^2  \textrm{cd}^2(v;a) + \textrm{cs}^2(w;b))(dv^2 + dw^2)\\
&t=\cosh{u}\, \nd(v;a)\, \ns(w;b), \ \ \ \ x=\cosh{u}\, \sd(v;a)\, \ds(w;b), \\
&y=\cosh{u}\, \cd(v;a)\, \cs(w;b), \ \ \ \ z=\sinh{u} \\ 
&-\infty < u < \infty, \ \ 0 < v < K(a), \ \ 0 < w < K(b), \ \ a^2 + b^2 = 1 \ \ \ \ \ \ \ \ \ \ \ \  \ \ \ \ \ \ \ \ \ \ \ \ \ \ \  \ \ \ \ \ \ \ \ \ \ \ \ \ \ \ \ \ \  \ \ \ \ \ \ \ \ \ \ \ \ \ \ \  \ \ \ \ \ \ \ \ \ \ \ \ \ \ \ \ \ \ \ \ \ \ \ \ \ \ \ \ \ \ \ \ \ \ \ \ \ \ \ \ \ \ \ \ \ \ \ \ \ \ \ \ \ \ \ \ \ \ \ \ \ \ \ \ \ \ \ \ \ \ \ \ \ \ \ \ \ \ \ \ \ \ \ \ \ \ \ \ \ \ \ \ \ \ \  \ \ \ \ \ \ \ \ \ \ \ \ \ \ \  \ \ \ \ \ \ \ \ \ \ \ \ \ \ \ \ \ \ \ \ \ \ \ \ \ \ \ \ \ \ \ \ \ \ \ \ \ \ \ \ \ \ \ \ \ \ \ \ \ \ \ \ \ \ \ \ \ \ \ \ \ \ \ \ \ \ \ \ \ \ \ \ \ \ \ \ \ \ \ \ \ 
\end{aligned}
\right.
\end{flalign*}

\vspace{10pt}

\noindent H-4. {\em Hyperbolic-elliptic web III}
\begin{flalign*}
\left \{
\begin{aligned}
&ds^2 = du^2 + \cosh^2{u}\, (\dc^2(v;a) + a^2 \sac^2(w;b))(dv^2 + dw^2) \\
&t=\cosh{u}\, \nc(v;a)\, \nc(w;b), \ \ \ \ x=\cosh{u}\, \sac(v;a)\, \dc(w;b), \\
&y=\cosh{u}\, \dc(v;a)\, \sac(w;b), \ \ \ \ z=\sinh{u} \\ 
&-\infty < u < \infty, \ \ 0 < v < K(a), \ \ 0 < w < K(b), \ \ a^2 + b^2 = 1 \ \ \ \ \ \ \ \ \ \ \ \  \ \ \ \ \ \ \ \ \ \ \ \ \ \ \  \ \ \ \ \ \ \ \ \ \ \ \ \ \ \ \ \ \  \ \ \ \ \ \ \ \ \ \ \ \ \ \ \  \ \ \ \ \ \ \ \ \ \ \ \ \ \ \ \ \ \ \ \ \ \ \ \ \ \ \ \ \ \ \ \ \ \ \ \ \ \ \ \ \ \ \ \ \ \ \ \ \ \ \ \ \ \ \ \ \ \ \ \ \ \ \ \ \ \ \ \ \ \ \ \ \ \ \ \ \ \ \ \ \ \ \ \ \ \ \ \ \ \ \ \ \ \ \  \ \ \ \ \ \ \ \ \ \ \ \ \ \ \  \ \ \ \ \ \ \ \ \ \ \ \ \ \ \ \ \ \ \ \ \ \ \ \ \ \ \ \ \ \ \ \ \ \ \ \ \ \ \ \ \ \ \ \ \ \ \ \ \ \ \ \ \ \ \ \ \ \ \ \ \ \ \ \ \ \ \ \ \ \ \ \ \ \ \ \ \ \ \ \ \ 
\end{aligned}
\right.
\end{flalign*}

\vspace{10pt}

\noindent H-5. {\em Spacelike rotational web II}
\begin{flalign*}
\left \{
\begin{aligned}
&ds^2 = du^2 + \cosh^2{u}\, (dv^2 + \sinh^2{v}\, dw^2) \\
&t=\cosh{u}\, \cosh{v}, \ \ \ \ x=\cosh{u}\, \sinh{v}\, \cos{w}, \\
&y=\cosh{u}\, \sinh{v}\, \sin{w}, \ \ \ \ z=\sinh{u} \\ 
&-\infty < u < \infty, \ \ 0 < v < \infty, \ \ 0 < w < 2\pi \ \ \ \ \ \ \ \ \ \ \ \ \  \ \ \ \ \ \ \ \ \ \ \ \ \ \ \  \ \ \ \ \ \ \ \ \ \ \ \ \ \ \ \ \ \  \ \ \ \ \ \ \ \ \ \ \ \ \ \ \  \ \ \ \ \ \ \ \ \ \ \ \ \ \ \ \ \ \ \ \ \ \ \ \ \ \ \ \ \ \ \ \ \ \ \ \ \ \ \ \ \ \ \ \ \ \ \ \ \ \ \ \ \ \ \ \ \ \ \ \ \ \ \ \ \ \ \ \ \ \ \ \ \ \ \ \ \ \ \ \ \ \ \ \ \ \ \ \ \ \ \ \ \ \ \  \ \ \ \ \ \ \ \ \ \ \ \ \ \ \  \ \ \ \ \ \ \ \ \ \ \ \ \ \ \ \ \ \ \ \ \ \ \ \ \ \ \ \ \ \ \ \ \ \ \ \ \ \ \ \ \ \ \ \ \ \ \ \ \ \ \ \ \ \ \ \ \ \ \ \ \ \ \ \ \ \ \ \ \ \ \ \ \ \ \ \ \ \ \ \ \ 
\end{aligned}
\right.
\end{flalign*}

\vspace{10pt}

\noindent H-6. {\em Timelike rotational web I}
\begin{flalign*}
\left \{
\begin{aligned}
&ds^2 = du^2 + \cosh^2{u}\, (dv^2 + \cosh^2{v}\, dw^2) \\
&t=\cosh{u}\, \cosh{v}\, \cosh{w}, \ \ \ \ x=\cosh{u}\, \cosh{v}\, \sinh{w}, \\
&y=\cosh{u}\, \sinh{v}, \ \ \ \ z=\sinh{u} \\ 
&-\infty < u < \infty, \ \ -\infty < v < \infty, \ \ -\infty < w < \infty \ \ \ \ \ \ \ \ \ \ \ \ \  \ \ \ \ \ \ \ \ \ \ \ \ \ \ \ \ \ \ \ \ \ \ \ \ \ \ \ \ \ \ \ \ \  \ \ \ \ \ \ \ \ \ \ \ \ \ \ \  \ \ \ \ \ \ \ \ \ \ \ \ \ \ \ \ \ \ \ \ \ \ \ \ \ \ \ \ \ \ \ \ \ \ \ \ \ \ \ \ \ \ \ \ \ \ \ \ \ \ \ \ \ \ \ \ \ \ \ \ \ \ \ \ \ \ \ \ \ \ \ \ \ \ \ \ \ \ \ \ \ \ \ \ \ \ \ \ \ \ \ \ \ \ \  \ \ \ \ \ \ \ \ \ \ \ \ \ \ \  \ \ \ \ \ \ \ \ \ \ \ \ \ \ \ \ \ \ \ \ \ \ \ \ \ \ \ \ \ \ \ \ \ \ \ \ \ \ \ \ \ \ \ \ \ \ \ \ \ \ \ \ \ \ \ \ \ \ \ \ \ \ \ \ \ \ \ \ \ \ \ \ \ \ \ \ \ \ \ \ \ 
\end{aligned}
\right.
\end{flalign*}

\vspace{10pt}

\noindent H-7. {\em Hyperbolic-complex elliptic web}
\begin{flalign*}
\left \{
\begin{aligned}
&ds^2 = du^2 + \cosh^2{u}\, (\sn^2(v;a)\, \dc^2(v;a) + \sn^2(w;b)\, \dc^2(w;b))(dv^2 + dw^2) \\
&t^2 + x^2 = \frac{2 \cosh^2{u}\, \dn(2v;a)\, \dn(2w;b)}{ab(1+\cn(2v;a))(1+\cn(2w;b))}, \ \ \ \ t^2 - x^2 = \frac{2 \cosh^2{u}\, (1 + \cn(2v;a)\, \cn(2w;b))}{(1 + \cn(2v;a))(1 + \cn(2w;b))}, \\
&y = \cosh{u}\, \sn(v;a)\, \dc(v;a)\, \sn(w;b)\, \dc(w;b), \ \ \ \ z = \sinh{u} \\
&-\infty < u < \infty, \ \ 0 < v < K(a), \ \ 0 < w < K(b), \ \ a^2 + b^2 = 1 \ \ \ \ \ \ \ \ \ \ \ \ \  \ \ \ \ \ \ \ \ \ \ \ \ \ \ \ \ \ \ \ \ \ \ \ \ \ \ \ \ \ \ \ \ \  \ \ \ \ \ \ \ \ \ \ \ \ \ \ \  \ \ \ \ \ \ \ \ \ \ \ \ \ \ \ \ \ \ \ \ \ \ \ \ \ \ \ \ \ \ \ \ \ \ \ \ \ \ \ \ \ \ \ \ \ \ \ \ \ \ \ \ \ \ \ \ \ \ \ \ \ \ \ \ \ \ \ \ \ \ \ \ \ \ \ \ \ \ \ \ \ \ \ \ \ \ \ \ \ \ \ \ \ \ \  \ \ \ \ \ \ \ \ \ \ \ \ \ \ \  \ \ \ \ \ \ \ \ \ \ \ \ \ \ \ \ \ \ \ \ \ \ \ \ \ \ \ \ \ \ \ \ \ \ \ \ \ \ \ \ \ \ \ \ \ \ \ \ \ \ \ \ \ \ \ \ \ \ \ \ \ \ \ \ \ \ \ \ \ \ \ \ \ \ \ \ \ \ \ \ \ 
\end{aligned}
\right.
\end{flalign*}

\vspace{10pt}

\noindent H-8. {\em Hyperbolic-null elliptic web I}
\begin{flalign*}
\left \{
\begin{aligned}
&ds^2 = du^2 + \cosh^2{u}\, (\sec^2{v} - \sech^2{w})(dv^2 + dw^2) \\
&t + x = \cosh{u}\, \sec{v}\, \sech{w}, \ \ \ \ t - x = \cosh{u}\, \cos{v}\, \cosh{w}\, (1 + \tan^2{v} \tanh^2{w}), \\
&y = \cosh{u}\, \tan{v}\, \tanh{w}, \ \ \ \ z = \sinh{u} \\
&-\infty < u < \infty, \ \ 0 < v < \frac{\pi}{2}, \ \ 0 < w < \infty\ \ \ \ \ \ \ \ \ \ \ \ \  \ \ \ \ \ \ \ \ \ \ \ \ \ \ \ \ \ \ \ \ \ \ \ \ \ \ \ \ \ \ \ \ \  \ \ \ \ \ \ \ \ \ \ \ \ \ \ \  \ \ \ \ \ \ \ \ \ \ \ \ \ \ \ \ \ \ \ \ \ \ \ \ \ \ \ \ \ \ \ \ \ \ \ \ \ \ \ \ \ \ \ \ \ \ \ \ \ \ \ \ \ \ \ \ \ \ \ \ \ \ \ \ \ \ \ \ \ \ \ \ \ \ \ \ \ \ \ \ \ \ \ \ \ \ \ \ \ \ \ \ \ \ \  \ \ \ \ \ \ \ \ \ \ \ \ \ \ \  \ \ \ \ \ \ \ \ \ \ \ \ \ \ \ \ \ \ \ \ \ \ \ \ \ \ \ \ \ \ \ \ \ \ \ \ \ \ \ \ \ \ \ \ \ \ \ \ \ \ \ \ \ \ \ \ \ \ \ \ \ \ \ \ \ \ \ \ \ \ \ \ \ \ \ \ \ \ \ \ \ 
\end{aligned}
\right.
\end{flalign*}

\vspace{10pt}

\noindent H-9. {\em Hyperbolic-null elliptic web II}
\begin{flalign*}
\left \{
\begin{aligned}
&ds^2 = du^2 + \cosh^2{u}\, (\textrm{csch}^2v + \textrm{sec}^2w)(dv^2 + dw^2) \\
&t + x = \cosh{u}\, \csch{v}\, \sec{w}, \ \ t - x = \cosh{u}\, \sinh{v}\, \cos{w}\, (1 + \textrm{coth}^2v \ \textrm{tan}^2w), \\
&y = \cosh{u}\, \coth{v}\, \tan{w}, \ \ \ \ z = \sinh{u} \\
&-\infty < u < \infty, \ \ 0 < v < \infty, \ \ 0 < w < \frac{\pi}{2} \ \ \ \ \ \ \ \ \ \ \ \  \ \ \ \ \ \ \ \ \ \ \ \ \ \ \ \ \ \ \ \ \ \ \ \ \ \ \ \ \ \ \ \ \  \ \ \ \ \ \ \ \ \ \ \ \ \ \ \  \ \ \ \ \ \ \ \ \ \ \ \ \ \ \ \ \ \ \ \ \ \ \ \ \ \ \ \ \ \ \ \ \ \ \ \ \ \ \ \ \ \ \ \ \ \ \ \ \ \ \ \ \ \ \ \ \ \ \ \ \ \ \ \ \ \ \ \ \ \ \ \ \ \ \ \ \ \ \ \ \ \ \ \ \ \ \ \ \ \ \ \ \ \ \  \ \ \ \ \ \ \ \ \ \ \ \ \ \ \  \ \ \ \ \ \ \ \ \ \ \ \ \ \ \ \ \ \ \ \ \ \ \ \ \ \ \ \ \ \ \ \ \ \ \ \ \ \ \ \ \ \ \ \ \ \ \ \ \ \ \ \ \ \ \ \ \ \ \ \ \ \ \ \ \ \ \ \ \ \ \ \ \ \ \ \ \ \ \ \ \ 
\end{aligned}
\right.
\end{flalign*}

\vspace{10pt}

\noindent H-10. {\em Null rotational web I}
\begin{flalign*}
\left \{
\begin{aligned}
&ds^2 = du^2 + \cosh^2{u}\, (dv^2 + e^{2v}dw^2) \\
&t+x = \cosh{u}\, e^v, \ \ t-x = \cosh{u}\, (e^{-v} + w^2e^v), \\
&y = \cosh{u}\, we^v, \ \ \ \ z = \sinh{u} \\
&-\infty < u < \infty, \ \ -\infty < v < \infty, \ \ -\infty < w < \infty \ \ \ \ \ \ \ \ \ \ \ \  \ \ \ \ \ \ \ \ \ \ \ \ \ \ \ \ \ \ \ \ \ \ \ \ \ \ \ \ \ \ \ \ \  \ \ \ \ \ \ \ \ \ \ \ \ \ \ \  \ \ \ \ \ \ \ \ \ \ \ \ \ \ \ \ \ \ \ \ \ \ \ \ \ \ \ \ \ \ \ \ \ \ \ \ \ \ \ \ \ \ \ \ \ \ \ \ \ \ \ \ \ \ \ \ \ \ \ \ \ \ \ \ \ \ \ \ \ \ \ \ \ \ \ \ \ \ \ \ \ \ \ \ \ \ \ \ \ \ \ \ \ \ \  \ \ \ \ \ \ \ \ \ \ \ \ \ \ \  \ \ \ \ \ \ \ \ \ \ \ \ \ \ \ \ \ \ \ \ \ \ \ \ \ \ \ \ \ \ \ \ \ \ \ \ \ \ \ \ \ \ \ \ \ \ \ \ \ \ \ \ \ \ \ \ \ \ \ \ \ \ \ \ \ \ \ \ \ \ \ \ \ \ \ \ \ \ \ \ \ 
\end{aligned}
\right.
\end{flalign*}

\vspace{10pt}

\noindent H-11. {\em Hyperbolic-null elliptic web III}
\begin{flalign*}
\left \{
\begin{aligned}
&ds^2 = du^2 + \cosh^2{u}\, (v^{-2} + w^{-2})(dv^2 + dw^2) \\
&t+x = \frac{\cosh{u}}{vw}, \ \ t-x = \frac{\cosh{u}(v^2 + w^2)^2}{4vw}, \\
&y = \frac{\cosh{u}(w^2 - v^2)}{2vw}, \ \ \ \ z = \sinh{u} \\
&-\infty < u < \infty, \ \ 0 < v < \infty, \ \ 0 < w < \infty \ \ \ \ \ \ \ \ \ \ \ \  \ \ \ \ \ \ \ \ \ \ \ \ \ \ \ \ \ \ \ \ \ \ \ \ \ \ \ \ \ \ \ \ \  \ \ \ \ \ \ \ \ \ \ \ \ \ \ \  \ \ \ \ \ \ \ \ \ \ \ \ \ \ \ \ \ \ \ \ \ \ \ \ \ \ \ \ \ \ \ \ \ \ \ \ \ \ \ \ \ \ \ \ \ \ \ \ \ \ \ \ \ \ \ \ \ \ \ \ \ \ \ \ \ \ \ \ \ \ \ \ \ \ \ \ \ \ \ \ \ \ \ \ \ \ \ \ \ \ \ \ \ \ \  \ \ \ \ \ \ \ \ \ \ \ \ \ \ \  \ \ \ \ \ \ \ \ \ \ \ \ \ \ \ \ \ \ \ \ \ \ \ \ \ \ \ \ \ \ \ \ \ \ \ \ \ \ \ \ \ \ \ \ \ \ \ \ \ \ \ \ \ \ \ \ \ \ \ \ \ \ \ \ \ \ \ \ \ \ \ \ \ \ \ \ \ \ \ \ \ 
\end{aligned}
\right.
\end{flalign*}

\vspace{10pt}

\noindent {\em Restriction to }dS$_3$ \\

\noindent For $\psi_1$, we have that $N_0(1)$ is isometric to an open subset of dS$_1$, and the restriction of $\psi_1$ gives a warped product in the region where $-t^2 + x^2 + y^2 < 0$. For $\psi_2$, we have that $N_0(1)$ is isometric to an open subset of $\mathbb{S}^1$, and the restriction of $\psi_2$ gives a warped product in the region where $-t^2 + x^2 + y^2 > 0$. The restriction of $A$ induces the standard coordinate on dS$_1$ or $\mathbb{S}^1$ respectively. We then get nine separable webs, corresponding to the nine possible webs we may lift from dS$_2$. For a list of the nine webs and their adapted coordinates on dS$_2$, see \cite{Rajaratnam2016}. \\

\noindent dS-3. {\em de Sitter-elliptic web I}

\vspace{10pt}
\noindent for $-t^2 + x^2 + y^2 < 0$
\begin{flalign*}
\left \{
\begin{aligned}
&ds^2 = -du^2 + \sinh^2{u}\, (a^2 \cd^2(v;a) + \cs^2(w;b))(dv^2 + dw^2) \\
&t = \sinh{u}\, \nd(v;a)\, \ns(w;b), \ \ \ \ x = \sinh{u}\, \sd(v;a)\, \ds(w;b), \\
&y = \sinh{u}\, \cd(v;a)\, \cs(w;b), \ \ \ \ z = \cosh{u} \\
&0 < u < \infty, \ \ \ \ 0 < v < K(a), \ \ \ \ 0 < w < K(b), \ \ \ \ a^2 + b^2 = 1 \ \ \ \ \ \ \ \ \ \ \ \ \ \ \ \ \ \ \ \ \ \ \ \ \ \ \ \ \ \ \ \ \ \ \ \ \ \ \ \ \ \ \ \ \ \ \ \ \ \ \ \ \ \ \ \ \ \ \ \ \ \ \ \ \ \ \ \ \ \ \ \ \ \ \  \ \ \ \ \ \ \ \ \ \ \ \ \ \ \  \ \ \ \ \ \ \ \ \ \ \ \ \ \ \ \ \ \ \ \ \ \ \ \ \ \ \ \ \ \ \ \ \ \ \ \ \ \ \ \ \ \ \ \ \ \ \ \ \ \ \ \ \ \ \ \ \ \ \ \ \ \ \ \ \ \ \ \ \ \ \ \ \ \ \ \ \ \ \ \ \ 
\end{aligned}
\right.
\end{flalign*}

\noindent for $-t^2 + x^2 + y^2 > 0$
\begin{flalign*}
\left \{
\begin{aligned}
&ds^2 = du^2 + \sin^2{u}\, (\dc^2(v;a) - a^2 \sn^2(w;a))(-dv^2 + dw^2) \\
&t = \sin{u}\, \sac(v;a)\, \dn(w;a), \ \ \ \ x = \sin{u}\, \nc(v;a)\, \cn(w;a), \\
&y = \sin{u}\, \dc(v;a)\, \sn(w;a), \ \ \ \ z = \cos{u} \\
&0 < u < \pi, \ \ \ \ 0 < v < K(a), \ \ \ \ 0 < w < K(a), \ \ \ \ a^2 + b^2 = 1 \ \ \ \ \ \ \ \ \ \ \ \ \ \ \ \ \ \ \ \ \ \ \ \ \ \ \ \ \ \ \ \ \ \ \ \ \ \ \ \ \ \ \ \ \ \ \ \ \ \ \ \ \ \ \ \ \ \ \ \ \ \ \ \ \ \ \ \ \ \ \ \ \ \ \  \ \ \ \ \ \ \ \ \ \ \ \ \ \ \  \ \ \ \ \ \ \ \ \ \ \ \ \ \ \ \ \ \ \ \ \ \ \ \ \ \ \ \ \ \ \ \ \ \ \ \ \ \ \ \ \ \ \ \ \ \ \ \ \ \ \ \ \ \ \ \ \ \ \ \ \ \ \ \ \ \ \ \ \ \ \ \ \ \ \ \ \ \ \ \ \ 
\end{aligned}
\right.
\end{flalign*}

\vspace{10pt}

\noindent dS-4. {\em de Sitter-elliptic web II}

\vspace{10pt}
\noindent for $-t^2 + x^2 + y^2 < 0$
\begin{flalign*}
\left \{
\begin{aligned}
&ds^2 = -du^2 + \sinh^2{u}\, (\dc^2(v;a) + a^2 \sac^2(w;b))(dv^2 + dw^2) \\
&t = \sinh{u}\, \nc(v;a)\, \nc(w;b), \ \ \ \ x = \sinh{u}\, \sac(v;a)\, \dc(w;b), \\
&y = \sinh{u}\, \dc(v;a)\, \sac(w;b), \ \ \ \ z = \cosh{u} \\
&0 < u < \infty, \ \ \ \ 0 < v < K(a), \ \ \ \ 0 < w < K(b), \ \ \ \ a^2 + b^2 = 1 \ \ \ \ \ \ \ \ \ \ \ \ \ \ \ \ \ \ \ \ \ \ \ \ \ \ \ \ \ \ \ \ \ \ \ \ \ \ \ \ \ \ \ \ \ \ \ \ \ \ \ \ \ \ \ \ \ \ \ \ \ \ \ \ \ \ \ \ \ \ \ \ \ \ \  \ \ \ \ \ \ \ \ \ \ \ \ \ \ \  \ \ \ \ \ \ \ \ \ \ \ \ \ \ \ \ \ \ \ \ \ \ \ \ \ \ \ \ \ \ \ \ \ \ \ \ \ \ \ \ \ \ \ \ \ \ \ \ \ \ \ \ \ \ \ \ \ \ \ \ \ \ \ \ \ \ \ \ \ \ \ \ \ \ \ \ \ \ \ \ \ 
\end{aligned}
\right.
\end{flalign*}

\noindent for $-t^2 + x^2 + y^2 > 0, \ \ a|t| - |x| > b\sqrt{-t^2 + x^2 + y^2}$
\begin{flalign*}
\left \{
\begin{aligned}
&ds^2 = du^2 + \sin^2{u}\, (\dc^2(v;a) - \dc^2(w;a))(-dv^2 + dw^2) \\
&t = a^{-1}b\, \sin{u}\, \nc(v;a)\, \nc(w;a), \ \ \ \ x = b\, \sin{u}\, \sac(v;a)\, \sac(w;a), \\
&y = a^{-1}\, \sin{u}\, \dc(v;a)\, \dc(w;a), \ \ \ \ z = \cos{u} \\
&0 < u < \pi, \ \ \ \ 0 < w < v < K(a), \ \ \ \ a^2 + b^2 = 1 \ \ \ \ \ \ \ \ \ \ \ \ \ \ \ \ \ \ \ \ \ \ \ \ \ \ \ \ \ \ \ \ \ \ \ \ \ \ \ \ \ \ \ \ \ \ \ \ \ \ \ \ \ \ \ \ \ \ \ \ \ \ \ \ \ \ \ \ \ \ \ \ \ \ \  \ \ \ \ \ \ \ \ \ \ \ \ \ \ \  \ \ \ \ \ \ \ \ \ \ \ \ \ \ \ \ \ \ \ \ \ \ \ \ \ \ \ \ \ \ \ \ \ \ \ \ \ \ \ \ \ \ \ \ \ \ \ \ \ \ \ \ \ \ \ \ \ \ \ \ \ \ \ \ \ \ \ \ \ \ \ \ \ \ \ \ \ \ \ \ \ 
\end{aligned}
\right.
\end{flalign*}

\noindent for $-t^2 + x^2 + y^2 > 0, \ \ a|t| + |x| < b\sqrt{-t^2 + x^2 + y^2}$
\begin{flalign*}
\left \{
\begin{aligned}
&ds^2 = du^2 + a^2 \sin^2{u}\, (\nd^2(v;b) - \nd^2(w;b))(dv^2 - dw^2) \\
&t = ab\, \sin{u}\, \sd(v;b)\, \sd(w;b), \ \ \ \ x = b\, \sin{u}\, \cd(v;a)\, \cd(w;a), \\
&y = a\, \sin{u}\, \nd(v;a)\, \nd(w;a), \ \ \ \ z = \cos{u} \\
&0 < u < \pi, \ \ \ \ 0 < w < v < K(a), \ \ \ \ a^2 + b^2 = 1 \ \ \ \ \ \ \ \ \ \ \ \ \ \ \ \ \ \ \ \ \ \ \ \ \ \ \ \ \ \ \ \ \ \ \ \ \ \ \ \ \ \ \ \ \ \ \ \ \ \ \ \ \ \ \ \ \ \ \ \ \ \ \ \ \ \ \ \ \ \ \ \ \ \ \  \ \ \ \ \ \ \ \ \ \ \ \ \ \ \  \ \ \ \ \ \ \ \ \ \ \ \ \ \ \ \ \ \ \ \ \ \ \ \ \ \ \ \ \ \ \ \ \ \ \ \ \ \ \ \ \ \ \ \ \ \ \ \ \ \ \ \ \ \ \ \ \ \ \ \ \ \ \ \ \ \ \ \ \ \ \ \ \ \ \ \ \ \ \ \ \ 
\end{aligned}
\right.
\end{flalign*}

\vspace{10pt}

\noindent dS-5. {\em Spacelike rotational web II}

\vspace{10pt}
\noindent for $-t^2 + x^2 + y^2 < 0$
\begin{flalign*}
\left \{
\begin{aligned}
&ds^2 = -du^2 + \sinh^2{u}\, (dv^2 + \sinh^2{v}\, dw^2) \\
&t = \sin{u}\, \cosh{v}, \ \ \ \ x = \sin{u}\, \sinh{v}\, \cos{w}, \\
&y = \sin{u}\, \sinh{v}\, \sin{w}, \ \ \ \ z = \cos{u} \\
&0 < u < \infty, \ \ \ \ 0 < v < \infty, \ \ \ \ 0 < w < 2\pi \ \ \ \ \ \ \ \ \ \ \ \ \ \ \ \ \ \ \ \ \ \ \ \ \ \ \ \ \ \ \ \ \ \ \ \ \ \ \ \ \ \ \ \ \ \ \ \ \ \ \ \ \ \ \ \ \ \ \ \ \ \ \ \ \ \ \ \ \ \ \ \ \ \ \  \ \ \ \ \ \ \ \ \ \ \ \ \ \ \  \ \ \ \ \ \ \ \ \ \ \ \ \ \ \ \ \ \ \ \ \ \ \ \ \ \ \ \ \ \ \ \ \ \ \ \ \ \ \ \ \ \ \ \ \ \ \ \ \ \ \ \ \ \ \ \ \ \ \ \ \ \ \ \ \ \ \ \ \ \ \ \ \ \ \ \ \ \ \ \ \ 
\end{aligned}
\right.
\end{flalign*}

\noindent for $-t^2 + x^2 + y^2 > 0$
\begin{flalign*}
\left \{
\begin{aligned}
&ds^2 = du^2 + \sin^2{u}\, (-dv^2 + \cosh^2{v}\, dw^2) \\
&t = \sin{u}\, \sinh{v}, \ \ \ \ x = \sin{u}\, \cosh{v}\, \cos{w}, \\
&y = \sin{u}\, \cosh{v}\, \sin{w}, \ \ \ \ z = \cos{u} \\
&0 < u < \pi, \ \ \ \ 0 < v < \infty, \ \ \ \ 0 < w < 2\pi \ \ \ \ \ \ \ \ \ \ \ \ \ \ \ \ \ \ \ \ \ \ \ \ \ \ \ \ \ \ \ \ \ \ \ \ \ \ \ \ \ \ \ \ \ \ \ \ \ \ \ \ \ \ \ \ \ \ \ \ \ \ \ \ \ \ \ \ \ \ \ \ \ \ \  \ \ \ \ \ \ \ \ \ \ \ \ \ \ \  \ \ \ \ \ \ \ \ \ \ \ \ \ \ \ \ \ \ \ \ \ \ \ \ \ \ \ \ \ \ \ \ \ \ \ \ \ \ \ \ \ \ \ \ \ \ \ \ \ \ \ \ \ \ \ \ \ \ \ \ \ \ \ \ \ \ \ \ \ \ \ \ \ \ \ \ \ \ \ \ \ 
\end{aligned}
\right.
\end{flalign*}

\vspace{10pt}

\noindent dS-6. {\em Timelike rotational web I}

\vspace{10pt}
\noindent for $-t^2 + x^2 + y^2 < 0$
\begin{flalign*}
\left \{
\begin{aligned}
&ds^2 = -du^2 + \sinh^2{u}\, (dv^2 + \cosh^2{v}\, dw^2) \\
&t = \sinh{u}\, \cosh{v}\, \cosh{w}, \ \ \ \ x = \sinh{u}\, \cosh{v}\, \sinh{w}, \\
&y = \sinh{u}\, \sinh{v}, \ \ \ \ z = \cosh{u} \\
&0 < u < \infty, \ \ \ \ -\infty < v < \infty, \ \ \ \ -\infty < w < \infty \ \ \ \ \ \ \ \ \ \ \ \ \ \ \ \ \ \ \ \ \ \ \ \ \ \ \ \ \ \ \ \ \ \ \ \ \ \ \ \ \ \ \ \ \ \ \ \ \ \ \ \ \ \ \ \ \ \ \ \ \ \ \ \ \ \ \ \ \ \ \ \ \ \ \ \ \ \ \ \ \ \ \ \ \ \ \ \ \ \ \ \ \ \ \ \ \ \ \ \ \ \ \ \ \ \ \ \ \ \ \ \ \ \ \ \ \ \ \ \ \ \  
\end{aligned}
\right.
\end{flalign*}

\noindent for $-t^2 + x^2 + y^2 > 0, \ \ -t^2 + x^2 > 0$
\begin{flalign*}
\left \{
\begin{aligned}
&ds^2 = du^2 + \sin^2{u}\, (dv^2 - \sin^2{v}\, dw^2) \\
&t = \sin{u}\, \sin{v}\, \sinh{w}, \ \ \ \ x = \sin{u}\, \sin{v}\, \cosh{w}, \\
&y = \sin{u}\, \cos{v}, \ \ \ \ z = \cos{u} \\
&0 < u < \pi, \ \ \ \ 0 < v < \pi, \ \ \ \ - \infty < w < \infty \ \ \ \ \ \ \ \ \ \ \ \ \ \ \ \ \ \ \ \ \ \ \ \ \ \ \ \ \ \ \ \ \ \ \ \ \ \ \ \ \ \ \ \ \ \ \ \ \ \ \ \ \ \ \ \ \ \ \ \ \ \ \ \ \ \ \ \ \ \ \ \ \ \ \ \ \ \ \ \ \ \ \ \ \ \ \ \ \ \ \ \ \ \ \ \ \ \ \ \ \ \ \ \ \ \ \ \ \ \ \ \ \ \ \ \ \ \ \ \ \ \ 
\end{aligned}
\right.
\end{flalign*}

\noindent for $-t^2 + x^2 + y^2 > 0, \ \ -t^2 + x^2 < 0$
\begin{flalign*}
\left \{
\begin{aligned}
&ds^2 = du^2 + \sin^2{u}\, (-dv^2 + \sinh^2{v}\, dw^2) \\
&t = \sin{u}\, \sinh{v}\, \cosh{w}, x = \sin{u}\, \sinh{v}\, \sinh{w}, \\
&y = \sin{u}\, \cosh{v}, \ \ \ \ z = \cos{u} \\
&0 < u < \pi, \ \ \ \ 0 < v < \infty, \ \ \ \ - \infty < w < \infty \ \ \ \ \ \ \ \ \ \ \ \ \ \ \ \ \ \ \ \ \ \ \ \ \ \ \ \ \ \ \ \ \ \ \ \ \ \ \ \ \ \ \ \ \ \ \ \ \ \ \ \ \ \ \ \ \ \ \ \ \ \ \ \ \ \ \ \ \ \ \ \ \ \ \ \ \ \ \ \ \ \ \ \ \ \ \ \ \ \ \ \ \ \ \ \ \ \ \ \ \ \ \ \ \ \ \ \ \ \ \ \ \ \ \ \ \ \ \ \ \ \ \ \ \ \ \ \ \ \ \ \ \ \ \ \ \ \ \ \ \ \ \ \ \ \ \ \ \ \ \ \ \ \ \ \ \ \ \ \ \ \ \ \ \ \ \ \ \ \ \ \ \ \ \ \ \ \
\end{aligned}
\right.
\end{flalign*}

\vspace{10pt}

\noindent dS-7. {\em de Sitter-complex elliptic web}

\vspace{10pt}
\noindent for $-t^2 + x^2 + y^2 < 0$
\begin{flalign*}
\left \{
\begin{aligned}
&ds^2 = -du^2 + \sinh^2{u}\, (\sn^2(v;a)\, \dc^2(v;a) + \sn^2(w;b)\, \dc^2(w;b))(dv^2 + dw^2) \\
&t^2 + x^2 = \frac{2 \sinh^2{u}\, \dn(2v;a)\, \dn(2w;b)}{ab(1+\cn(2v;a))(1+\cn(2w;b))}, \ \ \ \ t^2 - x^2 = \frac{2 \sinh^2{u}\, (1+\cn(2v;a)\, \cn(2w;b))}{(1+\cn(2v;a))(1+\cn(2w;a))}, \\
&y = \sin{u}\, \sn(v;a)\, \dc(v;a)\, \sn(w;a)\, \dc(w;a), \ \ \ \ z = \cos{u} \\
&0 < u < \infty, \ \ \ \ 0 < v < K(a), \ \ \ \ 0 < w < K(b) \ \ \ \ \ \ \ \ \ \ \ \ \ \ \ \ \ \ \ \ \ \ \ \ \ \ \ \ \ \ \ \ \ \ \ \ \ \ \ \ \ \ \ \ \ \ \ \ \ \ \ \ \ \ \ \ \ \ \ \ \ \ \ \ \ \ \ \ \ \ \ \ \ \ \ \ \ \ \ \ \ \ \ \ \ \ \ \ \ \ \ \ \ \ \ \ \ \ \ \ \ \ \ \ \ \ \ \ \ \ \ \ \ \ \ \ \
\end{aligned}
\right.
\end{flalign*}

\noindent for $-t^2 + x^2 + y^2 > 0$
\begin{flalign*}
\left \{
\begin{aligned}
&ds^2 = du^2 + \sin^2{u}\, (\sn^2(v;a)\, \dc^2(v;a) - \sn^2(w;a)\ \dc^2(w;a))(-dv^2 + dw^2) \\
&t^2 + x^2 = \frac{2 \sin^2{u}\, \dn(2v;a)\, \dn(2w;a)}{ab(1+\cn(2v;a))(1+\cn(2w;a))}, \ \ \ \ -t^2 + x^2 = \frac{2 \sin^2{u}\, (\cn(2v;a) + \cn(2w;a))}{(1+\cn(2v;a))(1+\cn(2w;a))}, \\
&y = \sin{u}\, \sn(v;a)\ \dc(v;a)\, \sn(w;a)\, \dc(w;a), \ \ \ \ z = \cos{u} \\
&0 < u < \pi, \ \ \ \ 0 < w < v < K(a) \ \ \ \ \ \ \ \ \ \ \ \ \ \ \ \ \ \ \ \ \ \ \ \ \ \ \ \ \ \ \ \ \ \ \ \ \ \ \ \ \ \ \ \ \ \ \ \ \ \ \ \ \ \ \ \ \ \ \ \ \ \ \ \ \ \ \ \ \ \ \ \ \ \ \ \ \ \ \ \ \ \ \ \ \ \ \ \ \ \ \ \ \ \ \ \ \ \ \ \ \ \ \ \ \ \ \ \ \ \ \ \ \ \ \ \ \ \ \ \ \ \ \ \ \ \ \ \ \ \ \ \ \ \ \ \ \ \ \ \ \ \ \ \ \ \ \ \ \ \ \ \ \ \ \ \ \ \ \ \ \ \ \ \ \ \ \ \ \ \ \ \ \ \ \ \ \ \ \ \ \ \ \ 
\end{aligned}
\right.
\end{flalign*}

\vspace{10pt}

\noindent dS-8. {\em de Sitter-null elliptic web I}

\vspace{10pt}
\noindent for $-t^2 + x^2 + y^2 < 0$
\begin{flalign*}
\left \{
\begin{aligned}
&ds^2 = -du^2 + \sinh^2{u}\, (\sec^2{v} - \sech^2{w})(dv^2 + dw^2) \\
&t + x = \sinh{u}\, \sec{v}\, \sech{w}, \ \ \ \ t - x = \sinh{u}\, \cos{v}\, \cosh{w}\, (1 + \tan^2{v}\, \tanh^2{w}), \\
&y = \sinh{u}\, \tan{v}\, \tanh{w}, \ \ \ \ z = \cosh{u} \\
&0 < u < \infty, \ \ \ \ 0 < v < \frac{\pi}{2}, \ \ \ \ 0 < w < \infty \ \ \ \ \ \ \ \ \ \ \ \ \ \ \ \ \ \ \ \ \ \ \ \ \ \ \ \ \ \ \ \ \ \ \ \ \ \ \ \ \ \ \ \ \ \ \ \ \ \ \ \ \ \ \ \ \ \ \ \ \ \ \ \ \ \ \ \ \ \ \ \ \ \ \ \ \ \ \ \ \ \ \ \ \ \ \ \ \ \ \ \ \ \ \ \ \ \ \ \ \ \ \ \ \ \ \ \ \ \ \ \ \ \ \ \ \ \ \
\end{aligned}
\right.
\end{flalign*}

\noindent for $-t^2 + x^2 + y^2 > 0$
\begin{flalign*}
\left \{
\begin{aligned}
&ds^2 = du^2 + \sin^2{u}\, (\sech^2{v} + \csch^2{w})(dv^2 - dw^2) \\
&t + x = \sin{u}\, \sech{v}\, \csch{w}, \ \ \ \ t - x = - \sin{u}\, \cosh{v}\, \sinh{w}\, (1 - \tanh^2{v}\, \coth^2{w}), \\
&y = \sin{u}\, \tanh{v}\, \coth{w}, \ \ \ \ z = \cos{u} \\
&0 < u < \pi, \ \ \ \ 0 < v < \infty, \ \ \ \ 0 < w < \infty \ \ \ \ \ \ \ \ \ \ \ \ \ \ \ \ \ \ \ \ \ \ \ \ \ \ \ \ \ \ \ \ \ \ \ \ \ \ \ \ \ \ \ \ \ \ \ \ \ \ \ \ \ \ \ \ \ \ \ \ \ \ \ \ \ \ \ \ \ \ \ \ \ \ \ \ \ \ \ \ \ \ \ \ \ \ \ \ \ \ \ \ \ \ \ \ \ \ \ \ \ \ \ \ \ \ \ \ \ \ \ \ \ \ \ \ \ \ \
\end{aligned}
\right.
\end{flalign*}

\vspace{10pt}

\noindent dS-9. {\em de Sitter-null elliptic web II}

\vspace{10pt}
\noindent for $-t^2 + x^2 + y^2 < 0$
\begin{flalign*}
\left \{
\begin{aligned}
&ds^2 = -du^2 + \sinh^2{u}\, (\csch^2{v} + \sec^2{w})(dv^2 + dw^2) \\
&t + x = \sinh{u}\, \csch{v}\, \sec{w}, \ \ \ \ t - x = \sinh{u}\, \sinh{v}\, \cos{w}\, (1 + \coth^2{v}\, \tan^2{w}), \\
&y = \sinh{u}\, \coth{v}\, \tan{w}, \ \ \ \ z = \cosh{u} \\
&0 < u < \infty, \ \ \ \ 0 < v < \infty, \ \ \ \ 0 < w < \frac{\pi}{2} \ \ \ \ \ \ \ \ \ \ \ \ \ \ \ \ \ \ \ \ \ \ \ \ \ \ \ \ \ \ \ \ \ \ \ \ \ \ \ \ \ \ \ \ \ \ \ \ \ \ \ \ \ \ \ \ \ \ \ \ \ \ \ \ \ \ \ \ \ \ \ \ \ \ \ \ \ \ \ \ \ \ \ \ \ \ \ \ \ \ \ \ \ \ \ \ \ \ \ \ \ \ \ \ \ \ \ \ \ \ \ \ \ \ \ \ \ \ \ \ \ \ \ \ \ \ \ \ \ \ \ \ \ \ \ \ \ \ \ \ \ \ \ \ \ \ \ \ \ \ \ \ \ \ \ 
\end{aligned}
\right.
\end{flalign*}

\noindent for $-t^2 + x^2 + y^2 > 0, \ \ |x| > \sqrt{-t^2 + x^2 + y^2}, \ \ tx > 0$
\begin{flalign*}
\left \{
\begin{aligned}
&ds^2 = du^2 + \sin^2{u}\, (\sec^2{v} - \sec^2{w})(-dv^2 + dw^2) \\
&t + x = \sin{u}\, \sec{v}\, \sec{w}, \ \ \ \ t - x = - \sin{u}\, \cos{v}\, \cos{w}\, (1 - \tan^2{v}\, \tan^2{w}), \\
&y = \sin{u}\, \tan{v}\, \tan{w}, \ \ \ \ z = \cos{u} \\
&0 < u < \pi, \ \ \ \ 0 < w < v < \frac{\pi}{2} \ \ \ \ \ \ \ \ \ \ \ \ \ \ \ \ \ \ \ \ \ \ \ \ \ \ \ \ \ \ \ \ \ \ \ \ \ \ \ \ \ \ \ \ \ \ \ \ \ \ \ \ \ \ \ \ \ \ \ \ \ \ \ \ \ \ \ \ \ \ \ \ \ \ \ \ \ \ \ \ \ \ \ \ \ \ \ \ \ \ \ \ \ \ \ \ \ \ \ \ \ \ \ \ \ \ \ \ \ \ \ \ \ \ \ \ \ \ \ \ \ \ \ \ \ \ \ \ \ \ \ \ \ \ \ \ \ \ \ \ \ \ \ \ \ \ \ \ \ \ \ \ \ \ \ 
\end{aligned}
\right.
\end{flalign*}

\noindent for $-t^2 + x^2 + y^2 > 0, \ \ |x| > \sqrt{-t^2 + x^2 + y^2}, \ \ tx < 0, \ \ |y| > \sqrt{-t^2 + x^2 + y^2}$
\begin{flalign*}
\left \{
\begin{aligned}
&ds^2 = du^2 + \sin^2{u}\, (\csch^2{v} - \csch^2{w})(dv^2 - dw^2) \\
&t + x = \sin{u}\, \csch{v}\, \csch{w}, \ \ \ \ t - x = - \sin{u}\, \sinh{v}\, \sinh{w}\, (1 - \coth^2{v}\, \coth^2{w}), \\
&y = \sin{u}\, \coth{v}\, \coth{w}, \ \ \ \ z = \cos{u} \\
&0 < u < \pi, \ \ \ \ 0 < w < v < \infty \ \ \ \ \ \ \ \ \ \ \ \ \ \ \ \ \ \ \ \ \ \ \ \ \ \ \ \ \ \ \ \ \ \ \ \ \ \ \ \ \ \ \ \ \ \ \ \ \ \ \ \ \ \ \ \ \ \ \ \ \ \ \ \ \ \ \ \ \ \ \ \ \ \ \ \ \ \ \ \ \ \ \ \ \ \ \ \ \ \ \ \ \ \ \ \ \ \ \ \ \ \ \ \ \ \ \ \ \ \ \ \ \ \ \ \ \ \ \ \ \ \ \ \ \ \ \ \ \ \ \ \ \ \ \ \ \ \ \ \ \ \ \ \ \ \ \ \ \ \ \ \ \ \ \ 
\end{aligned}
\right.
\end{flalign*}

\noindent for $-t^2 + x^2 + y^2 > 0, \ \ |x| > \sqrt{-t^2 + x^2 + y^2}, \ \ tx < 0, \ \ |y| < \sqrt{-t^2 + x^2 + y^2}$
\begin{flalign*}
\left \{
\begin{aligned}
&ds^2 = du^2 + \sin^2{u}\, (\sech^2{v} - \sech^2{w})(dv^2 - dw^2) \\
&t + x = \sin{u}\, \sech{v}\, \sech{w}, \ \ \ \ t - x = - \sin{u}\, \cosh{v}\, \cosh{w}\, (1 - \tanh^2{v}\, \tanh^2{w}), \\
&y = \sin{u}\, \tanh{v}\, \tanh{w}, \ \ \ \ z = \cos{u} \\
&0 < u < \pi, \ \ \ \ 0 < v < w < \infty \ \ \ \ \ \ \ \ \ \ \ \ \ \ \ \ \ \ \ \ \ \ \ \ \ \ \ \ \ \ \ \ \ \ \ \ \ \ \ \ \ \ \ \ \ \ \ \ \ \ \ \ \ \ \ \ \ \ \ \ \ \ \ \ \ \ \ \ \ \ \ \ \ \ \ \ \ \ \ \ \ \ \ \ \ \ \ \ \ \ \ \ \ \ \ \ \ \ \ \ \ \ \ \ \ \ \ \ \ \ \ \ \ \ \ \ \ \ \ \ \ \ \ \ \ \ \ \ \ \ \ \ \ \ \ \ \ \ \ \ \ \ \ \ \ \ \ \ \ \ \ \ \ \ \ 
\end{aligned}
\right.
\end{flalign*}

\vspace{10pt}

\noindent dS-10. {\em Null rotational web I}

\vspace{10pt}
\noindent for $-t^2 + x^2 + y^2 < 0$
\begin{flalign*}
\left \{
\begin{aligned}
&ds^2 = -du^2 + \sinh^2{u}\, (dv^2 + e^{2v} dw^2) \\
&t + x = e^v \sinh{u}, \ \ \ \ t - x = \sinh{u}\, (e^{-v} + w^2e^v) \\
&y = w e^v \sinh{u}, \ \ \ \ z = \cosh{u} \\
&0 < u < \infty, \ \ \ \ -\infty < v < \infty, \ \ \ \ - \infty < w < \infty \ \ \ \ \ \ \ \ \ \ \ \ \ \ \ \ \ \ \ \ \ \ \ \ \ \ \ \ \ \ \ \ \ \ \ \ \ \ \ \ \ \ \ \ \ \ \ \ \ \ \ \ \ \ \ \ \ \ \ \ \ \ \ \ \ \ \ \ \ \ \ \ \ \ \ \ \ \ \ \ \ \ \ \ \ \ \ \ \ \ \ \ \ \ \ \ \ \ \ \ \ \ \ \ \ \ \ \ \ \ \ \ \ \ \ \ \
\end{aligned}
\right.
\end{flalign*}

\noindent for $-t^2 + x^2 + y^2 > 0$
\begin{flalign*}
\left \{
\begin{aligned}
&ds^2 = du^2 + \sin^2{u}\, (- dv^2 + e^{2v} dw^2) \\
&t + x = \sin{u}\, (e^{-v} - w^2 e^v), \ \ \ \ t - x = - e^v \sin{u}, \\
&y = w e^v \sin{u}, \ \ \ \ z = \cos{u} \\
&0 < u < \pi, \ \ \ \ -\infty < v < \infty, \ \ \ \ - \infty < w < \infty \ \ \ \ \ \ \ \ \ \ \ \ \ \ \ \ \ \ \ \ \ \ \ \ \ \ \ \ \ \ \ \ \ \ \ \ \ \ \ \ \ \ \ \ \ \ \ \ \ \ \ \ \ \ \ \ \ \ \ \ \ \ \ \ \ \ \ \ \ \ \ \ \ \ \ \ \ \ \ \ \ \ \ \ \ \ \ \ \ \ \ \ \ \ \ \ \ \ \ \ \ \ \ \ \ \ \ \ \ \ \ \ \ \ \ \ \
\end{aligned}
\right.
\end{flalign*}

\vspace{10pt}

\noindent dS-11. {\em de Sitter-null elliptic web III}

\vspace{10pt}
\noindent for $-t^2 + x^2 + y^2 > 0$
\begin{flalign*}
\left \{
\begin{aligned}
&ds^2 = du^2 + \sin^2{u}\, (v^{-2} - w^{-2})(-dv^2 + dw^2) \\
&t + x = \frac{\sin{u}}{vw}, \ \ \ \ t - x = \frac{\sin{u}\, (v^2 - w^2)^2}{4vw}, \\
&y = \frac{\sin{u}\, (v^2 + w^2)}{2vw}, \ \ \ \ z = \cos{u} \\
&0 < u < \pi, \ \ \ \ 0 < v < w < \infty \ \ \ \ \ \ \ \ \ \ \ \ \ \ \ \ \ \ \ \ \ \ \ \ \ \ \ \ \ \ \ \ \ \ \ \ \ \ \ \ \ \ \ \ \ \ \ \ \ \ \ \ \ \ \ \ \ \ \ \ \ \ \ \ \ \ \ \ \ \ \ \ \ \ \ \ \ \ \ \ \ \ \ \ \ \ \ \ \ \ \ \ \ \ \ \ \ \ \ \ \ \ \ \ \ \ \ \ \ \ \ \ \ \ \ \ \ \ \ \ \ \ \ \ \ \ \ \ \ \
\end{aligned}
\right.
\end{flalign*}

\noindent for $-t^2 + x^2 + y^2 < 0$
\begin{flalign*}
\left \{
\begin{aligned}
&ds^2 = -du^2 + \sinh^2{u}\, (v^{-2} + w^{-2})(dv^2 + dw^2) \\
&t + x = \frac{\sinh{u}}{vw}, \ \ \ \ t - x = \frac{\sinh{u}\, (v^2 + w^2)^2}{4vw}, \\
&y = \frac{\sinh{u}\, (w^2 - v^2)}{2vw}, \ \ \ \ z = \cosh{u} \\
&0 < u < \pi, \ \ \ \ 0 < v < \infty, \ \ \ \ 0 < w < \infty \ \ \ \ \ \ \ \ \ \ \ \ \ \ \ \ \ \ \ \ \ \ \ \ \ \ \ \ \ \ \ \ \ \ \ \ \ \ \ \ \ \ \ \ \ \ \ \ \ \ \ \ \ \ \ \ \ \ \ \ \ \ \ \ \ \ \ \ \ \ \ \ \ \ \ \ \ \ \ \ \ \ \ \ \ \ \ \ \ \ \ \ \ \ \ \ \ \ \ \ \ \ \ \ \ \ 
\end{aligned}
\right.
\end{flalign*}

\vspace{10pt}

\noindent {\bf 4.3} \ $A = J_{-1}(0) \oplus J_1(0) \oplus J_1(1) \oplus J_1(1)$
\vspace{10pt}

\noindent If $A$ has a two-dimensional Lorentzian eigenspace and a two-dimensional spacelike eigenspace, then the associated CT is reducible, and we may choose pseudo-Cartesian coordinates such that $A$ takes the above form. Algorithm \ref{WP} again gives two different warped products which decompose $A + r \odot r$, depending on our choice of eigenvector in the Lorentzian eigenspace of $A$. As in section 4.2, these two warped products will map into different regions of $\mathbb{E}^4_1$.

Let us first choose a timelike eigenvector of $A$, say $-\partial_t$, and a spacelike eigenvector, say $\partial_z$, in the construction of algorithm \ref{WP}. By equation \eqref{non-nullWP}, the resulting warped product $\psi_1$ is
\begin{align*}
\psi_1 &: N_0 \times_{\rho_1} \mathbb{H}^1 \times_{\rho_2} \mathbb{S}^1 \rightarrow \mathbb{E}^4_1 \\
							&(-\tilde{t} \partial_t + \tilde{z} \partial_z, p_1,p_2) \mapsto \tilde{t}p_1 + \tilde{z}p_2
\end{align*}
where $N_0 = \{-\tilde{t} \partial_t + \tilde{z} \partial_z \in \mathbb{E}^4_1 \ | \ \tilde{t} > 0, \ \tilde{z} > 0 \}$, $\rho_1(-\tilde{t} \partial_t + \tilde{z} \partial_z) = \tilde{t}$ and $\rho_2(-\tilde{t} \partial_t + \tilde{z} \partial_z) = \tilde{z}$. By equation \eqref{nonnullimage} and the remarks following it, the image of $\psi_1$ consists of all points $(t,x,y,z)$ such that $-t^2 + x^2 < 0$ and $t > 0$. If instead we choose two spacelike eigenvectors of $A$, say $\partial_x$ and $\partial_z$, in algorithm \ref{WP}, we find that the resulting warped product $\psi_2$ is
\begin{align*}
\psi_2 &: N_0 \times_{\rho_1} \mathrm{dS}_1 \times_{\rho_2} \mathbb{S}^1 \rightarrow \mathbb{E}^4_1 \\
							&(\tilde{x} \partial_x + \tilde{z} \partial_z, p_1,p_2) \mapsto \tilde{x}p_1 + \tilde{z}p_2
\end{align*}
where $N_0 = \{\tilde{x} \partial_x + \tilde{z} \partial_z \in \mathbb{E}^4_1 \ | \ \tilde{x} > 0, \ \tilde{z} > 0 \}$, $\rho_1(\tilde{x} \partial_x + \tilde{z} \partial_z) = \tilde{x}$ and $\rho_2(\tilde{x} \partial_x + \tilde{z} \partial_z) = \tilde{z}$. By equation \eqref{nonnullimage}, the image of $\psi_2$ consists of all points $(t,x,y,z)$ such that $-t^2 + x^2 > 0$. \\

\noindent {\em Restriction to $\mathbb{H}^3$} \\

\noindent First note that $\psi_2$ does not restrict to a warped product of $\mathbb{H}^3$. Now, for $\psi_1$, $N_0$ is isometric to an open subset of $\mathbb{E}^2_1$, and so $N_0(-1)$ is isometric to an open subset of $\mathbb{H}^1$. The restriction of $A$ induces the standard coordinate $u$ on $\mathbb{H}^1$. We then get the following web, upon lifting the standard coordinates $v$ and $w$ from $\mathbb{H}^1$ and $\mathbb{S}^1$ respectively. \\

\noindent H-12. {\em Spacelike-timelike rotational web}
\begin{flalign*}
\left \{
\begin{aligned}
&ds^2 = du^2 + \cosh^2{u}\, dv^2 + \sinh^2{u}\, dw^2 \\
&t=\cosh{u}\, \cosh{v}, \ \ \ \ x=\cosh{u}\, \sinh{v}, \\
&y=\sinh{u}\, \sin{w}, \ \ \ \ z=\sinh{u}\, \cos{w} \\ 
&0 < u < \infty, \ \ \ \ -\infty < v < \infty, \ \ \ \ 0 < w < 2\pi, \ \ \ \ \ \ \ \ \ \ \ \  \ \ \ \ \ \ \ \ \ \ \ \ \ \ \  \ \ \ \ \ \ \ \ \ \ \ \ \ \ \ \ \ \  \ \ \ \ \ \ \ \ \ \ \ \ \ \ \  \ \ \ \ \ \ \ \ \ \ \ \ \ \ \ \ \ \ \ \ \ \ \ \ \ \ \ \ \ \ \ \ \ \ \ \ \ \ \ \ \ \ \ \ \ \ \ \ \ \ \ \ \ \ \ \ \ \ \ \ \ \ \ \ \ \ \ \ \ \ \ \ \ \ \ \ \ \ \ \ \ \ \ \ \ \ \ \ \ \ \ \ \ \ \  \ \ \ \ \ \ \ \ \ \ \ \ \ \ \  \ \ \ \ \ \ \ \ \ \ \ \ \ \ \ \ \ \ \ \ \ \ \ \ \ \ \ \ \ \ \ \ \ \ \ \ \ \ \ \ \ \ \ \ \ \ \ \ \ \ \ \ \ \ \ \ \ \ \ \ \ \ \ \ \ \ \ \ \ \ \ \ \ \ \ \ \ \ \ \ \ 
\end{aligned}
\right.
\end{flalign*}

\vspace{10pt}

\noindent {\em Restriction to }dS$_3$ \\

\noindent For $\psi_1$, we have that $N_0(1)$ is isometric to an open subset of dS$_1$, and the restriction of $\psi_1$ gives a warped product in the region where $-t^2 + x^2 < 0$ and $t > 0$. For $\psi_2$, we have that $N_0(1)$ is isometric to an open subset of $\mathbb{S}^1$, and the restriction of $\psi_2$ gives a warped product in the region where $-t^2 + x^2 > 0$. The restriction of $A$ induces the standard coordinate on dS$_1$ or $\mathbb{S}^1$ respectively. Upon lifting the standard coordinates from the other factors, we get \\

\noindent dS-12. {\em Spacelike-timelike rotational web}

\vspace{10pt}
\noindent for $-t^2 + x^2 < 0$
\begin{flalign*}
\left \{
\begin{aligned}
&ds^2 = -du^2 + \sinh^2{u}\, dv^2 + \cosh^2{u}\, dw^2 \\
&t = \sinh{u}\, \cosh{v}, \ \ \ \ x = \sinh{u}\, \sinh{v}, \\
&y = \cosh{u}\, \sin{w}, \ \ \ \ z = \cosh{u}\, \cos{w} \\
&0 < u < \infty, \ \ \ \ -\infty < v < \infty, \ \ \ \ 0 < w < 2\pi, \ \ \ \ \ \ \ \ \ \ \ \ \ \ \ \ \ \ \ \ \ \ \ \ \ \ \ \ \ \ \ \ \ \ \ \ \ \ \ \ \ \ \ \ \ \ \ \ \ \ \ \ \ \ \ \ \ \ \ \ \ \ \ \ \ \ \ \ \ \ \ \ \ \ \ \  \ \ \ \ \ \ \ \ \ \ \ \ \ \ \  \ \ \ \ \ \ \ \ \ \ \ \ \ \ \ \ \ \ \ \ \ \ \ \ \ \ \ \ \ \ \ \ \ \ \ \ \ \ \ \ \ \ \ \ \ \ \ \ \ \ \ \ \ \ \ \ \ \ \ \ \ \ \ \ \ \ \ \ \ \ \ \ \ \ \ \ \ \ \ \ \ 
\end{aligned}
\right.
\end{flalign*}

\noindent for $-t^2 + x^2 > 0$
\begin{flalign*}
\left \{
\begin{aligned}
&ds^2 = du^2 - \cos^2{u}\, dv^2 + \sin^2{u}\, dw^2 \\
&t = \cos{u}\, \sinh{v}, \ \ \ \ x = \cos{u}\, \cosh{v}, \\
&y = \sin{u}\, \sin{w}, \ \ \ \ z = \sin{u}\, \cos{w} \\
&0 < u < \frac{\pi}{2}, \ \ \ \ -\infty < v < \infty, \ \ \ \ 0 < w < 2\pi, \ \ \ \ \ \ \ \ \ \ \ \ \ \ \ \ \ \ \ \ \ \ \ \ \ \ \ \ \ \ \ \ \ \ \ \ \ \ \ \ \ \ \ \ \ \ \ \ \ \ \ \ \ \ \ \ \ \ \ \ \ \ \ \ \ \ \ \ \ \ \ \ \ \ \ \  \ \ \ \ \ \ \ \ \ \ \ \ \ \ \  \ \ \ \ \ \ \ \ \ \ \ \ \ \ \ \ \ \ \ \ \ \ \ \ \ \ \ \ \ \ \ \ \ \ \ \ \ \ \ \ \ \ \ \ \ \ \ \ \ \ \ \ \ \ \ \ \ \ \ \ \ \ \ \ \ \ \ \ \ \ \ \ \ \ \ \ \ \ \ \ \ 
\end{aligned}
\right.
\end{flalign*}

\vspace{10pt}

\noindent {\bf 4.4} \ $A = J_{-1}(0) \oplus J_1(0) \oplus J_1(a^2) \oplus J_1(1), \ \ \ \ 0 < a < 1$
\vspace{10pt}

\noindent Let us now consider the case where $A$ takes the above form up to geometric equivalence. As $A$ has a two-dimensional Lorentzian eigenspace, the associated CT is reducible. We consider the construction given by algorithm \ref{WP} for the timelike and spacelike cases. 

Let us first choose a timelike eigenvector of $A$, say $-\partial_t$ in the construction of algorithm \ref{WP}. By equation \eqref{non-nullWP}, the resulting warped product $\psi_1$ is
\begin{align*}
\psi_1 &: N_0 \times_{\rho} \mathbb{H}^1 \rightarrow \mathbb{E}^4_1 \\
							&(-\tilde{t} \partial_t + y \partial_y + z \partial_z, p) \mapsto y\partial_y + z\partial_z + \tilde{t}p
\end{align*}
where $N_0 = \{-\tilde{t} \partial_t + y\partial_y + z \partial_z \in \mathbb{E}^4_1 \ | \ \tilde{t} > 0 \}$ and $\rho(-\tilde{t} \partial_t + y\partial_y + z\partial_z) = \tilde{t}$. By equation \eqref{nonnullimage} and the remarks following it, the image of $\psi_1$ consists of all points $(t,x,y,z)$ such that $-t^2 + x^2 < 0$ and $t > 0$. If instead we choose a spacelike eigenvector of $A$, say $\partial_x$, in algorithm \ref{WP}, then we have that the resulting warped product $\psi_2$ is
\begin{align*}
\psi_2 &: N_0 \times_{\rho_1} \mathrm{dS}_1 \rightarrow \mathbb{E}^4_1 \\
							&(\tilde{x} \partial_x + y\partial_y + z \partial_z, p) \mapsto y\partial_y + z\partial_z + \tilde{x}p
\end{align*}
where $N_0 = \{\tilde{x} \partial_x + y\partial_y + z \partial_z \in \mathbb{E}^4_1 \ | \ \tilde{x} > 0 \}$ and $\rho(\tilde{x} \partial_x + y\partial_y + z \partial_z) = \tilde{x}$. By equation \eqref{nonnullimage}, the image of $\psi_2$ consists of all points $(t,x,y,z)$ such that $-t^2 + x^2 > 0$. \\

\noindent {\em Restriction to $\mathbb{H}^3$} \\

\noindent First note that $\psi_2$ does not restrict to a warped product of $\mathbb{H}^3$. Now, for $\psi_1$, $N_0$ is isometric to an open subset of $\mathbb{E}^3_1$, and so $N_0(-1)$ is isometric to an open subset of $\mathbb{H}^2$. The restriction of $A$ to $N_0$ is given by $\diag(0,a^2,1)$, which induces on $\mathbb{H}^2$ the elliptic web of type I (see appendix \ref{appB}). Upon lifting these coordinates to $\mathbb{H}^3$ via $\psi$ we get \\

\noindent H-13. {\em Timelike rotational web II}
\begin{flalign*}
\left \{
\begin{aligned}
&ds^2 = (a^2 \cd^2(v;a) + \cs^2(w;b))(dv^2 + dw^2) + \nd^2(v;a)\, \ns^2(w;b)\, du^2 \\
&t=\nd(v;a)\, \ns(w;b)\, \cosh{u}, \ \ \ \ x=\nd(v;a)\, \ns(w;b)\, \sinh{u}, \\
&y=\sd(v;a)\, \ds(w;b), \ \ \ \ z=\cd(v;a)\, \cs(w;b)\, \\ 
&-\infty < u < \infty, \ \ \ \ 0 < v < K(a), \ \ \ \ 0 < w < K(b), \ \ \ \ \ \ \ \ \ \ \ \  \ \ \ \ \ \ \ \ \ \ \ \ \ \ \  \ \ \ \ \ \ \ \ \ \ \ \ \ \ \ \ \ \  \ \ \ \ \ \ \ \ \ \ \ \ \ \ \  \ \ \ \ \ \ \ \ \ \ \ \ \ \ \ \ \ \ \ \ \ \ \ \ \ \ \ \ \ \ \ \ \ \ \ \ \ \ \ \ \ \ \ \ \ \ \ \ \ \ \ \ \ \ \ \ \ \ \ \ \ \ \ \ \ \ \ \ \ \ \ \ \ \ \ \ \ \ \ \ \ \ \ \ \ \ \ \ \ \ \ \ \ \ \  \ \ \ \ \ \ \ \ \ \ \ \ \ \ \  \ \ \ \ \ \ \ \ \ \ \ \ \ \ \ \ \ \ \ \ \ \ \ \ \ \ \ \ \ \ \ \ \ \ \ \ \ \ \ \ \ \ \ \ \ \ \ \ \ \ \ \ \ \ \ \ \ \ \ \ \ \ \ \ \ \ \ \ \ \ \ \ \ \ \ \ \ \ \ \ \ 
\end{aligned}
\right.
\end{flalign*}

\vspace{10pt}

\noindent {\em Restriction to }dS$_3$ \\

\noindent For $\psi_1$, $N_0(1)$ is isometric to an open subset of dS$_2$, and the restriction of $\psi_1$ gives a warped product in the region where $-t^2 + x^2 < 0$ and $t > 0$. For $\psi_2$, $N_0(1)$ is isometric to an open subset of $\mathbb{S}^2$, and the restriction of $\psi_2$ gives a warped product in the region where $-t^2 + x^2 > 0$. The restriction of $A$ to $N_0$ is given by $\diag(0,a^2,1)$, which induces the Neumann web on $\mathbb{S}^2$, and the elliptic web of type I on dS$_2$. We therefore get \\

\noindent dS-13. {\em Timelike rotational web II}

\vspace{10pt}
\noindent for $-t^2 + x^2 < 0$
\begin{flalign*}
\left \{
\begin{aligned}
&ds^2 = (\dc^2(u;a) - a^2 \sn^2(v;a))(-du^2 + dv^2) + \sac^2(u;a)\, \dn^2(v;a)\, dw^2 \\
&t = \sac(u;a)\, \dn(v;a)\, \cosh{w}, \ \ \ \ x = \sac(u;a)\, \dn(v;a)\, \sinh{w}, \\
&y = \nc(u;a)\, \cn(v;a), \ \ \ \ z = \dc(u;a)\, \sn(v;a) \\
&0 < u < K(a), \ \ \ \ 0 < v < K(a), \ \ \ \ -\infty < w < \infty, \ \ \ \ 0 < a < 1 \ \ \ \ \ \ \ \ \ \ \ \ \ \ \ \ \ \ \ \ \ \ \ \ \ \ \ \ \ \ \ \ \ \ \ \ \ \ \ \ \ \ \ \ \ \ \ \ \ \ \ \ \ \ \ \ \ \ \ \ \ \ \ \ \ \ \ \ \ \ \ \  \ \ \ \ \ \ \ \ \ \ \ \ \ \ \  \ \ \ \ \ \ \ \ \ \ \ \ \ \ \ \ \ \ \ \ \ \ \ \ \ \ \ \ \ \ \ \ \ \ \ \ \ \ \ \ \ \ \ \ \ \ \ \ \ \ \ \ \ \ \ \ \ \ \ \ \ \ \ \ \ \ \ \ \ \ \ \ \ \ \ \ \ \ \ \ \ 
\end{aligned}
\right.
\end{flalign*}

\noindent for $-t^2 + x^2 > 0$
\begin{flalign*}
\left \{
\begin{aligned}
&ds^2 = (a^2 \cn^2(u;a) + b^2 \cn^2(v;b))(du^2 + dv^2) - \sn^2(u;a)\, \dn^2(v;b)\, dw^2  \\
&t = \sn(u;a)\, \dn(v;b)\, \sinh{w}, \ \ \ \ x = \sn(u;a)\, \dn(v;b)\, \cosh{w}, \\
&y = \cn(v;a)\, \cn(w;b), \ \ \ \ z = \dn(v;a)\, \sn(w;b) \\
&0 < u < 2K(a), \ \ \ \ -K(b) < v < K(b), \ \ \ \ -\infty < w < \infty, \ \ \ \ a^2 + b^2 = 1 \ \ \ \ \ \ \ \ \ \ \ \ \ \ \ \ \ \ \ \ \ \ \ \ \ \ \ \ \ \ \ \ \ \ \ \ \ \ \ \ \ \ \ \ \ \ \ \ \ \ \ \ \ \ \ \ \ \ \ \ \ \ \ \ \ \ \ \ \ \ \ \ \ \ \ \  \ \ \ \ \ \ \ \ \ \ \ \ \ \ \  \ \ \ \ \ \ \ \ \ \ \ \ \ \ \ \ \ \ \ \ \ \ \ \ \ \ \ \ \ \ \ \ \ \ \ \ \ \ \ \ \ \ \ \ \ \ \ \ \ \ \ \ \ \ \ \ \ \ \ \ \ \ \ \ \ \ \ \ \ \ \ \ \ \ \ \ \ \ \ \ \ 
\end{aligned}
\right.
\end{flalign*}

\vspace{10pt}

\noindent {\bf 4.5} \ $A = J_{-1}(0) \oplus J_1(0) \oplus J_1(-a^2) \oplus J_1(1), \ \ \ \ 0 < a < 1$
\vspace{10pt}

\noindent Let us now consider the case where $A$ takes the above form up to geometric equivalence. As $A$ has a two-dimensional Lorentzian eigenspace, the associated CT is reducible. Note that the construction given in algorithm \ref{WP} yields precisely the same warped products $\psi_1$ and $\psi_2$ as in section 4.4. Indeed this case differs from the one above only in the restrictions of $A$ to $N_0$. \\

\noindent {\em Restriction to $\mathbb{H}^3$} \\

\noindent As above, $\psi_2$ does not restrict to a warped product of $\mathbb{H}^3$. For $\psi_1$ the restriction of $A$ to $N_0$ has the form $\diag(0,-a^2,1)$, which is geometrically equivalent (in $N_0$) to $\diag(\tilde{a}^2,0,1)$, where $\tilde{a}^2 = a^2(1+a^2)^{-1}$ and $0 < \tilde{a} < 1$. This $A$ induces the elliptic web of type II on $N_0(-1) \cong \mathbb{H}^2$. So, dropping the tilde on $\tilde{a}$ and lifting the induced coordinates to $\mathbb{H}^3$, we have \\

\noindent H-14. {\em Timelike rotational web III}
\begin{flalign*}
\left \{
\begin{aligned}
&ds^2 = (\dc^2(v;a) + a^2 \sac^2(w;b))(dv^2 + dw^2) + \nc^2(v;a)\, \nc^2(w;b)\, du^2 \\
&t=\nc(v;a)\, \nc(w;b)\, \cosh{u}, \ \ \ \ x=\nc(v;a)\, \nc(w;b)\, \sinh{u}, \\
&y=\sac(v;a)\, \dc(w;b), \ \ \ \ z=\dc(v;a)\, \sac(w;b)\, \\ 
&-\infty < u < \infty, \ \ \ \ 0 < v < K(a), \ \ \ \ 0 < w < K(b), \ \ \ \ \ \ \ \ \ \ \ \  \ \ \ \ \ \ \ \ \ \ \ \ \ \ \  \ \ \ \ \ \ \ \ \ \ \ \ \ \ \ \ \ \  \ \ \ \ \ \ \ \ \ \ \ \ \ \ \  \ \ \ \ \ \ \ \ \ \ \ \ \ \ \ \ \ \ \ \ \ \ \ \ \ \ \ \ \ \ \ \ \ \ \ \ \ \ \ \ \ \ \ \ \ \ \ \ \ \ \ \ \ \ \ \ \ \ \ \ \ \ \ \ \ \ \ \ \ \ \ \ \ \ \ \ \ \ \ \ \ \ \ \ \ \ \ \ \ \ \ \ \ \ \  \ \ \ \ \ \ \ \ \ \ \ \ \ \ \  \ \ \ \ \ \ \ \ \ \ \ \ \ \ \ \ \ \ \ \ \ \ \ \ \ \ \ \ \ \ \ \ \ \ \ \ \ \ \ \ \ \ \ \ \ \ \ \ \ \ \ \ \ \ \ \ \ \ \ \ \ \ \ \ \ \ \ \ \ \ \ \ \ \ \ \ \ \ \ \ \ 
\end{aligned}
\right.
\end{flalign*}

\vspace{10pt}

\noindent {\em Restriction to }dS$_3$ \\

\noindent For $\psi_1$, $\psi_2$, $N_0(1)$ is isometric to an open subset of dS$_2$, $\mathbb{S}^2$ respectively. The restriction of $A$ to $N_0$ has the form $\diag(0,-a^2,1)$, which as above is geometrically equivalent to $\diag(\tilde{a}^2, 0,1)$, with $\tilde{a}^2 = a^2(1+a^2)^{-1}$ and $0 < \tilde{a} < 1$. This $A$ induces the elliptic web of type II on dS$_2$, and the Neumann web on $\mathbb{S}^2$. Again dropping the tilde on $\tilde{a}$, we have \\

\noindent dS-14. {\em Timelike rotational web III}

\vspace{10pt}
\noindent for $-t^2 + x^2 < 0, \ \ a\sqrt{t^2 - x^2} - |y| > b$
\begin{flalign*}
\left \{
\begin{aligned}
&ds^2 = (\dc^2(u;a) - \dc^2(v;a))(-du^2 + dv^2) + a^{-2}b^2 \nc^2(u;a)\, \nc^2(v;a)\, dw^2 \\
&t = a^{-1}b\, \nc(u;a)\, \nc(v;a)\, \cosh{w}, \ \ \ \ x = \nc(u;a)\, \nc(v;a)\, \sinh{w}, \\
&y = b\, \sac(u;a)\, \sac(v;a), \ \ \ \ z = a^{-1} \dc(u;a)\, \dc(v;a) \\
&0 < v < u < K(a), \ \ \ \ -\infty < w < \infty, \ \ \ \ a^2 + b^2 = 1 \ \ \ \ \ \ \ \ \ \ \ \ \ \ \ \ \ \ \ \ \ \ \ \ \ \ \ \ \ \ \ \ \ \ \ \ \ \ \ \ \ \ \ \ \ \ \ \ \ \ \ \ \ \ \ \ \ \ \ \ \ \ \ \ \ \ \ \ \ \ \ \  \ \ \ \ \ \ \ \ \ \ \ \ \ \ \  \ \ \ \ \ \ \ \ \ \ \ \ \ \ \ \ \ \ \ \ \ \ \ \ \ \ \ \ \ \ \ \ \ \ \ \ \ \ \ \ \ \ \ \ \ \ \ \ \ \ \ \ \ \ \ \ \ \ \ \ \ \ \ \ \ \ \ \ \ \ \ \ \ \ \ \ \ \ \ \ \ 
\end{aligned}
\right.
\end{flalign*}

\noindent for $-t^2 + x^2 < 0, \ \ a\sqrt{t^2 - x^2} + |y| < b$
\begin{flalign*}
\left \{
\begin{aligned}
&ds^2 = a^2(\nd^2(u;b) - \nd^2(v;b))(du^2 - dv^2) + a^{2}b^2 \sd^2(u;b)\, \sd^2(v;b)\, dw^2 \\
&t = ab\, \sd(u;b)\, \sd(v;b)\, \cosh{w}, \ \ \ \ x = \sd(u;b)\, \sd(v;b)\, \sinh{w}, \\
&y = b\, \cd(u;b)\, \cd(v;b), \ \ \ \ z = a\, \nd(u;b)\, \nd(v;b) \\
&0 < v < u < K(b), \ \ \ \ -\infty < w < \infty, \ \ \ \ a^2 + b^2 = 1 \ \ \ \ \ \ \ \ \ \ \ \ \ \ \ \ \ \ \ \ \ \ \ \ \ \ \ \ \ \ \ \ \ \ \ \ \ \ \ \ \ \ \ \ \ \ \ \ \ \ \ \ \ \ \ \ \ \ \ \ \ \ \ \ \ \ \ \ \ \ \ \  \ \ \ \ \ \ \ \ \ \ \ \ \ \ \  \ \ \ \ \ \ \ \ \ \ \ \ \ \ \ \ \ \ \ \ \ \ \ \ \ \ \ \ \ \ \ \ \ \ \ \ \ \ \ \ \ \ \ \ \ \ \ \ \ \ \ \ \ \ \ \ \ \ \ \ \ \ \ \ \ \ \ \ \ \ \ \ \ \ \ \ \ \ \ \ \ 
\end{aligned}
\right.
\end{flalign*}

\noindent for $-t^2 + x^2 > 0$
\begin{flalign*}
\left \{
\begin{aligned}
&ds^2 = (a^2 \cn^2(u;a) +  b^2 \cn^2(v;b))(du^2 + dv^2) - \cn^2(u;a)\, \cn^2(v;b)\, dw^2 \\
&t = \cn(u;a)\, \cn(v;b)\, \sinh{w}, \ \ \ \ x = \cn(u;a)\, \cn(v;b)\, \cosh{w}, \\
&y = \sn(u;a)\, \dn(v;b), \ \ \ \ z = \dn(u;a)\, \sn(v;b) \\
&-K(a) < u < K(a), \ \ \ \ -K(b) < v < K(b), \ \ \ \ -\infty < w < \infty, \ \ \ \ a^2 + b^2 = 1 \ \ \ \ \ \ \ \ \ \ \ \ \ \ \ \ \ \ \ \ \ \ \ \ \ \ \ \ \ \ \ \ \ \ \ \ \ \ \ \ \ \ \ \ \ \ \ \ \ \ \ \ \ \ \ \ \ \ \ \ \ \ \ \ \ \ \ \ \ \ \ \  \ \ \ \ \ \ \ \ \ \ \ \ \ \ \  \ \ \ \ \ \ \ \ \ \ \ \ \ \ \ \ \ \ \ \ \ \ \ \ \ \ \ \ \ \ \ \ \ \ \ \ \ \ \ \ \ \ \ \ \ \ \ \ \ \ \ \ \ \ \ \ \ \ \ \ \ \ \ \ \ \ \ \ \ \ \ \ \ \ \ \ \ \ \ \ \ 
\end{aligned}
\right.
\end{flalign*}
\vspace{10pt}

\vspace{10pt}

\noindent {\bf 4.6} \ $A = J_{-1}(1) \oplus J_1(a^2) \oplus J_1(0) \oplus J_1(0), \ \ \ \ 0 < a < 1$
\vspace{10pt}

\noindent Let us now consider the case where $A$ takes the above form up to geometric equivalence. As $A$ has a two-dimensional spacelike eigenspace, the associated CT is reducible. Choosing a unit vector in this eigenspace, say $\partial_y$, algorithm \ref{WP} yields a warped product $\psi$ which decomposes $A + r \odot r$ in the ambient space. By equation \eqref{non-nullWP}, $\psi$ is given by
\begin{align*}
\psi &: N_0 \times_{\rho} \mathbb{S}^1 \rightarrow \mathbb{E}^4_1 \\
							&(t \partial_t + x \partial_x + \tilde{y} \partial_y, p) \mapsto t\partial_t + x\partial_x + \tilde{y}p
\end{align*}
where $N_0 = \{t \partial_t + x\partial_x + \tilde{y} \partial_y \in \mathbb{E}^4_1 \ | \ \tilde{y} > 0 \}$ and $\rho(t \partial_t + x\partial_x + \tilde{y} \partial_y ) = \tilde{y}$. By equation \eqref{nonnullimage}, the image of $\psi$ is dense in $\mathbb{E}^4_1$. We consider the restrictions of $\psi$ to $N_0(-1)$ and $N_0(1)$ respectively. \\

\noindent {\em Restriction to $\mathbb{H}^3$} \\

\noindent $N_0(-1)$ is isometric to an open subset of $\mathbb{H}^2$. The restriction of $A$ to $N_0$ has the form $\diag(1,a^2,0)$, which induces the elliptic web of type I on $\mathbb{H}^2$. We therefore have  \\

\noindent H-15. {\em Spacelike rotational web III}
\begin{flalign*}
\left \{
\begin{aligned}
&ds^2 = (a^2\cd^2(v;a) + \cs^2(w;b))(dv^2 + dw^2) + \cd^2(v;a)\, \cs^2(w;b)\, du^2 \\
&t=\nd(v;a)\, \ns(w;b), \ \ \ \ x=\sd(v;a)\, \ds(w;b), \\
&y=\cd(v;a)\, \cs(w;b)\, \sin{u}, \ \ \ \ z=\cd(v;a)\, \cs(w;b)\, \cos{u} \\ 
&0 < u < 2\pi, \ \ \ \ 0 < v < K(a), \ \ \ \ 0 < w < K(b), \ \ \ \ a^2 + b^2 = 1 \ \ \ \ \ \ \ \  \ \ \ \ \ \ \ \ \ \ \ \ \ \ \  \ \ \ \ \ \ \ \ \ \ \ \ \ \ \ \ \ \  \ \ \ \ \ \ \ \ \ \ \ \ \ \ \  \ \ \ \ \ \ \ \ \ \ \ \ \ \ \ \ \ \ \ \ \ \ \ \ \ \ \ \ \ \ \ \ \ \ \ \ \ \ \ \ \ \ \ \ \ \ \ \ \ \ \ \ \ \ \ \ \ \ \ \ \ \ \ \ \ \ \ \ \ \ \ \ \ \ \ \ \ \ \ \ \ \ \ \ \ \ \ \ \ \ \ \ \ \ \  \ \ \ \ \ \ \ \ \ \ \ \ \ \ \  \ \ \ \ \ \ \ \ \ \ \ \ \ \ \ \ \ \ \ \ \ \ \ \ \ \ \ \ \ \ \ \ \ \ \ \ \ \ \ \ \ \ \ \ \ \ \ \ \ \ \ \ \ \ \ \ \ \ \ \ \ \ \ \ \ \ \ \ \ \ \ \ \ \ \ \ \ \ \ \ \ 
\end{aligned}
\right.
\end{flalign*}

\vspace{10pt}

\noindent {\em Restriction to }dS$_3$ \\

\noindent $N_0(1)$ is isometric to an open subset of dS$_2$. The restriction of $A$ to $N_0$ again has the form $\diag(1,a^2,0)$, which induces the elliptic web of type I on dS$_2$. We therefore have  \\

\noindent dS-15. {\em Spacelike rotational web III}
\begin{flalign*}
\left \{
\begin{aligned}
&ds^2 = (\dc^2(u;a) - a^2 \sn^2(v;a))(-du^2 + dv^2) + \dc^2(u;a)\, \sn^2(v;a)\, du^2 \\
&t=\sac(u;a)\, \dn(v;a), \ \ \ \ x=\nc(u;a)\, \cn(v;a), \\
&y=\dc(u;a)\, \sn(v;a)\, \sin{w}, \ \ \ \ z=\dc(u;a)\, \sn(v;a)\, \cos{w} \\ 
&0 < u < K(a), \ \ \ \ 0 < v < K(a), \ \ \ \ 0 < w < 2\pi \ \ \ \ \ \ \ \ \ \ \ \  \ \ \ \ \ \ \ \ \ \ \ \ \ \ \  \ \ \ \ \ \ \ \ \ \ \ \ \ \ \ \ \ \  \ \ \ \ \ \ \ \ \ \ \ \ \ \ \  \ \ \ \ \ \ \ \ \ \ \ \ \ \ \ \ \ \ \ \ \ \ \ \ \ \ \ \ \ \ \ \ \ \ \ \ \ \ \ \ \ \ \ \ \ \ \ \ \ \ \ \ \ \ \ \ \ \ \ \ \ \ \ \ \ \ \ \ \ \ \ \ \ \ \ \ \ \ \ \ \ \ \ \ \ \ \ \ \ \ \ \ \ \ \  \ \ \ \ \ \ \ \ \ \ \ \ \ \ \  \ \ \ \ \ \ \ \ \ \ \ \ \ \ \ \ \ \ \ \ \ \ \ \ \ \ \ \ \ \ \ \ \ \ \ \ \ \ \ \ \ \ \ \ \ \ \ \ \ \ \ \ \ \ \ \ \ \ \ \ \ \ \ \ \ \ \ \ \ \ \ \ \ \ \ \ \ \ \ \ \ 
\end{aligned}
\right.
\end{flalign*}

\vspace{10pt}

\noindent {\bf 4.7} \ $A = J_{-1}(a^2) \oplus J_1(1) \oplus J_1(0) \oplus J_1(0), \ \ \ \ 0 < a < 1$
\vspace{10pt}

\noindent Let us now consider the case where $A$ takes the above form up to geometric equivalence. Note that the construction of algorithm \ref{WP} for this case yields precisely the same warped product as above in section 4.6. Indeed this case differs from that one only in the restrictions of $A$ to $N_0$. \\

\noindent {\em Restriction to $\mathbb{H}^3$} \\

\noindent $N_0(-1)$ is isometric to an open subset of $\mathbb{H}^2$. The restriction of $A$ to $N_0$ has the form $\diag(a^2,1,0)$, which induces the elliptic web of type II on $\mathbb{H}^2$. We therefore have  \\

\noindent H-16. {\em Spacelike rotational web IV}
\begin{flalign*}
\left \{
\begin{aligned}
&ds^2 = (\dc^2(v;a) + a^2 \sac^2(w;b))(dv^2 + dw^2) + \dc^2(v;a)\, \sac^2(w;b)\, du^2 \\
&t=\nc(v;a)\, \nc(w;b), \ \ \ \ x=\sac(v;a)\, \dc(w;b), \\
&y=\dc(v;a)\, \sac(w;b)\, \sin{u}, \ \ \ \ z=\dc(v;a)\, \sac(w;b)\, \cos{u} \\ 
&0 < u < 2\pi, \ \ \ \ 0 < v < K(a), \ \ \ \ 0 < w < K(b), \ \ \ \ a^2 + b^2 = 1 \ \ \ \ \ \ \ \  \ \ \ \ \ \ \ \ \ \ \ \ \ \ \  \ \ \ \ \ \ \ \ \ \ \ \ \ \ \ \ \ \  \ \ \ \ \ \ \ \ \ \ \ \ \ \ \  \ \ \ \ \ \ \ \ \ \ \ \ \ \ \ \ \ \ \ \ \ \ \ \ \ \ \ \ \ \ \ \ \ \ \ \ \ \ \ \ \ \ \ \ \ \ \ \ \ \ \ \ \ \ \ \ \ \ \ \ \ \ \ \ \ \ \ \ \ \ \ \ \ \ \ \ \ \ \ \ \ \ \ \ \ \ \ \ \ \ \ \ \ \ \  \ \ \ \ \ \ \ \ \ \ \ \ \ \ \  \ \ \ \ \ \ \ \ \ \ \ \ \ \ \ \ \ \ \ \ \ \ \ \ \ \ \ \ \ \ \ \ \ \ \ \ \ \ \ \ \ \ \ \ \ \ \ \ \ \ \ \ \ \ \ \ \ \ \ \ \ \ \ \ \ \ \ \ \ \ \ \ \ \ \ \ \ \ \ \ \ 
\end{aligned}
\right.
\end{flalign*}

\vspace{10pt}

\noindent {\em Restriction to }dS$_3$ \\

\noindent $N_0(1)$ is isometric to an open subset of dS$_2$. The restriction of $A$ to $N_0$ again has the form $\diag(a^2,1,0)$, which induces the elliptic web of type II on dS$_2$. We therefore have  \\

\noindent dS-16. {\em Spacelike rotational web IV}

\vspace{10pt}
\noindent for $a|t| - |x| > b$
\begin{flalign*}
\left \{
\begin{aligned}
&ds^2 = (\dc^2(u;a) - \dc^2(v;a))(-du^2 + dv^2) + a^{-2} \dc^2(u;a)\, \dc^2(v;a)\, dw^2 \\
&t = a^{-1}b\, \nc(u;a)\, \nc(v;a), \ \ \ \ x = b\, \sac(u;a)\, \sac(v;a), \\
&y = a^{-1}\, \dc(u;a)\, \dc(v;a)\, \sin{w}, \ \ \ \ z = a^{-1} \dc(u;a)\, \dc(v;a)\, \cos{w} \\
&0 < v < u < K(a), \ \ \ \ 0 < w < 2\pi, \ \ \ \ a^2 + b^2 = 1 \ \ \ \ \ \ \ \ \ \ \ \ \ \ \ \ \ \ \ \ \ \ \ \ \ \ \ \ \ \ \ \ \ \ \ \ \ \ \ \ \ \ \ \ \ \ \ \ \ \ \ \ \ \ \ \ \ \ \ \ \ \ \ \ \ \ \ \ \ \ \ \  \ \ \ \ \ \ \ \ \ \ \ \ \ \ \  \ \ \ \ \ \ \ \ \ \ \ \ \ \ \ \ \ \ \ \ \ \ \ \ \ \ \ \ \ \ \ \ \ \ \ \ \ \ \ \ \ \ \ \ \ \ \ \ \ \ \ \ \ \ \ \ \ \ \ \ \ \ \ \ \ \ \ \ \ \ \ \ \ \ \ \ \ \ \ \ \ 
\end{aligned}
\right.
\end{flalign*}

\noindent for $a|t| + |x| < b$
\begin{flalign*}
\left \{
\begin{aligned}
&ds^2 = a^2(\nd^2(u;b) - \nd^2(v;b))(du^2 - dv^2) + a^{2} \nd^2(u;b)\, \nd^2(v;b)\, dw^2 \\
&t = ab\, \sd(u;b)\, \sd(v;b), \ \ \ \ x = b\, \cd(u;b)\, \cd(v;b), \\
&y = a\, \nd(u;b)\, \nd(v;b)\, \sin{w}, \ \ \ \ z = a\, \nd(u;b)\, \nd(v;b)\, \cos{w} \\
&0 < v < u < K(b), \ \ \ \ 0 < w < 2\pi, \ \ \ \ a^2 + b^2 = 1 \ \ \ \ \ \ \ \ \ \ \ \ \ \ \ \ \ \ \ \ \ \ \ \ \ \ \ \ \ \ \ \ \ \ \ \ \ \ \ \ \ \ \ \ \ \ \ \ \ \ \ \ \ \ \ \ \ \ \ \ \ \ \ \ \ \ \ \ \ \ \ \  \ \ \ \ \ \ \ \ \ \ \ \ \ \ \  \ \ \ \ \ \ \ \ \ \ \ \ \ \ \ \ \ \ \ \ \ \ \ \ \ \ \ \ \ \ \ \ \ \ \ \ \ \ \ \ \ \ \ \ \ \ \ \ \ \ \ \ \ \ \ \ \ \ \ \ \ \ \ \ \ \ \ \ \ \ \ \ \ \ \ \ \ \ \ \ \ 
\end{aligned}
\right.
\end{flalign*}

\vspace{10pt}

\noindent {\bf 4.8} \ $A = J_{-1}(1) \oplus J_1(-a^2) \oplus J_1(0) \oplus J_1(0), \ \ \ \ 0 < a < 1$
\vspace{10pt}

\noindent Let us now consider the case where $A$ takes the above form up to geometric equivalence. Note that the construction of algorithm \ref{WP} for this case yields precisely the same warped product as in the last two sections. Again, this case differs from that one only in the restrictions of $A$ to $N_0$. \\

\noindent {\em Restriction to $\mathbb{H}^3$} \\

\noindent $N_0(-1)$ is isometric to an open subset of $\mathbb{H}^2$. The restriction of $A$ to $N_0$ has the form $\diag(1,-a^2,0)$, which is geometrically equivalent (in $N_0$) to $\diag(1,0,\tilde{a}^2)$ where $\tilde{a}^2 = a^2(1+a^2)^{-1}$ and $0 < \tilde{a} < 1$. This $A$ induces the elliptic web of type I on $\mathbb{H}^2$. Note that this case differs from H-15 above since the warping functions are different. Dropping the tilde on $\tilde{a}$, we get \\

\noindent H-17. {\em Spacelike rotational web V}
\begin{flalign*}
\left \{
\begin{aligned}
&ds^2 = (a^2\cd^2(v;a) + \cs^2(w;b))(dv^2 + dw^2) + \sd^2(v;a)\, \ds^2(w;b)\, du^2 \\
&t=\nd(v;a)\, \ns(w;b), \ \ \ \ x=\cd(v;a)\, \cs(w;b), \\
&y=\sd(v;a)\, \ds(w;b)\, \sin{u}, \ \ \ \ z=\sd(v;a)\, \ds(w;b)\, \cos{u} \\ 
&0 < u < 2\pi, \ \ \ \ 0 < v < K(a), \ \ \ \ 0 < w < K(b), \ \ \ \ a^2 + b^2 = 1 \ \ \ \ \ \ \ \  \ \ \ \ \ \ \ \ \ \ \ \ \ \ \  \ \ \ \ \ \ \ \ \ \ \ \ \ \ \ \ \ \  \ \ \ \ \ \ \ \ \ \ \ \ \ \ \  \ \ \ \ \ \ \ \ \ \ \ \ \ \ \ \ \ \ \ \ \ \ \ \ \ \ \ \ \ \ \ \ \ \ \ \ \ \ \ \ \ \ \ \ \ \ \ \ \ \ \ \ \ \ \ \ \ \ \ \ \ \ \ \ \ \ \ \ \ \ \ \ \ \ \ \ \ \ \ \ \ \ \ \ \ \ \ \ \ \ \ \ \ \ \  \ \ \ \ \ \ \ \ \ \ \ \ \ \ \  \ \ \ \ \ \ \ \ \ \ \ \ \ \ \ \ \ \ \ \ \ \ \ \ \ \ \ \ \ \ \ \ \ \ \ \ \ \ \ \ \ \ \ \ \ \ \ \ \ \ \ \ \ \ \ \ \ \ \ \ \ \ \ \ \ \ \ \ \ \ \ \ \ \ \ \ \ \ \ \ \ 
\end{aligned}
\right.
\end{flalign*}

\vspace{10pt}

\noindent {\em Restriction to }dS$_3$ \\

\noindent $N_0(1)$ is isometric to an open subset of dS$_2$. The restriction of $A$ to $N_0$ again has the form $\diag(1,-a^2,0)$, which is equivalent to $\diag(1,0,\tilde{a}^2)$ with $\tilde{a}^2 = a^2(1+a^2)^{-1}$ and $0 < \tilde{a} < 1$. This $A$ induces the elliptic web of type I on dS$_2$. Again, this case differs from dS-15 above since the warping functions are different. Dropping the tilde on $\tilde{a}$, we get \\

\noindent dS-17. {\em Spacelike rotational web V}
\begin{flalign*}
\left \{
\begin{aligned}
&ds^2 = (\dc^2(u;a) - a^2 \sn^2(v;a))(-du^2 + dv^2) + \nc^2(u;a)\, \cn^2(v;a)\, du^2 \\
&t=\sac(u;a)\, \dn(v;a), \ \ \ \ x=\dc(u;a)\, \sn(v;a), \\
&y=\nc(u;a)\, \cn(v;a)\, \sin{w}, \ \ \ \ z=\nc(u;a)\, \cn(v;a)\, \cos{w} \\ 
&0 < u < K(a), \ \ \ \ 0 < v < K(a), \ \ \ \ 0 < w < 2\pi \ \ \ \ \ \ \ \ \ \ \ \  \ \ \ \ \ \ \ \ \ \ \ \ \ \ \  \ \ \ \ \ \ \ \ \ \ \ \ \ \ \ \ \ \  \ \ \ \ \ \ \ \ \ \ \ \ \ \ \  \ \ \ \ \ \ \ \ \ \ \ \ \ \ \ \ \ \ \ \ \ \ \ \ \ \ \ \ \ \ \ \ \ \ \ \ \ \ \ \ \ \ \ \ \ \ \ \ \ \ \ \ \ \ \ \ \ \ \ \ \ \ \ \ \ \ \ \ \ \ \ \ \ \ \ \ \ \ \ \ \ \ \ \ \ \ \ \ \ \ \ \ \ \ \  \ \ \ \ \ \ \ \ \ \ \ \ \ \ \  \ \ \ \ \ \ \ \ \ \ \ \ \ \ \ \ \ \ \ \ \ \ \ \ \ \ \ \ \ \ \ \ \ \ \ \ \ \ \ \ \ \ \ \ \ \ \ \ \ \ \ \ \ \ \ \ \ \ \ \ \ \ \ \ \ \ \ \ \ \ \ \ \ \ \ \ \ \ \ \ \ 
\end{aligned}
\right.
\end{flalign*}

\vspace{10pt}

\noindent {\bf 4.9} \ $A = J_{1}(i) \oplus J_1(-i) \oplus J_1(c) \oplus J_1(c), \ \ c \in \mathbb{R}$
\vspace{10pt}

\noindent Consider the case where $A$ takes the above form (in a complex basis) up to geometric equivalence. Since $A$ has a two-dimensional spacelike eigenspace, the induced CT is reducible. Note that the construction of algorithm \ref{WP} yields the same warped product as in sections 4.6-4.8 above. Again, the only difference in this case is the restriction of $A$. \\

\noindent {\em Restriction to $\mathbb{H}^3$} \\

\noindent $N_0(-1)$ is isometric to an open subset of $\mathbb{H}^2$. The restriction of $A$ to $N_0$ has the form $\diag(i,-i,c)$, which induces the complex elliptic web on $\mathbb{H}^2$. We therefore get \\

\noindent H-18. {\em Spacelike rotational web VI}
\begin{flalign*}
\left \{
\begin{aligned}
&ds^2 = (\sn^2(v;a)\, \dc^2(v;a) + \sn^2(w;b)\, \dc^2(w;b))(dv^2 + dw^2) + \rho^2(v,w)\, du^2 \\
&\rho(v,w) = \sn(v;a)\, \dc(v;a)\, \sn(w;b)\, \dc(w;b) \\
&t^2 + x^2 = \frac{2\, \dn(2v;a)\, \dn(2w;b)}{ab(1+\cn(2v;a))(1+\cn(2w;b))}, \ \ \ \ t^2 - x^2= \frac{2(1+\cn(2v;a)\cn(2w;b))}{(1+\cn(2v;a)(1+\cn(2w;b)}, \\
&y = \sn(v;a)\, \dc(v;a)\, \sn(w;b)\, \dc(w;b)\, \sin{u}, \ \ \ \ z = \sn(v;a)\, \dc(v;a)\, \sn(w;b)\, \dc(w;b)\, \cos{u} \\ 
&0 < u < 2\pi, \ \ \ \ 0 < v < K(a), \ \ \ \ 0 < w < K(b), \ \ \ \ a^2 + b^2 = 1 \ \ \ \ \ \ \ \ \ \ \ \  \ \ \ \ \ \ \ \ \ \ \ \ \ \ \  \ \ \ \ \ \ \ \ \ \ \ \ \ \ \ \ \ \ \ \ \ \ \ \ \ \ \ \ \ \ \ \ \ \ \ \ \ \ \ \ \ \ \ \ \ \ \ \ \ \ \ \ \  \ \ \ \ \ \ \ \ \ \ \ \ \ \ \  \ \ \ \ \ \ \ \ \ \ \ \ \ \ \ \ \ \ \ \ \ \ \ \ \ \ \ \ \ \ \ \ \ \ \ \ \ \ \ \ \ \ \ \ \ \ \ \ \ \ \ \ \ \ \ \ \ \ \ \ \ \ \ \ \ \ \ \ \ \ \ \ \ \ \ \ \ \ \ \ \ 
\end{aligned}
\right.
\end{flalign*}

\vspace{5pt}

\noindent {\em Restriction to }dS$_3$ \\

\noindent $N_0(1)$ is isometric to an open subset of dS$_2$. The restriction of $A$ to $N_0$ again has the form $\diag(i,-i,c)$, which induces the complex elliptic web on dS$_2$. Again, this case differs from dS-15 above since the warping functions are different. Dropping the tilde on $\tilde{a}$, we get \\

\noindent dS-18. {\em Spacelike rotational web VI}
\begin{flalign*}
\left \{
\begin{aligned}
&ds^2 = (\sn^2(u;a)\, \dc^2(u;a) - \sn^2(v;a)\, \dc^2(v;a))(-du^2 + dv^2) + \rho^2(u,v)\, dw^2 \\
&\rho(u,v) = \sn(u;a)\, \dc(u;a)\, \sn(v;a)\, \dc(v;a) \\
&t^2 + x^2 =\frac{2\, \dn(2u;a)\, \dn(2v;a)}{ab(1 + \cn(2u;a))(1 + \cn(2v;a))}, \ \ \ \ -t^2 + x^2= \frac{2(\cn(2u;a) + \cn(2v;a))}{(1 + \cn(2u;a))(1 + \cn(2v;a))}, \\
&y = \sn(u;a)\, \dc(u;a)\, \sn(v;a)\, \dc(v;a)\, \sin{w}, \ \ \ \ z = \sn(u;a)\, \dc(u;a)\, \sn(v;a)\, \dc(v;a)\, \cos{w} \\ 
&0 < u < K(a), \ \ \ \ 0 < v < K(a), \ \ \ \ 0 < w < 2\pi, \ \ \ \ a^2 + b^2 = 1 \ \ \ \ \ \ \ \ \ \ \ \  \ \ \ \ \ \ \ \ \ \ \ \ \ \ \  \ \ \ \ \ \ \ \ \ \ \ \ \ \ \ \ \ \  \ \ \ \ \ \ \ \ \ \ \ \ \ \ \  \ \ \ \ \ \ \ \ \ \ \ \ \ \ \ \ \ \ \ \ \ \ \ \ \ \ \ \ \ \ \ \ \ \ \ \ \ \ \ \ \ \ \ \ \ \ \ \ \ \ \ \ \ \ \ \ \ \ \ \ \ \ \ \ \ \ \ \ \ \ \ \ \ \ \ \ \ \ \ \ \ \ \ \ \ \ \ \ \ \ \ \ \ \ \  \ \ \ \ \ \ \ \ \ \ \ \ \ \ \  \ \ \ \ \ \ \ \ \ \ \ \ \ \ \ \ \ \ \ \ \ \ \ \ \ \ \ \ \ \ \ \ \ \ \ \ \ \ \ \ \ \ \ \ \ \ \ \ \ \ \ \ \ \ \ \ \ \ \ \ \ \ \ \ \ \ \ \ \ \ \ \ \ \ \ \ \ \ \ \ \ 
\end{aligned}
\right.
\end{flalign*}

\vspace{10pt}

\noindent {\bf 4.10} \ $A = J_{-1}(0) \oplus J_1(a) \oplus J_1(b) \oplus J_1(1), \ \ \ \ 0 < a < b < 1$
\vspace{10pt}

\noindent Let us now consider the case where $A$ takes the above form up to geometric equivalence. In this case, since $A$ has no multidimensional eigenspaces, the induced CT $L$ is irreducible. If we denote the eigenfunctions of $L$ by $u,v,w$, then in the notation of equations \eqref{ICTform}-\eqref{ICTmetric}, we have $k = 0$, and the characteristic polynomials of $L$ and $A$ are, respectively,
$$p(\zeta) = (\zeta - u)(\zeta - v)(\zeta - w), \ \ \ \ \ \ \ \ \ B(\zeta) = \zeta(\zeta - a)(\zeta - b)(\zeta - 1)$$

\vspace{5pt}

\noindent {\em Restriction to $\mathbb{H}^3$} \\

\noindent For $\mathbb{H}^3$, equation \eqref{ICTb} immediately yields the transformation equations to pseudo-Cartesian coordinates in $\mathbb{E}^4_1$, while equation \eqref{ICTmetric} gives the metric. We may, without loss of generality, let $w < v < u$, in which case the coordinate ranges can be inferred from the condition that the metric be positive definite, and that the pseudo-Cartesian coordinates be real. We therefore get \\

\noindent H-19. {\em Real ellipsoidal web I}
\begin{flalign*}
\left \{
\begin{aligned}
&ds^2 = \frac{(u-v)(u-w)}{4u(u-a)(u-b)(u-1)}\, du^2 + \frac{(u-v)(v-w)}{4v(v-a)(v-b)(1-v)}\, dv^2 + \frac{(u-w)(v-w)}{4w(w-a)(b-w)(1-w)}\, dw^2 \\
&t^2=\frac{uvw}{ab}, \ \ \ \ x^2= \frac{(u-a)(v-a)(w-a)}{a(b-a)(1-a)}, \\
&y^2 = \frac{(u-b)(v-b)(b-w)}{b(b-a)(1-b)}, \ \ \ \ z^2 = \frac{(u-1)(1-v)(1-w)}{(1-a)(1-b)} \\ 
&0 < a < w < b < v < 1 < u \ \ \ \ \ \ \ \ \ \ \ \  \ \ \ \ \ \ \ \ \ \ \ \ \ \ \  \ \ \ \ \ \ \ \ \ \ \ \ \ \ \ \ \ \  \ \ \ \ \ \ \ \ \ \ \ \ \ \ \  \ \ \ \ \ \ \ \ \ \ \ \ \ \ \ \ \ \ \ \ \ \ \ \ \ \ \ \ \ \ \ \ \ \ \ \ \ \ \ \ \ \ \ \ \ \ \ \ \ \ \ \ \ \ \ \ \ \ \ \ \ \ \ \ \ \ \ \ \ \ \ \ \ \ \ \ \ \ \ \ \ \ \ \ \ \ \ \ \ \ \ \ \ \ \  \ \ \ \ \ \ \ \ \ \ \ \ \ \ \  \ \ \ \ \ \ \ \ \ \ \ \ \ \ \ \ \ \ \ \ \ \ \ \ \ \ \ \ \ \ \ \ \ \ \ \ \ \ \ \ \ \ \ \ \ \ \ \ \ \ \ \ \ \ \ \ \ \ \ \ \ \ \ \ \ \ \ \ \ \ \ \ \ \ \ \ \ \ \ \ \ 
\end{aligned}
\right.
\end{flalign*}
Therefore there is only one isometrically inequivalent coordinate chart. \\

\vspace{5pt}

\noindent {\em Restriction to }dS$_3$ \\

\noindent Again, for dS$_3$ equations \eqref{ICTb} and \eqref{ICTmetric} give the transformation equations and metric respectively. To determine the admissible coordinate ranges, we let $w<v<u$ and impose the conditions that the metric be Lorentzian and that the pseudo-Cartesian coordinates be real. We get \\

\noindent dS-19. {\em Real ellipsoidal web I}
\begin{flalign*}
\left \{
\begin{aligned}
&ds^2 = \frac{(u-v)(u-w)}{4u(u-a)(u-b)(1 - u)}\, du^2 + \frac{(u-v)(v-w)}{4v(v-a)(b - v)(1-v)}\, dv^2 + \frac{(u-w)(v-w)}{4w(a-w)(b-w)(1-w)}\, dw^2 \\
&t^2=-\frac{uvw}{ab}, \ \ \ \ x^2= \frac{(a-u)(a-v)(a-w)}{a(b-a)(1-a)}, \\
&y^2 = \frac{(u-b)(b-v)(b-w)}{b(b-a)(1-b)}, \ \ \ \ z^2 = \frac{(1-u)(1-v)(1-w)}{(1-a)(1-b)} \\ 
&w < 0 < a < v < b < u < 1, \ \ w \ \mathrm{timelike} \ \ \ \ \ \ \ \ \ \ \ \  \ \ \ \ \ \ \ \ \ \ \ \ \ \ \  \ \ \ \ \ \ \ \ \ \ \ \ \ \ \ \ \ \  \ \ \ \ \ \ \ \ \ \ \ \ \ \ \  \ \ \ \ \ \ \ \ \ \ \ \ \ \ \ \ \ \ \ \ \ \ \ \ \ \ \ \ \ \ \ \ \ \ \ \ \ \ \ \ \ \ \ \ \ \ \ \ \ \ \ \ \ \ \ \ \ \ \ \ \ \ \ \ \ \ \ \ \ \ \ \ \ \ \ \ \ \ \ \ \ \ \ \ \ \ \ \ \ \ \ \ \ \ \  \ \ \ \ \ \ \ \ \ \ \ \ \ \ \  \ \ \ \ \ \ \ \ \ \ \ \ \ \ \ \ \ \ \ \ \ \ \ \ \ \ \ \ \ \ \ \ \ \ \ \ \ \ \ \ \ \ \ \ \ \ \ \ \ \ \ \ \ \ \ \ \ \ \ \ \ \ \ \ \ \ \ \ \ \ \ \ \ \ \ \ \ \ \ \ \ 
\end{aligned}
\right.
\end{flalign*}
Therefore there is only one isometrically inequivalent coordinate chart. \\

\vspace{10pt}

\noindent {\bf 4.11} \ $A = J_{-1}(a) \oplus J_1(0) \oplus J_1(b) \oplus J_1(1), \ \ \ \ 0 < a < b < 1$
\vspace{10pt}

\noindent Consider the case where $A$ takes the above form up to geometric equivalence. The induced CT $L$ is irreducible. If we denote the eigenfunctions of $L$ by $u,v,w$, then in the notation of equations \eqref{ICTform}-\eqref{ICTmetric}, $k=0$, and $p(z)$ and $B(z)$ are precisely the same as in 4.9 above. \\

\noindent {\em Restriction to $\mathbb{H}^3$} \\

\noindent For $\mathbb{H}^3$, equations \eqref{ICTb} and \eqref{ICTmetric} respectively give the transformation to pseudo-Cartesian coordinates, and the form of the metric in coordinates $(u,v,w)$. Letting $w <v < u$, we impose the signature of the metric and the reality of the coordinates to obtain the coordinate ranges. We get\\

\noindent H-20. {\em Real ellipsoidal web II}
\begin{flalign*}
\left \{
\begin{aligned}
&ds^2 = \frac{(u-v)(u-w)}{4u(u-a)(u-b)(u-1)}\, du^2 + \frac{(u-v)(v-w)}{4v(v-a)(v-b)(1-v)}\, dv^2 + \frac{(u-w)(v-w)}{4w(w-a)(b-w)(1-w)}\, dw^2 \\
&t^2=\frac{(u-a)(v-a)(a-w)}{a(b-a)(1-a)}, \ \ \ \ x^2= \frac{uvw}{ab}, \\
&y^2 = \frac{(u-b)(v-b)(b-w)}{b(b-a)(1-b)}, \ \ \ \ z^2 = \frac{(u-1)(1-v)(1-w)}{(1-a)(1-b)} \\ 
&0 < w < a < b < v < 1 < u \ \ \ \ \ \ \ \ \ \ \ \  \ \ \ \ \ \ \ \ \ \ \ \ \ \ \  \ \ \ \ \ \ \ \ \ \ \ \ \ \ \ \ \ \  \ \ \ \ \ \ \ \ \ \ \ \ \ \ \  \ \ \ \ \ \ \ \ \ \ \ \ \ \ \ \ \ \ \ \ \ \ \ \ \ \ \ \ \ \ \ \ \ \ \ \ \ \ \ \ \ \ \ \ \ \ \ \ \ \ \ \ \ \ \ \ \ \ \ \ \ \ \ \ \ \ \ \ \ \ \ \ \ \ \ \ \ \ \ \ \ \ \ \ \ \ \ \ \ \ \ \ \ \ \  \ \ \ \ \ \ \ \ \ \ \ \ \ \ \  \ \ \ \ \ \ \ \ \ \ \ \ \ \ \ \ \ \ \ \ \ \ \ \ \ \ \ \ \ \ \ \ \ \ \ \ \ \ \ \ \ \ \ \ \ \ \ \ \ \ \ \ \ \ \ \ \ \ \ \ \ \ \ \ \ \ \ \ \ \ \ \ \ \ \ \ \ \ \ \ \ 
\end{aligned}
\right.
\end{flalign*}
Therefore there is only one isometrically inequivalent coordinate chart. \\

\vspace{5pt}

\noindent {\em Restriction to }dS$_3$ \\

\noindent For dS$_3$ equations \eqref{ICTb} and \eqref{ICTmetric} give the transformation equations and metric respectively. To determine the admissible coordinate ranges, we impose the conditions that the metric be Lorentzian and that the pseudo-Cartesian coordinates be real. We hence obtain \\

\noindent dS-20. {\em Real ellipsoidal web II}
\begin{flalign*}
\left \{
\begin{aligned}
&ds^2 = \frac{(u-v)(u-w)}{4u(u-a)(u-b)(1-u)}\, du^2 + \frac{(u-v)(v-w)}{4v(v-a)(b - v)(1-v)}\, dv^2 + \frac{(u-w)(v-w)}{4w(a-w)(b-w)(1-w)}\, dw^2 \\
&t^2=\frac{(u-a)(v-a)(w-a)}{a(b-a)(1-a)}, \ \ \ \ x^2= \frac{uvw}{ab}, \\
&y^2 = \frac{(u-b)(b-v)(b-w)}{b(b-a)(1-b)}, \ \ \ \ z^2 = \frac{(1-u)(1-v)(1-w)}{(1-a)(1-b)} \\ 
&0 < a < b < w < 1 < v < u, \ \ u \ \mathrm{timelike} \\
&0 < a < b < w < v < u < 1, \ \ v \ \mathrm{timelike} \\
&0 < a < w < v < b < u < 1, \ \ w \ \mathrm{timelike} \\
&0 < w < v < a < b < u < 1, \ \ v \ \mathrm{timelike} \\
&w < v < 0 < a < b < u < 1, \ \ w \ \mathrm{timelike} \ \ \ \ \ \ \ \ \ \ \ \  \ \ \ \ \ \ \ \ \ \ \ \ \ \ \  \ \ \ \ \ \ \ \ \ \ \ \ \ \ \ \ \ \  \ \ \ \ \ \ \ \ \ \ \ \ \ \ \  \ \ \ \ \ \ \ \ \ \ \ \ \ \ \ \ \ \ \ \ \ \ \ \ \ \ \ \ \ \ \ \ \ \ \ \ \ \ \ \ \ \ \ \ \ \ \ \ \ \ \ \ \ \ \ \ \ \ \ \ \ \ \ \ \ \ \ \ \ \ \ \ \ \ \ \ \ \ \ \ \ \ \ \ \ \ \ \ \ \ \ \ \ \ \  \ \ \ \ \ \ \ \ \ \ \ \ \ \ \  \ \ \ \ \ \ \ \ \ \ \ \ \ \ \ \ \ \ \ \ \ \ \ \ \ \ \ \ \ \ \ \ \ \ \ \ \ \ \ \ \ \ \ \ \ \ \ \ \ \ \ \ \ \ \ \ \ \ \ \ \ \ \ \ \ \ \ \ \ \ \ \ \ \ \ \ \ \ \ \ \ 
\end{aligned}
\right.
\end{flalign*}
Therefore there are five isometrically inequivalent coordinate charts, each one corresponding to one of the five admissible coordinate ranges above. \\

\vspace{10pt}

\noindent {\bf 4.12} \ $A = J_{1}(i) \oplus J_1(-i) \oplus J_1(a) \oplus J_1(b), \ \ \ \ a < b$
\vspace{10pt}

\noindent Consider the case where $A$ takes the above form (in a complex basis) up to geometric equivalence. As $A$ has no multidimensional eigenspaces, the induced CT $L$ is irreducible. If we denote the eigenfunctions of $L$ by $u,v,w$, then in the notation of equations \eqref{ICTform}-\eqref{ICTmetric}, $k=0$ and the characteristic polynomial of $A$ is $B(\zeta) = (\zeta^2 + 1)(\zeta - a)(\zeta - b)$. Note that in applying equation \eqref{ICTb}, we must use the complex coordinates $\chi, \bar{\chi}$ in which $A$ takes the above form, where
$$\chi := \frac{1}{\sqrt{2}}(t - ix)$$
We can obtain transformation equations to real coordinates by noting that 
$$|\chi^2| = \frac{1}{2}(t^2+ x^2), \ \ \ \ \ \ \ \mathrm{Re}(\chi^2) = \frac{1}{2}(-t^2 + x^2)$$

\vspace{5pt}

\noindent {\em Restriction to $\mathbb{H}^3$} \\

\noindent For $\mathbb{H}^3$, equations \eqref{ICTb} and the above relations give the transformation to pseudo-Cartesian coordinates, while equation \eqref{ICTmetric} gives the metric. Letting $w <v < u$, we impose the signature of the metric and the reality of the coordinates to obtain the coordinate ranges. We get\\

\noindent H-21. {\em Complex ellipsoidal web}
\begin{flalign*}
\left \{
\begin{aligned}
&ds^2 = \frac{(u-v)(u-w)}{4(u^2+1)(u-a)(u-b)}\, du^2 + \frac{(u-v)(v-w)}{4(v^2 + 1)(v - a)(b-v)}\, dv^2 + \frac{(u-w)(v-w)}{4(w^2+1)(a-w)(b-w)}\, dw^2 \\
&t^2 + x^2 = \frac{\sqrt{u^2 + 1}\sqrt{v^2 + 1}\sqrt{w^2 + 1}}{\sqrt{a^2 + 1}\sqrt{b^2 + 1}}, \ \ \ \ y^2 = \frac{(u-a)(v-a)(a-w)}{(b-a)(a^2 + 1)}, \\
&t^2 - x^2= \frac{(a+b)(u + v + w - uvw) + (ab - 1)(uv + uw + vw - 1)}{(a^2 + 1)(b^2 + 1)}, \ \ \ \ z^2 = \frac{(u-b)(b-v)(b-w)}{(b-a)(b^2 + 1)} \\ 
&w < a < v < b < u \ \ \ \ \ \ \ \ \ \ \ \  \ \ \ \ \ \ \ \ \ \ \ \ \ \ \  \ \ \ \ \ \ \ \ \ \ \ \ \ \ \ \ \ \  \ \ \ \ \ \ \ \ \ \ \ \ \ \ \  \ \ \ \ \ \ \ \ \ \ \ \ \ \ \ \ \ \ \ \ \ \ \ \ \ \ \ \ \ \ \ \ \ \ \ \ \ \ \ \ \ \ \ \ \ \ \ \ \ \ \ \ \ \ \ \ \ \ \ \ \ \ \ \ \ \ \ \ \ \ \ \ \ \ \ \ \ \ \ \ \ \ \ \ \ \ \ \ \ \ \ \ \ \ \  \ \ \ \ \ \ \ \ \ \ \ \ \ \ \  \ \ \ \ \ \ \ \ \ \ \ \ \ \ \ \ \ \ \ \ \ \ \ \ \ \ \ \ \ \ \ \ \ \ \ \ \ \ \ \ \ \ \ \ \ \ \ \ \ \ \ \ \ \ \ \ \ \ \ \ \ \ \ \ \ \ \ \ \ \ \ \ \ \ \ \ \ \ \ \ \ 
\end{aligned}
\right.
\end{flalign*}
Therefore there is only one isometrically inequivalent coordinate chart. \\

\vspace{5pt}

\noindent {\em Restriction to }dS$_3$ \\

\noindent Again, equation \eqref{ICTb} and the above relations give the transformation to pseudo-Cartesian coordinates, while equation \eqref{ICTmetric} gives the metric. Letting $w < v < u$ and imposing the signature of the metric and the reality of the coordinates, we obtain the admissible coordinate ranges. \\

\noindent dS-21. {\em Complex ellipsoidal web}
\begin{flalign*}
\left \{
\begin{aligned}
&ds^2 = \frac{(u-v)(u-w)}{4(u^2 + 1)(u-a)(b-u)}\, du^2 + \frac{(u-v)(v-w)}{4(v^2+1)(v-a)(v-b)}\, dv^2 + \frac{(u-w)(v-w)}{4(w^2+1)(w-a)(b-w)}\, dw^2 \\
&t^2 + x^2 = \frac{\sqrt{u^2 + 1}\sqrt{v^2 + 1}\sqrt{w^2 + 1}}{\sqrt{a^2 + 1}\sqrt{b^2 + 1}}, \ \ \ \ y^2 = \frac{(u-a)(a-v)(a-w)}{(b-a)(a^2 + 1)}, \\
&x^2 - t^2= \frac{(a+b)(u + v + w - uvw) + (1 - ab)(uv + uw + vw - 1)}{(a^2+1)(b^2+1)}, \ \ \ \ z^2 = \frac{(b-u)(b-v)(b-w)}{(b-a)(b^2 + 1)} \\ 
&w < v < a < u < b, \ \ w \ \mathrm{timelike} \\
&a < w < b < v < u, \ \ u \ \mathrm{timelike} \\
&a < w < v < u < b, \ \ v \ \mathrm{timelike} \ \ \ \ \ \ \ \ \ \ \ \  \ \ \ \ \ \ \ \ \ \ \ \ \ \ \  \ \ \ \ \ \ \ \ \ \ \ \ \ \ \ \ \ \  \ \ \ \ \ \ \ \ \ \ \ \ \ \ \  \ \ \ \ \ \ \ \ \ \ \ \ \ \ \ \ \ \ \ \ \ \ \ \ \ \ \ \ \ \ \ \ \ \ \ \ \ \ \ \ \ \ \ \ \ \ \ \ \ \ \ \ \ \ \ \ \ \ \ \ \ \ \ \ \ \ \ \ \ \ \ \ \ \ \ \ \ \ \ \ \ \ \ \ \ \ \ \ \ \ \ \ \ \ \  \ \ \ \ \ \ \ \ \ \ \ \ \ \ \  \ \ \ \ \ \ \ \ \ \ \ \ \ \ \ \ \ \ \ \ \ \ \ \ \ \ \ \ \ \ \ \ \ \ \ \ \ \ \ \ \ \ \ \ \ \ \ \ \ \ \ \ \ \ \ \ \ \ \ \ \ \ \ \ \ \ \ \ \ \ \ \ \ \ \ \ \ \ \ \ \ 
\end{aligned}
\right.
\end{flalign*}
Therefore there are three isometrically inequivalent coordinate charts, each one corresponding to one of the three admissible coordinate ranges above. \\

\vspace{10pt}

\noindent {\bf 4.13} \ $A = J_{2}(0)^T \oplus J_1(0) \oplus J_1(0)$
\vspace{10pt}

\noindent Let us now consider the case where $A$ has a three-dimensional degenerate eigenspace. In this case, we may choose coordinates $(\eta,\xi,y,z)$ such that $A = \partial_\xi \odot \partial_\xi$, where $\eta$ and $\xi$ are null Cartesian coordinates such that $\langle \partial_\eta, \partial_\xi \rangle = 1$. The CT induced by $A$ is reducible, and algorithm \ref{WP} gives a warped product $\psi$ which decomposes $A + r \odot r$. By equation \eqref{nullWP}, $\psi$ is given by
\begin{align*}
\psi &: N_0 \times_{\rho} \mathbb{E}^2 \rightarrow \mathbb{E}^4_1 \\
							&(\eta \partial_\eta + \tilde{\xi} \partial_\xi, p) \mapsto (\tilde{\xi} - \frac{1}{2} \eta\, (Pp)^2)\partial_{\xi} + \eta \partial_\eta + \eta (Pp)
\end{align*}
where $P : \mathbb{E}^4_1 \rightarrow \mathrm{span}\{\partial_y, \partial_z\}$ is the orthogonal projection, $N_0 = \{ \eta \partial_\eta + \tilde{\xi} \partial_\xi \in \mathbb{E}^4_1 \ | \ \eta > 0 \}$ and $\rho(\eta \partial_\eta + \tilde{\xi} \partial_\xi) = \eta$. By equation \eqref{nullimage}, the image of $\psi$ consists of all points $(\eta,\xi,y,z)$ such that $\eta > 0$. Note also that $P$ is an isometry from the spherical factor $\mathbb{E}^2$ to $\mathrm{span}\{\partial_y, \partial_z\}$.

Note that we elect to write the coordinate web below in terms of the orthogonal pseudo-Cartesian coordinates $(t,x)$ associated with $(\eta,\xi)$, which by our convention are given by
$$\eta = t + x , \ \ \ \ \ \ \ \ \ \xi = \frac{1}{2}(-t+x)$$

\noindent {\em Restriction to $\mathbb{H}^3$} \\

\noindent $N_0(-1)$ is isometric to $\mathbb{H}^1$. The restriction of $A$ induces the standard coordinate $u$ on $\mathbb{H}^1$. We then get four separable webs, corresponding to the four possible webs we may lift from $\mathbb{E}^2$. These webs may be found in any standard reference; see \cite{Rajaratnam2016} for example. \\

\noindent H-22. {\em Parabolically-embedded translational web}
\begin{flalign*}
\left \{
\begin{aligned}
&ds^2 = du^2 + e^{2u}(dv^2 + dw^2) \\
&t-x = e^{-u} + e^u(v^2 + w^2), \ \ \ \ t + x = e^u, \\
&y = e^uv, \ \ \ \ z = e^uw \\ 
&-\infty < u < \infty, \ \ \ \ -\infty < v < \infty, \ \ \ \ -\infty < w < \infty \ \ \ \ \ \ \ \ \ \ \ \  \ \ \ \ \ \ \ \ \ \ \ \ \ \ \  \ \ \ \ \ \ \ \ \ \ \ \ \ \ \ \ \ \ \ \ \ \ \ \ \ \ \ \ \ \ \ \ \ \ \ \ \ \ \ \ \ \ \ \ \ \ \ \ \ \ \ \ \  \ \ \ \ \ \ \ \ \ \ \ \ \ \ \  \ \ \ \ \ \ \ \ \ \ \ \ \ \ \ \ \ \ \ \ \ \ \ \ \ \ \ \ \ \ \ \ \ \ \ \ \ \ \ \ \ \ \ \ \ \ \ \ \ \ \ \ \ \ \ \ \ \ \ \ \ \ \ \ \ \ \ \ \ \ \ \ \ \ \ \ \ \ \ \ \ 
\end{aligned}
\right.
\end{flalign*}

\vspace{10pt}

\noindent H-23. {\em Parabolically-embedded polar web}
\begin{flalign*}
\left \{
\begin{aligned}
&ds^2 = du^2 + e^{2u}(dv^2 + v^2\, dw^2) \\
&t-x = e^{-u} + e^uv^2, \ \ \ \ t + x = e^u, \\
&y = e^uv\sin{w}, \ \ \ \ z = e^u\cos{w} \\ 
&-\infty < u < \infty, \ \ \ \ 0 < v < \infty, \ \ \ \ 0 < w < 2\pi \ \ \ \ \ \ \ \ \ \ \ \  \ \ \ \ \ \ \ \ \ \ \ \ \ \ \  \ \ \ \ \ \ \ \ \ \ \ \ \ \ \ \ \ \ \ \ \ \ \ \ \ \ \ \ \ \ \ \ \ \ \ \ \ \ \ \ \ \ \ \ \ \ \ \ \ \ \ \ \  \ \ \ \ \ \ \ \ \ \ \ \ \ \ \  \ \ \ \ \ \ \ \ \ \ \ \ \ \ \ \ \ \ \ \ \ \ \ \ \ \ \ \ \ \ \ \ \ \ \ \ \ \ \ \ \ \ \ \ \ \ \ \ \ \ \ \ \ \ \ \ \ \ \ \ \ \ \ \ \ \ \ \ \ \ \ \ \ \ \ \ \ \ \ \ \ 
\end{aligned}
\right.
\end{flalign*}

\vspace{10pt}

\noindent H-24. {\em Parabolically-embedded elliptic web}
\begin{flalign*}
\left \{
\begin{aligned}
&ds^2 = du^2 + a^2e^{2u}(\cosh^2{v} - \cos^2{w})(dv^2 + dw^2) \\
&t-x = e^{-u} + a^2e^u(\cosh^2{v} - \sin^2{w}), \ \ \ \ t + x = e^u, \\
&y = ae^u\cosh{v}\, \cos{w}, \ \ \ \ z = ae^u \sinh{v}\, \sin{w} \\ 
&-\infty < u < \infty, \ \ \ \ 0 < v < \infty, \ \ \ \ 0 < w < 2\pi, \ \ \ \ a > 0 \ \ \ \ \ \ \ \ \ \ \ \  \ \ \ \ \ \ \ \ \ \ \ \ \ \ \  \ \ \ \ \ \ \ \ \ \ \ \ \ \ \ \ \ \ \ \ \ \ \ \ \ \ \ \ \ \ \ \ \ \ \ \ \ \ \ \ \ \ \ \ \ \ \ \ \ \ \ \ \  \ \ \ \ \ \ \ \ \ \ \ \ \ \ \  \ \ \ \ \ \ \ \ \ \ \ \ \ \ \ \ \ \ \ \ \ \ \ \ \ \ \ \ \ \ \ \ \ \ \ \ \ \ \ \ \ \ \ \ \ \ \ \ \ \ \ \ \ \ \ \ \ \ \ \ \ \ \ \ \ \ \ \ \ \ \ \ \ \ \ \ \ \ \ \ \ 
\end{aligned}
\right.
\end{flalign*}

\vspace{10pt}

\noindent H-25. {\em Parabolically-embedded parabolic web}
\begin{flalign*}
\left \{
\begin{aligned}
&ds^2 = du^2 + e^{2u}(v^2 + w^2)(dv^2 + dw^2) \\
&t-x = e^{-u} + \frac{1}{4}e^u(v^2 + w^2)^2, \ \ \ \ t + x = e^u, \\
&y = \frac{1}{2}e^u(v^2 - w^2), \ \ \ \ z = e^uvw \\ 
&-\infty < u < \infty, \ \ \ \ 0 < v < \infty, \ \ \ \ 0 < w < 2\pi \ \ \ \ \ \ \ \ \ \ \ \  \ \ \ \ \ \ \ \ \ \ \ \ \ \ \  \ \ \ \ \ \ \ \ \ \ \ \ \ \ \ \ \ \ \ \ \ \ \ \ \ \ \ \ \ \ \ \ \ \ \ \ \ \ \ \ \ \ \ \ \ \ \ \ \ \ \ \ \  \ \ \ \ \ \ \ \ \ \ \ \ \ \ \  \ \ \ \ \ \ \ \ \ \ \ \ \ \ \ \ \ \ \ \ \ \ \ \ \ \ \ \ \ \ \ \ \ \ \ \ \ \ \ \ \ \ \ \ \ \ \ \ \ \ \ \ \ \ \ \ \ \ \ \ \ \ \ \ \ \ \ \ \ \ \ \ \ \ \ \ \ \ \ \ \ 
\end{aligned}
\right.
\end{flalign*}

\vspace{10pt}

\noindent {\em Restriction to }dS$_3$ \\

\noindent $N_0(1)$ is isometric to an open subset of dS$_1$. The restriction of $A$ to $N_0$ induces the standard coordinate $u$ on dS$_1$, and as above we get four separable webs, upon lifting the four possible webs from the spherical factor $\mathbb{E}^2$. These are \\

\noindent dS-22. {\em Parabolically-embedded translational web}
\begin{flalign*}
\left \{
\begin{aligned}
&ds^2 = -du^2 + e^{2u}(dv^2 + dw^2) \\
&t-x = -e^{-u} + e^u(v^2 + w^2), \ \ \ \ t + x = e^u, \\
&y = e^uv, \ \ \ \ z = e^uw \\ 
&-\infty < u < \infty, \ \ \ \ -\infty < v < \infty, \ \ \ \ -\infty < w < \infty \ \ \ \ \ \ \ \ \ \ \ \  \ \ \ \ \ \ \ \ \ \ \ \ \ \ \  \ \ \ \ \ \ \ \ \ \ \ \ \ \ \ \ \ \ \ \ \ \ \ \ \ \ \ \ \ \ \ \ \ \ \ \ \ \ \ \ \ \ \ \ \ \ \ \ \ \ \ \ \  \ \ \ \ \ \ \ \ \ \ \ \ \ \ \  \ \ \ \ \ \ \ \ \ \ \ \ \ \ \ \ \ \ \ \ \ \ \ \ \ \ \ \ \ \ \ \ \ \ \ \ \ \ \ \ \ \ \ \ \ \ \ \ \ \ \ \ \ \ \ \ \ \ \ \ \ \ \ \ \ \ \ \ \ \ \ \ \ \ \ \ \ \ \ \ \ 
\end{aligned}
\right.
\end{flalign*}

\vspace{10pt}

\noindent dS-23. {\em Parabolically-embedded polar web}
\begin{flalign*}
\left \{
\begin{aligned}
&ds^2 = -du^2 + e^{2u}(dv^2 + v^2\, dw^2) \\
&t-x = -e^{-u} + e^uv^2, \ \ \ \ t + x = e^u, \\
&y = e^uv\sin{w}, \ \ \ \ z = e^u\cos{w} \\ 
&-\infty < u < \infty, \ \ \ \ 0 < v < \infty, \ \ \ \ 0 < w < 2\pi \ \ \ \ \ \ \ \ \ \ \ \  \ \ \ \ \ \ \ \ \ \ \ \ \ \ \  \ \ \ \ \ \ \ \ \ \ \ \ \ \ \ \ \ \ \ \ \ \ \ \ \ \ \ \ \ \ \ \ \ \ \ \ \ \ \ \ \ \ \ \ \ \ \ \ \ \ \ \ \  \ \ \ \ \ \ \ \ \ \ \ \ \ \ \  \ \ \ \ \ \ \ \ \ \ \ \ \ \ \ \ \ \ \ \ \ \ \ \ \ \ \ \ \ \ \ \ \ \ \ \ \ \ \ \ \ \ \ \ \ \ \ \ \ \ \ \ \ \ \ \ \ \ \ \ \ \ \ \ \ \ \ \ \ \ \ \ \ \ \ \ \ \ \ \ \ 
\end{aligned}
\right.
\end{flalign*}

\vspace{10pt}

\noindent dS-24. {\em Parabolically-embedded elliptic web}
\begin{flalign*}
\left \{
\begin{aligned}
&ds^2 = -du^2 + a^2e^{2u}(\cosh^2{v} - \cos^2{w})(dv^2 + dw^2) \\
&t-x = -e^{-u} + a^2e^u(\cosh^2{v} - \sin^2{w}), \ \ \ \ t + x = e^u, \\
&y = ae^u\cosh{v}\, \cos{w}, \ \ \ \ z = ae^u \sinh{v}\, \sin{w} \\ 
&-\infty < u < \infty, \ \ \ \ 0 < v < \infty, \ \ \ \ 0 < w < 2\pi, \ \ \ \ a > 0 \ \ \ \ \ \ \ \ \ \ \ \  \ \ \ \ \ \ \ \ \ \ \ \ \ \ \  \ \ \ \ \ \ \ \ \ \ \ \ \ \ \ \ \ \ \ \ \ \ \ \ \ \ \ \ \ \ \ \ \ \ \ \ \ \ \ \ \ \ \ \ \ \ \ \ \ \ \ \ \  \ \ \ \ \ \ \ \ \ \ \ \ \ \ \  \ \ \ \ \ \ \ \ \ \ \ \ \ \ \ \ \ \ \ \ \ \ \ \ \ \ \ \ \ \ \ \ \ \ \ \ \ \ \ \ \ \ \ \ \ \ \ \ \ \ \ \ \ \ \ \ \ \ \ \ \ \ \ \ \ \ \ \ \ \ \ \ \ \ \ \ \ \ \ \ \ 
\end{aligned}
\right.
\end{flalign*}

\vspace{10pt}

\noindent dS-25. {\em Parabolically-embedded parabolic web}
\begin{flalign*}
\left \{
\begin{aligned}
&ds^2 = -du^2 + e^{2u}(v^2 + w^2)(dv^2 + dw^2) \\
&t-x = -e^{-u} + \frac{1}{4}e^u(v^2 + w^2)^2, \ \ \ \ t + x = e^u, \\
&y = \frac{1}{2}e^u(v^2 - w^2), \ \ \ \ z = e^uvw \\ 
&-\infty < u < \infty, \ \ \ \ 0 < v < \infty, \ \ \ \ 0 < w < 2\pi \ \ \ \ \ \ \ \ \ \ \ \  \ \ \ \ \ \ \ \ \ \ \ \ \ \ \  \ \ \ \ \ \ \ \ \ \ \ \ \ \ \ \ \ \ \ \ \ \ \ \ \ \ \ \ \ \ \ \ \ \ \ \ \ \ \ \ \ \ \ \ \ \ \ \ \ \ \ \ \  \ \ \ \ \ \ \ \ \ \ \ \ \ \ \  \ \ \ \ \ \ \ \ \ \ \ \ \ \ \ \ \ \ \ \ \ \ \ \ \ \ \ \ \ \ \ \ \ \ \ \ \ \ \ \ \ \ \ \ \ \ \ \ \ \ \ \ \ \ \ \ \ \ \ \ \ \ \ \ \ \ \ \ \ \ \ \ \ \ \ \ \ \ \ \ \ 
\end{aligned}
\right.
\end{flalign*}

\vspace{10pt}

\noindent {\bf 4.14} \ $A = J_{2}(0)^T \oplus J_1(0) \oplus J_1(c), \ \ \ \ c > 0$
\vspace{10pt}

\noindent Let us now consider the case where $A$ takes the above form, in coordinates $(\eta,\xi,y,z)$, up to geometric equivalence. By rescaling our null coordinates $(\eta,\xi)$, we may assume that $c = 1$. Since $A$ has a degenerate two-dimensional eigenspace, the CT induced by $A$ is reducible, and algorithm \ref{WP} yields a warped product $\psi$ which decomposes $A + r \odot r$. By equation \eqref{nullWP}, $\psi$ is given by
\begin{align*}
\psi &: N_0 \times_{\rho} \mathbb{E}^1 \rightarrow \mathbb{E}^4_1 \\
							&(\eta \partial_\eta + \tilde{\xi} \partial_\xi + z \partial_z, p) \mapsto (\tilde{\xi} - \frac{1}{2} \eta\, (Pp)^2)\partial_{\xi} + \eta \partial_\eta + z\partial_z + \eta (Pp)
\end{align*}
where $P : \mathbb{E}^4_1 \rightarrow \mathrm{span}\{\partial_y\}$ is the orthogonal projection, $N_0 = \{ \eta \partial_\eta + \tilde{\xi} \partial_\xi + z\partial_z \in \mathbb{E}^4_1 \ | \ \eta > 0 \}$ and $\rho(\eta \partial_\eta + \tilde{\xi} \partial_\xi + z\partial_z) = \eta$. By equation \eqref{nullimage}, the image of $\psi$ consists of all points $(\eta,\xi,y,z)$ such that $\eta > 0$. Note also that $P$ gives an isometry between the spherical factor $\mathbb{E}^2$ and its image. \\

\noindent {\em Restriction to $\mathbb{H}^3$} \\

\noindent $N_0(-1)$ is isometric to $\mathbb{H}^2$. The restriction of $A$ to $N_0$ is $J_2(0)^T \oplus J_1(1)$, which induces the null elliptic web of type I on $\mathbb{H}^2$. Writing these out in terms of the pseudo-Cartesian coordinates $(t,x)$ associated with $\eta, \xi$, we obtain \\

\noindent H-26. {\em Null rotational web II}
\begin{flalign*}
\left \{
\begin{aligned}
&ds^2 = (\sec^2{v} - \sech^2{w})(dv^2 + dw^2) + \sec^2{v}\, \sech^2{w} \, du^2\\
&t-x = \cos{v}\, \cosh{w}\, (1 + \tan^2{v}\, \tanh^2{w}) - u^2\, \sec{v}\, \sech{w}, \ \ \ \ t + x = \sec{v}\, \sech{w}, \\
&y = u\, \sec{v}\, \sech{w}, \ \ \ \ z = \tan{v}\, \tanh{w} \\ 
&-\infty < u < \infty, \ \ \ \ 0 < v < \frac{\pi}{2}, \ \ \ \ -\infty < w < \infty \ \ \ \ \ \ \ \ \ \ \ \  \ \ \ \ \ \ \ \ \ \ \ \ \ \ \  \ \ \ \ \ \ \ \ \ \ \ \ \ \ \ \ \ \ \ \ \ \ \ \ \ \ \ \ \ \ \ \ \ \ \ \ \ \ \ \ \ \ \ \ \ \ \ \ \ \ \ \ \  \ \ \ \ \ \ \ \ \ \ \ \ \ \ \  \ \ \ \ \ \ \ \ \ \ \ \ \ \ \ \ \ \ \ \ \ \ \ \ \ \ \ \ \ \ \ \ \ \ \ \ \ \ \ \ \ \ \ \ \ \ \ \ \ \ \ \ \ \ \ \ \ \ \ \ \ \ \ \ \ \ \ \ \ \ \ \ \ \ \ \ \ \ \ \ \ 
\end{aligned}
\right.
\end{flalign*}

\vspace{10pt}

\noindent {\em Restriction to }dS$_3$ \\

\noindent $N_0(1)$ is isometric to an open subset of dS$_2$. The restriction of $A$ to $N_0$ is as above, and induces the null elliptic web of type I on dS$_2$. Writing these out in terms of the pseudo-Cartesian coordinates $(t,x)$ associated with $\eta, \xi$, we obtain \\

\noindent dS-26. {\em Null rotational web II}
\begin{flalign*}
\left \{
\begin{aligned}
&ds^2 = (\sech^2{u} + \csch^2{v})(du^2 - dv^2) + \sech^2{u}\, \csch^2{v}\, dw^2 \\
&t-x = w^2\, \sech{u}\, \csch{v} + \cosh{u}\, \sinh{v}\, (1 - \tanh^2{u}\, \coth^2{v}), \ \ \ \ t + x = \sech{u}\, \csch{v}, \\
&y = w\, \sech{u}\, \csch{v}, \ \ \ \ z = \tanh{u}\, \coth{v} \\ 
&-\infty < u < \infty, \ \ \ \ 0 < v < \infty, \ \ \ \ -\infty < w < \infty \ \ \ \ \ \ \ \ \ \ \ \  \ \ \ \ \ \ \ \ \ \ \ \ \ \ \  \ \ \ \ \ \ \ \ \ \ \ \ \ \ \ \ \ \ \ \ \ \ \ \ \ \ \ \ \ \ \ \ \ \ \ \ \ \ \ \ \ \ \ \ \ \ \ \ \ \ \ \ \  \ \ \ \ \ \ \ \ \ \ \ \ \ \ \  \ \ \ \ \ \ \ \ \ \ \ \ \ \ \ \ \ \ \ \ \ \ \ \ \ \ \ \ \ \ \ \ \ \ \ \ \ \ \ \ \ \ \ \ \ \ \ \ \ \ \ \ \ \ \ \ \ \ \ \ \ \ \ \ \ \ \ \ \ \ \ \ \ \ \ \ \ \ \ \ \ 
\end{aligned}
\right.
\end{flalign*}

\vspace{10pt}

\noindent {\bf 4.15} \ $A = J_{2}(0)^T \oplus J_1(0) \oplus J_1(c), \ \ \ \ c < 0$
\vspace{10pt}

\noindent Let us now consider the case where $A$ takes the above form, in coordinates $(\eta,\xi,y,z)$, up to geometric equivalence. By rescaling our null coordinates $(\eta,\xi)$, we may assume that $c = -1$. Since $A$ has a degenerate two-dimensional eigenspace, the CT induced by $A$ is reducible, and algorithm \ref{WP} yields precisely the same warped product $\psi$ as above in section 4.14. Indeed, this case differs from the one above only in the restriction of $A$ to $N_0$. \\

\noindent {\em Restriction to $\mathbb{H}^3$} \\

\noindent $N_0(-1)$ is isometric to $\mathbb{H}^2$. The restriction of $A$ to $N_0$ is $J_2(0)^T \oplus J_1(-1)$, which induces the null elliptic web of type II on $\mathbb{H}^2$. Writing these out in terms of the pseudo-Cartesian coordinates $(t,x)$ associated with $\eta, \xi$, we obtain \\

\noindent H-27. {\em Null rotational web III}
\begin{flalign*}
\left \{
\begin{aligned}
&ds^2 = (\csch^2{v} - \sec^2{w})(dv^2 + dw^2) + \csch^2{v}\, \sec^2{w} \, du^2\\
&t-x = \sinh{v}\, \cos{w}\, (1 + \coth^2{v}\, \tan^2{w}) + u^2\, \csch{v}\, \sec{w}, \ \ \ \ t + x = \sec{v}\, \sech{w}, \\
&y = u\, \csch{v}\, \sec{w}, \ \ \ \ z = \coth{v}\, \tan{w} \\ 
&-\infty < u < \infty, \ \ \ \ 0 < v < \infty, \ \ \ \ 0 < w < \frac{\pi}{2} \ \ \ \ \ \ \ \ \ \ \ \  \ \ \ \ \ \ \ \ \ \ \ \ \ \ \  \ \ \ \ \ \ \ \ \ \ \ \ \ \ \ \ \ \ \ \ \ \ \ \ \ \ \ \ \ \ \ \ \ \ \ \ \ \ \ \ \ \ \ \ \ \ \ \ \ \ \ \ \  \ \ \ \ \ \ \ \ \ \ \ \ \ \ \  \ \ \ \ \ \ \ \ \ \ \ \ \ \ \ \ \ \ \ \ \ \ \ \ \ \ \ \ \ \ \ \ \ \ \ \ \ \ \ \ \ \ \ \ \ \ \ \ \ \ \ \ \ \ \ \ \ \ \ \ \ \ \ \ \ \ \ \ \ \ \ \ \ \ \ \ \ \ \ \ \ 
\end{aligned}
\right.
\end{flalign*}

\vspace{10pt}

\noindent {\em Restriction to }dS$_3$ \\

\noindent $N_0(1)$ is isometric to an open subset of dS$_2$. The restriction of $A$ to $N_0$ is as above, and induces the null elliptic web of type II on dS$_2$. Writing these out in terms of the pseudo-Cartesian coordinates $(t,x)$ associated with $\eta, \xi$, we obtain \\

\noindent dS-27. {\em Null rotational web II}

\vspace{10pt}
\noindent for $|2x(t+x) + y^2| > 2|t+x|, \ \ (t+x)^2 > |1-z^2|$
\begin{flalign*}
\left \{
\begin{aligned}
&ds^2 = (\sec^2{u} + \sec^2{v})(-du^2 + dv^2) + \sec^2{u}\, \sec^2{v}\, dw^2 \\
&t - x = w^2\, \sec{u}\, \sec{v} - \cos{u}\, \cos{v}\, (1 - \tan^2{u}\, \tan^2{v}), \ \ \ \ t+x = \sec{u}\, \sec{v}, \\
&y = w\, \sec{u}\, \sec{v}, \ \ \ \ z = \tan{u}\, \tan{v} \\
&0 < v < u < \frac{\pi}{2}, \ \ \ \ -\infty < w < \infty \ \ \ \ \ \ \ \ \ \ \ \ \ \ \ \ \ \ \ \ \ \ \ \ \ \ \ \ \ \ \ \ \ \ \ \ \ \ \ \ \ \ \ \ \ \ \ \ \ \ \ \ \ \ \ \ \ \ \ \ \ \ \ \ \ \ \ \ \ \ \ \  \ \ \ \ \ \ \ \ \ \ \ \ \ \ \  \ \ \ \ \ \ \ \ \ \ \ \ \ \ \ \ \ \ \ \ \ \ \ \ \ \ \ \ \ \ \ \ \ \ \ \ \ \ \ \ \ \ \ \ \ \ \ \ \ \ \ \ \ \ \ \ \ \ \ \ \ \ \ \ \ \ \ \ \ \ \ \ \ \ \ \ \ \ \ \ \ 
\end{aligned}
\right.
\end{flalign*}

\noindent for $|2x(t+x) + y^2| > 2|t+x|, \ \ (t+x)^2 < |1-z^2|, \ \ |z| > 1$
\begin{flalign*}
\left \{
\begin{aligned}
&ds^2 = (\csch^2{v} + \csch^2{u})(du^2 - dv^2) + \csch^2{u}\, \csch^2{v}\, dw^2 \\
&t - x = w^2\, \csch{u}\, \csch{v} - \sinh{u}\, \sinh{v}\, (1 - \coth^2{u}\, \coth^2{v}), \ \ \ \ t+x = \csch{u}\, \csch{v}, \\
&y = w\, \csch{u}\, \csch{v}, \ \ \ \ z = \coth{u}\, \coth{v} \\
&0 < v < u < \infty, \ \ \ \ -\infty < w < \infty \ \ \ \ \ \ \ \ \ \ \ \ \ \ \ \ \ \ \ \ \ \ \ \ \ \ \ \ \ \ \ \ \ \ \ \ \ \ \ \ \ \ \ \ \ \ \ \ \ \ \ \ \ \ \ \ \ \ \ \ \ \ \ \ \ \ \ \ \ \ \ \ \ \ \  \ \ \ \ \ \ \ \ \ \ \ \ \ \ \  \ \ \ \ \ \ \ \ \ \ \ \ \ \ \ \ \ \ \ \ \ \ \ \ \ \ \ \ \ \ \ \ \ \ \ \ \ \ \ \ \ \ \ \ \ \ \ \ \ \ \ \ \ \ \ \ \ \ \ \ \ \ \ \ \ \ \ \ \ \ \ \ \ \ \ \ \ \ \ \ \ 
\end{aligned}
\right.
\end{flalign*}

\noindent for $|2x(t+x) + y^2| > 2|t+x|, \ \ (t+x)^2 < |1-z^2|, \ \ |z| < 1$
\begin{flalign*}
\left \{
\begin{aligned}
&ds^2 = (\sech^2{u} - \sech^2{v})(du^2 - dv^2) + \sech^2{u}\, \sech^2{v}\, dw^2 \\
&t - x = w^2\, \sech{u}\, \sech{v} - \cosh{u}\, \cosh{v}\, (1 - \tanh^2{u}\, \tanh^2{v}), \ \ \ \ t+x = \sech{u}\, \sech{v}, \\
&y = w\, \sech{u}\, \sech{v}, \ \ \ \ z = \tanh{u}\, \tanh{v} \\
&0 < v < u < \infty, \ \ \ \ -\infty < w < \infty \ \ \ \ \ \ \ \ \ \ \ \ \ \ \ \ \ \ \ \ \ \ \ \ \ \ \ \ \ \ \ \ \ \ \ \ \ \ \ \ \ \ \ \ \ \ \ \  \ \ \ \ \ \ \ \ \ \ \ \ \ \ \  \ \ \ \ \ \ \ \ \ \ \ \ \ \ \ \ \ \ \ \ \ \ \ \ \ \ \ \ \ \ \ \ \ \ \ \ \ \ \ \ \ \ \ \ \ \ \ \ \ \ \ \ \ \ \ \ \ \ \ \ \ \ \ \ \ \ \ \ \ \ \ \ \ \ \ \ \ \ \ \ \ 
\end{aligned}
\right.
\end{flalign*}
\vspace{10pt}

\vspace{10pt}

\noindent {\bf 4.16} \ $A = J_{2}(0)^T \oplus J_1(c) \oplus J_1(c), \ \ \ \ c > 0$
\vspace{10pt}

\noindent Let us now consider the case where $A$ takes the above form, in coordinates $(\eta,\xi,y,z)$, up to geometric equivalence. By rescaling our null coordinates $(\eta,\xi)$, we may assume that $c = 1$. Since $A$ has a spacelike two-dimensional eigenspace, the CT induced by $A$ is reducible, and algorithm \ref{WP} yields the same warped product $\psi$ as in sections 4.6-4.9. Again, this case differs only in the restriction of $A$ to $N_0$. \\

\noindent {\em Restriction to $\mathbb{H}^3$} \\

\noindent $N_0(-1)$ is isometric to an open subset of $\mathbb{H}^2$. The restriction of $A$ to $N_0$ is $J_2(0)^T \oplus J_1(1)$, which induces the null elliptic web of type I on $\mathbb{H}^2$. Writing these out in terms of the pseudo-Cartesian coordinates $(t,x)$ associated with $\eta, \xi$, we obtain \\

\noindent H-28. {\em Spacelike rotational web VII}
\begin{flalign*}
\left \{
\begin{aligned}
&ds^2 = (\sec^2{v} - \sech^2{w})(dv^2 + dw^2) + \tan^2{v}\, \tanh^2{w} \, du^2\\
&t-x = \cos{v}\, \cosh{w}\, (1 + \tan^2{v}\, \tanh^2{w}), \ \ \ \ t + x = \sec{v}\, \sech{w}, \\
&y = \tan{v}\, \tanh{w}\, \sin{u}, \ \ \ \ z = \tan{v}\, \tanh{w}\, \cos{u} \\ 
&0 < u < 2\pi, \ \ \ \ 0 < v < \frac{\pi}{2}, \ \ \ \ 0 < w < \infty \ \ \ \ \ \ \ \ \ \ \ \  \ \ \ \ \ \ \ \ \ \ \ \ \ \ \  \ \ \ \ \ \ \ \ \ \ \ \ \ \ \ \ \ \ \ \ \ \ \ \ \ \ \ \ \ \ \ \ \ \ \ \ \ \ \ \ \ \ \ \ \ \ \ \ \ \ \ \ \  \ \ \ \ \ \ \ \ \ \ \ \ \ \ \  \ \ \ \ \ \ \ \ \ \ \ \ \ \ \ \ \ \ \ \ \ \ \ \ \ \ \ \ \ \ \ \ \ \ \ \ \ \ \ \ \ \ \ \ \ \ \ \ \ \ \ \ \ \ \ \ \ \ \ \ \ \ \ \ \ \ \ \ \ \ \ \ \ \ \ \ \ \ \ \ \ 
\end{aligned}
\right.
\end{flalign*}

\vspace{10pt}

\noindent {\em Restriction to }dS$_3$ \\

\noindent $N_0(1)$ is isometric to an open subset of dS$_2$. The restriction of $A$ to $N_0$ is as above, and induces the null elliptic web of type I on dS$_2$. Writing these out in terms of the pseudo-Cartesian coordinates $(t,x)$ associated with $\eta, \xi$, we obtain \\

\noindent dS-28. {\em Spacelike rotational web VII}
\begin{flalign*}
\left \{
\begin{aligned}
&ds^2 = (\sech^2{u} + \csch^2{v})(du^2 - dv^2) + \tanh^2{u}\, \coth^2{v}\, dw^2 \\
&t-x = - \cosh{u}\, \sinh{v}\, (1 - \tanh^2{u}\, \coth^2{v}), \ \ \ \ t + x = \sech{u}\, \csch{v}, \\
&y = \tanh{u}\, \coth{v}\, \sin{w}, \ \ \ \ z = \tanh{u}\, \coth{v}\, \cos{w} \\ 
&0 < u < \infty, \ \ \ \ 0 < v < \infty, \ \ \ \ 0 < w < 2\pi \ \ \ \ \ \ \ \ \ \ \ \  \ \ \ \ \ \ \ \ \ \ \ \ \ \ \  \ \ \ \ \ \ \ \ \ \ \ \ \ \ \ \ \ \ \ \ \ \ \ \ \ \ \ \ \ \ \ \ \ \ \ \ \ \ \ \ \ \ \ \ \ \ \ \ \ \ \ \ \  \ \ \ \ \ \ \ \ \ \ \ \ \ \ \  \ \ \ \ \ \ \ \ \ \ \ \ \ \ \ \ \ \ \ \ \ \ \ \ \ \ \ \ \ \ \ \ \ \ \ \ \ \ \ \ \ \ \ \ \ \ \ \ \ \ \ \ \ \ \ \ \ \ \ \ \ \ \ \ \ \ \ \ \ \ \ \ \ \ \ \ \ \ \ \ \ 
\end{aligned}
\right.
\end{flalign*}

\vspace{10pt}

\noindent {\bf 4.17} \ $A = J_{2}(0)^T \oplus J_1(c) \oplus J_1(c), \ \ \ \ c < 0$
\vspace{10pt}

\noindent Let us now consider the case where $A$ takes the above form, in coordinates $(\eta,\xi,y,z)$, up to geometric equivalence. By rescaling our null coordinates $(\eta,\xi)$, we may assume that $c = -1$. Since $A$ has a spacelike two-dimensional eigenspace, the CT induced by $A$ is reducible, and algorithm \ref{WP} yields the same warped product $\psi$ as in sections 4.6-4.9 and 4.16 above. Again, this case differs only in the restriction of $A$ to $N_0$. \\

\noindent {\em Restriction to $\mathbb{H}^3$} \\

\noindent $N_0(-1)$ is isometric to an open subset of $\mathbb{H}^2$. The restriction of $A$ to $N_0$ is $J_2(0)^T \oplus J_1(-1)$, which induces the null elliptic web of type II on $\mathbb{H}^2$. Writing these out in terms of the pseudo-Cartesian coordinates $(t,x)$ associated with $\eta, \xi$, we obtain \\

\noindent H-29. {\em Spacelike rotational web VIII}
\begin{flalign*}
\left \{
\begin{aligned}
&ds^2 = (\csch^2{v} - \sec^2{w})(dv^2 + dw^2) + \coth^2{v}\, \tan^2{w} \, du^2\\
&t-x = \sinh{v}\, \cos{w}\, (1 + \coth^2{v}\, \tan^2{w}), \ \ \ \ t + x = \csch{v}\, \sec{w}, \\
&y = \coth{v}\, \tan{w}\, \sin{u}, \ \ \ \ z = \coth{v}\, \tan{w}\, \cos{u} \\ 
&0 < u < 2\pi, \ \ \ \ 0 < v < \infty, \ \ \ \ 0 < w < \frac{\pi}{2} \ \ \ \ \ \ \ \ \ \ \ \  \ \ \ \ \ \ \ \ \ \ \ \ \ \ \  \ \ \ \ \ \ \ \ \ \ \ \ \ \ \ \ \ \ \ \ \ \ \ \ \ \ \ \ \ \ \ \ \ \ \ \ \ \ \ \ \ \ \ \ \ \ \ \ \ \ \ \ \  \ \ \ \ \ \ \ \ \ \ \ \ \ \ \  \ \ \ \ \ \ \ \ \ \ \ \ \ \ \ \ \ \ \ \ \ \ \ \ \ \ \ \ \ \ \ \ \ \ \ \ \ \ \ \ \ \ \ \ \ \ \ \ \ \ \ \ \ \ \ \ \ \ \ \ \ \ \ \ \ \ \ \ \ \ \ \ \ \ \ \ \ \ \ \ \ 
\end{aligned}
\right.
\end{flalign*}

\vspace{10pt}

\noindent {\em Restriction to }dS$_3$ \\

\noindent $N_0(1)$ is isometric to an open subset of dS$_2$. The restriction of $A$ to $N_0$ is as above, and induces the null elliptic web of type II on dS$_2$. Writing these out in terms of the pseudo-Cartesian coordinates $(t,x)$ associated with $\eta, \xi$, we obtain \\

\noindent dS-29. {\em Spacelike rotational web VIII}

\vspace{10pt}
\noindent for $|x| > 1, \ \ tx > 0$
\begin{flalign*}
\left \{
\begin{aligned}
&ds^2 = (\sec^2{u} + \sec^2{v})(-du^2 + dv^2) + \tan^2{u}\, \tan^2{v}\, dw^2 \\
&t - x = - \cos{u}\, \cos{v}\, (1 - \tan^2{u}\, \tan^2{v}), \ \ \ \ t+x = \sec{u}\, \sec{v}, \\
&y = \tan{u}\, \tan{v}\, \sin{w}, \ \ \ \ z = \tan{u}\, \tan{v}\, \cos{w} \\
&0 < v < u < \frac{\pi}{2}, \ \ \ \ 0 < w < \frac{\pi}{2} \ \ \ \ \ \ \ \ \ \ \ \ \ \ \ \ \ \ \ \ \ \ \ \ \ \ \ \ \ \ \ \ \ \ \ \ \ \ \ \ \ \ \ \ \ \ \ \ \ \ \ \ \ \ \ \ \ \ \ \ \ \ \ \ \ \ \ \ \ \ \ \  \ \ \ \ \ \ \ \ \ \ \ \ \ \ \  \ \ \ \ \ \ \ \ \ \ \ \ \ \ \ \ \ \ \ \ \ \ \ \ \ \ \ \ \ \ \ \ \ \ \ \ \ \ \ \ \ \ \ \ \ \ \ \ \ \ \ \ \ \ \ \ \ \ \ \ \ \ \ \ \ \ \ \ \ \ \ \ \ \ \ \ \ \ \ \ \ 
\end{aligned}
\right.
\end{flalign*}

\noindent for $|x| > 1, \ \ tx < 0, \ \ \sqrt{y^2 + z^2} > 1$
\begin{flalign*}
\left \{
\begin{aligned}
&ds^2 = (\csch^2{v} + \csch^2{u})(du^2 - dv^2) + \coth^2{u}\, \coth^2{v}\, dw^2 \\
&t - x = - \sinh{u}\, \sinh{v}\, (1 - \coth^2{u}\, \coth^2{v}), \ \ \ \ t+x = \csch{u}\, \csch{v}, \\
&y = \coth{u}\, \coth{v}\, \sin{w}, \ \ \ \ z = \coth{u}\, \coth{v}\, \cos{w} \\
&0 < v < u < \infty, \ \ \ \ 0 < w < 2\pi \ \ \ \ \ \ \ \ \ \ \ \ \ \ \ \ \ \ \ \ \ \ \ \ \ \ \ \ \ \ \ \ \ \ \ \ \ \ \ \ \ \ \ \ \ \ \ \ \ \ \ \ \ \ \ \ \ \ \ \ \ \ \ \ \ \ \ \ \ \ \ \ \ \ \  \ \ \ \ \ \ \ \ \ \ \ \ \ \ \  \ \ \ \ \ \ \ \ \ \ \ \ \ \ \ \ \ \ \ \ \ \ \ \ \ \ \ \ \ \ \ \ \ \ \ \ \ \ \ \ \ \ \ \ \ \ \ \ \ \ \ \ \ \ \ \ \ \ \ \ \ \ \ \ \ \ \ \ \ \ \ \ \ \ \ \ \ \ \ \ \ 
\end{aligned}
\right.
\end{flalign*}

\noindent for $|x| > 1, \ \ tx < 0, \ \ \sqrt{y^2 + z^2} < 1$
\begin{flalign*}
\left \{
\begin{aligned}
&ds^2 = (\sech^2{u} - \sech^2{v})(du^2 - dv^2) + \tanh^2{u}\, \tanh^2{v}\, dw^2 \\
&t - x = - \cosh{u}\, \cosh{v}\, (1 - \tanh^2{u}\, \tanh^2{v}), \ \ \ \ t+x = \sech{u}\, \sech{v}, \\
&y = \tanh{u}\, \tanh{v}\, \sin{w}, \ \ \ \ z = \tanh{u}\, \tanh{v}\, \cos{w} \\
&0 < v < u < \infty, \ \ \ \ 0 < w < 2\pi \ \ \ \ \ \ \ \ \ \ \ \ \ \ \ \ \ \ \ \ \ \ \ \ \ \ \ \ \ \ \ \ \ \ \ \ \ \ \ \ \ \ \ \ \ \ \ \  \ \ \ \ \ \ \ \ \ \ \ \ \ \ \  \ \ \ \ \ \ \ \ \ \ \ \ \ \ \ \ \ \ \ \ \ \ \ \ \ \ \ \ \ \ \ \ \ \ \ \ \ \ \ \ \ \ \ \ \ \ \ \ \ \ \ \ \ \ \ \ \ \ \ \ \ \ \ \ \ \ \ \ \ \ \ \ \ \ \ \ \ \ \ \ \ 
\end{aligned}
\right.
\end{flalign*}
\vspace{10pt}

\vspace{10pt}

\noindent {\bf 4.18} \ $A = J_2(0)^T \oplus J_1(a) \oplus J_1(b), \ \ \ \ 0 < a < b$
\vspace{10pt}

\noindent We now consider the case where $A$ takes the above form, in coordinates $(\eta,\xi,y,z)$, up to geometric equivalence. Note that by rescaling our null coordinates $(\eta,\xi)$, we may assume that $b = 1$. In this case, since $A$ has no multidimensional eigenspaces, the induced CT $L$ is irreducible. So, in the notation of equations \eqref{ICTform}-\eqref{ICTmetric}, we have $k = 2$, and
$$B(\zeta) = \zeta^2(\zeta - a)(\zeta - 1), \ \ \ \ \ \ \ \ \ B_{U^{\bot}}(\zeta) = (\zeta - a)(\zeta - 1), \ \ \ \ \ \ \ \ \ p(\zeta) = (\zeta - u)(\zeta - v)(\zeta - w) $$
where $u,v,w$ are the eigenfunctions of $L$. Application of equations \eqref{ICTa} and \eqref{ICTb} yield the transformation equations between $(u,v,w)$ and $(\eta,\xi,y,z)$ on $\mathbb{H}^3$ and dS$_3$, as appropriate. As usual, we write out the transformation equations below in terms of pseudo-Cartesian coordinates $(t,x,y,z)$ associated with our null Cartesian coordinates $(\eta,\xi,y,z)$. \\

\vspace{5pt}

\noindent {\em Restriction to $\mathbb{H}^3$} \\

\noindent The metric is given by equation \eqref{ICTmetric}. Letting $w < v < u$, we impose the signature of the metric and the reality of the coordinates. This yields the coordinate ranges. Hence we get \\

\noindent H-30. {\em Null ellipsoidal web I}
\begin{flalign*}
\left \{
\begin{aligned}
&ds^2 = \frac{(u-v)(u-w)}{4u^2(u-a)(u-1)}\, du^2 + \frac{(u-v)(v-w)}{4v^2(v-a)(1-v)}\, dv^2 + \frac{(u-w)(v-w)}{4w^2(a-w)(1-w)}\, dw^2 \\
&(t+x)^2=\frac{uvw}{a}, \ \ \ \ -t^2 + x^2= \frac{1}{a^2}((1+a)uvw-a(uv + uw + vw)), \\
&y^2 = \frac{(u-a)(v-a)(a-w)}{a^2(1-a)}, \ \ \ \ z^2 = \frac{(u-1)(1-v)(1-w)}{1-a} \\ 
&0 < w < a < v < 1 < u \ \ \ \ \ \ \ \ \ \ \ \  \ \ \ \ \ \ \ \ \ \ \ \ \ \ \  \ \ \ \ \ \ \ \ \ \ \ \ \ \ \ \ \ \  \ \ \ \ \ \ \ \ \ \ \ \ \ \ \  \ \ \ \ \ \ \ \ \ \ \ \ \ \ \ \ \ \ \ \ \ \ \ \ \ \ \ \ \ \ \ \ \ \ \ \ \ \ \ \ \ \ \ \ \ \ \ \ \ \ \ \ \ \ \ \ \ \ \ \ \ \ \ \ \ \ \ \ \ \ \ \ \ \ \ \ \ \ \ \ \ \ \ \ \ \ \ \ \ \ \ \ \ \ \  \ \ \ \ \ \ \ \ \ \ \ \ \ \ \  \ \ \ \ \ \ \ \ \ \ \ \ \ \ \ \ \ \ \ \ \ \ \ \ \ \ \ \ \ \ \ \ \ \ \ \ \ \ \ \ \ \ \ \ \ \ \ \ \ \ \ \ \ \ \ \ \ \ \ \ \ \ \ \ \ \ \ \ \ \ \ \ \ \ \ \ \ \ \ \ \ 
\end{aligned}
\right.
\end{flalign*}
Therefore there is only one isometrically inequivalent coordinate chart. \\

\vspace{5pt}

\noindent {\em Restriction to }dS$_3$ \\

\noindent The metric is given by equation \eqref{ICTmetric}. Letting $w < v < u$, we impose the signature of the metric and the reality of the coordinates. This yields the coordinate ranges. Thus we obtain \\

\noindent dS-30. {\em Null ellipsoidal web I}
\begin{flalign*}
\left \{
\begin{aligned}
&ds^2 = \frac{(u-v)(u-w)}{4u^2(u-a)(1 - u)}\, du^2 + \frac{(u-v)(v-w)}{4v^2(a - v)(1-v)}\, dv^2 - \frac{(u-w)(v-w)}{4w^2(a-w)(1-w)}\, dw^2 \\
&(t+x)^2=-\frac{uvw}{a}, \ \ \ \ -t^2 + x^2= \frac{1}{a^2}(a(uv + uw + vw) - (1+a)uvw), \\
&y^2 = \frac{(u-a)(v-a)(w-a)}{a^2(1-a)}, \ \ \ \ z^2 = \frac{(1-u)(1-v)(1-w)}{1-a} \\ 
&w < 0 < v < a < u < 1, \ \ w \ \mathrm{timelike} \ \ \ \ \ \ \ \ \ \ \ \  \ \ \ \ \ \ \ \ \ \ \ \ \ \ \  \ \ \ \ \ \ \ \ \ \ \ \ \ \ \ \ \ \  \ \ \ \ \ \ \ \ \ \ \ \ \ \ \  \ \ \ \ \ \ \ \ \ \ \ \ \ \ \ \ \ \ \ \ \ \ \ \ \ \ \ \ \ \ \ \ \ \ \ \ \ \ \ \ \ \ \ \ \ \ \ \ \ \ \ \ \ \ \ \ \ \ \ \ \ \ \ \ \ \ \ \ \ \ \ \ \ \ \ \ \ \ \ \ \ \ \ \ \ \ \ \ \ \ \ \ \ \ \  \ \ \ \ \ \ \ \ \ \ \ \ \ \ \  \ \ \ \ \ \ \ \ \ \ \ \ \ \ \ \ \ \ \ \ \ \ \ \ \ \ \ \ \ \ \ \ \ \ \ \ \ \ \ \ \ \ \ \ \ \ \ \ \ \ \ \ \ \ \ \ \ \ \ \ \ \ \ \ \ \ \ \ \ \ \ \ \ \ \ \ \ \ \ \ \ 
\end{aligned}
\right.
\end{flalign*}
Therefore there is only one isometrically inequivalent coordinate chart. \\

\vspace{10pt}

\noindent {\bf 4.19} \ $A = J_2(0)^T \oplus J_1(a) \oplus J_1(b), \ \ \ \ b < 0 < a$
\vspace{10pt}

\noindent We now consider the case where $A$ takes the above form, in coordinates $(\eta,\xi,y,z)$, up to geometric equivalence. Note that by rescaling our null coordinates $(\eta,\xi)$, we may assume that $b = -1$. In this case, since $A$ has no multidimensional eigenspaces, the induced CT $L$ is irreducible. So, in the notation of equations \eqref{ICTform}-\eqref{ICTmetric}, we have $k = 2$, and
$$B(\zeta) = \zeta^2(\zeta - a)(\zeta - 1), \ \ \ \ \ \ \ \ \ B_{U^{\bot}}(\zeta) = (\zeta - a)(\zeta + 1), \ \ \ \ \ \ \ \ \ p(\zeta) = (\zeta - u)(\zeta - v)(\zeta - w) $$
where $u,v,w$ are the eigenfunctions of $L$. Application of equations \eqref{ICTa} and \eqref{ICTb} yield the transformation equations between $(u,v,w)$ and $(\eta,\xi,y,z)$ on $\mathbb{H}^3$ and dS$_3$, as appropriate. As usual, we write out the transformation equations below in terms of pseudo-Cartesian coordinates $(t,x,y,z)$ associated with our null Cartesian coordinates $(\eta,\xi,y,z)$. \\

\vspace{5pt}

\noindent {\em Restriction to $\mathbb{H}^3$} \\

\noindent The metric is given by equation \eqref{ICTmetric}. Letting $w < v < u$, we impose the signature of the metric and the reality of the coordinates to obtain the coordinate ranges. We thus get \\

\noindent H-31. {\em Null ellipsoidal web II}
\begin{flalign*}
\left \{
\begin{aligned}
&ds^2 = \frac{(u-v)(u-w)}{4u^2(u-a)(u+1)}\, du^2 + \frac{(u-v)(v-w)}{4v^2(a-v)(v+1)}\, dv^2 + \frac{(u-w)(v-w)}{4w^2(w-a)(w+1)}\, dw^2 \\
&(t+x)^2=-\frac{uvw}{a}, \ \ \ \ -t^2 + x^2= \frac{1}{a^2}((a-1)uvw + a(uv + uw + vw)), \\
&y^2 = \frac{(u-a)(v-a)(w-a)}{a^2(1+a)}, \ \ \ \ z^2 = -\frac{(u+1)(v+1)(w+1)}{1+a} \\ 
&w < -1 < 0 < v < a < u \ \ \ \ \ \ \ \ \ \ \ \  \ \ \ \ \ \ \ \ \ \ \ \ \ \ \  \ \ \ \ \ \ \ \ \ \ \ \ \ \ \ \ \ \  \ \ \ \ \ \ \ \ \ \ \ \ \ \ \  \ \ \ \ \ \ \ \ \ \ \ \ \ \ \ \ \ \ \ \ \ \ \ \ \ \ \ \ \ \ \ \ \ \ \ \ \ \ \ \ \ \ \ \ \ \ \ \ \ \ \ \ \ \ \ \ \ \ \ \ \ \ \ \ \ \ \ \ \ \ \ \ \ \ \ \ \ \ \ \ \ \ \ \ \ \ \ \ \ \ \ \ \ \ \  \ \ \ \ \ \ \ \ \ \ \ \ \ \ \  \ \ \ \ \ \ \ \ \ \ \ \ \ \ \ \ \ \ \ \ \ \ \ \ \ \ \ \ \ \ \ \ \ \ \ \ \ \ \ \ \ \ \ \ \ \ \ \ \ \ \ \ \ \ \ \ \ \ \ \ \ \ \ \ \ \ \ \ \ \ \ \ \ \ \ \ \ \ \ \ \ 
\end{aligned}
\right.
\end{flalign*}
Therefore there is only one isometrically inequivalent coordinate chart. \\

\vspace{5pt}

\noindent {\em Restriction to }dS$_3$ \\

\noindent The metric is given by equation \eqref{ICTmetric}. Assuming $w < v < u$, we impose the signature of the metric and the reality of the coordinates to obtain the coordinate ranges. Thus, we have \\

\noindent dS-31. {\em Null ellipsoidal web II}
\begin{flalign*}
\left \{
\begin{aligned}
&ds^2 = \frac{(u-v)(u-w)}{4u^2(a-u)(u+1)}\, du^2 + \frac{(u-v)(v-w)}{4v^2(v-a)(v+1)}\, dv^2 + \frac{(u-w)(v-w)}{4w^2(a-w)(w+1)}\, dw^2 \\
&(t+x)^2=\frac{uvw}{a}, \ \ \ \ -t^2 + x^2= -\frac{1}{a^2}(a(uv + uw + vw) + (a-1)uvw), \\
&y^2 = \frac{(a-u)(a-v)(a-w)}{a^2(1+a)}, \ \ \ \ z^2 = \frac{(u+1)(v+1)(w+1)}{1+a} \\ 
&w < v < -1 < 0 < u < a, \ \ w \ \mathrm{timelike} \\
&-1 < w < v < 0 < u < a, \ \ v \ \mathrm{timelike} \\
&-1 < 0 < w < v < u < a, \ \ v \ \mathrm{timelike} \\
&-1 < 0 < w < a < v < u, \ \ u \ \mathrm{timelike} \ \ \ \ \ \ \ \ \ \ \ \  \ \ \ \ \ \ \ \ \ \ \ \ \ \ \  \ \ \ \ \ \ \ \ \ \ \ \ \ \ \ \ \ \  \ \ \ \ \ \ \ \ \ \ \ \ \ \ \  \ \ \ \ \ \ \ \ \ \ \ \ \ \ \ \ \ \ \ \ \ \ \ \ \ \ \ \ \ \ \ \ \ \ \ \ \ \ \ \ \ \ \ \ \ \ \ \ \ \ \ \ \ \ \ \ \ \ \ \ \ \ \ \ \ \ \ \ \ \ \ \ \ \ \ \ \ \ \ \ \ \ \ \ \ \ \ \ \ \ \ \ \ \ \  \ \ \ \ \ \ \ \ \ \ \ \ \ \ \  \ \ \ \ \ \ \ \ \ \ \ \ \ \ \ \ \ \ \ \ \ \ \ \ \ \ \ \ \ \ \ \ \ \ \ \ \ \ \ \ \ \ \ \ \ \ \ \ \ \ \ \ \ \ \ \ \ \ \ \ \ \ \ \ \ \ \ \ \ \ \ \ \ \ \ \ \ \ \ \ \ 
\end{aligned}
\right.
\end{flalign*}
Therefore there are four isometrically inequivalent coordinate charts, each one corresponding to one of the four admissible coordinate ranges above. \\

\vspace{10pt}

\noindent {\bf 4.20} \ $A = J_2(0)^T \oplus J_1(-a) \oplus J_1(-b), \ \ \ \ 0 < a < b$
\vspace{10pt}

\noindent We now consider the case where $A$ takes the above form, in coordinates $(\eta,\xi,y,z)$, up to geometric equivalence. Note that by rescaling our null coordinates $(\eta,\xi)$, we may assume that $-b = -1$. In this case, since $A$ has no multidimensional eigenspaces, the induced CT $L$ is irreducible. So, in the notation of equations \eqref{ICTform}-\eqref{ICTmetric}, we have $k = 2$, and
$$B(\zeta) = \zeta^2(\zeta - a)(\zeta - 1), \ \ \ \ \ \ \ \ \ B_{U^{\bot}}(\zeta) = (\zeta - a)(\zeta + 1), \ \ \ \ \ \ \ \ \ p(\zeta) = (\zeta - u)(\zeta - v)(\zeta - w) $$
where $u,v,w$ are the eigenfunctions of $L$. Application of equations \eqref{ICTa} and \eqref{ICTb} yield the transformation equations between $(u,v,w)$ and $(\eta,\xi,y,z)$ on $\mathbb{H}^3$ and dS$_3$, as appropriate. As usual, we write out the transformation equations below in terms of pseudo-Cartesian coordinates $(t,x,y,z)$ associated with our null Cartesian coordinates $(\eta,\xi,y,z)$. \\

\vspace{5pt}

\noindent {\em Restriction to $\mathbb{H}^3$} \\

\noindent The metric is given by equation \eqref{ICTmetric}. Letting $w < v < u$, we impose the signature of the metric and the reality of the coordinates to obtain the coordinate ranges. \\

\noindent H-32. {\em Null ellipsoidal web III}
\begin{flalign*}
\left \{
\begin{aligned}
&ds^2 = \frac{(u-v)(u-w)}{4u^2(u+a)(u+1)}\, du^2 - \frac{(u-v)(v-w)}{4v^2(v+a)(v+1)}\, dv^2 + \frac{(u-w)(v-w)}{4w^2(w+a)(w+1)}\, dw^2 \\
&(t+x)^2=\frac{uvw}{a}, \ \ \ \ -t^2 + x^2= -\frac{1}{a^2}((1+a)uvw + a(uv + uw + vw)), \\
&y^2 = \frac{(u+a)(v+a)(w+a)}{a^2(1-a)}, \ \ \ \ z^2 = -\frac{(u+1)(v+1)(w+1)}{1-a} \\ 
&w < -1 < v < -a < 0 < u \ \ \ \ \ \ \ \ \ \ \ \  \ \ \ \ \ \ \ \ \ \ \ \ \ \ \  \ \ \ \ \ \ \ \ \ \ \ \ \ \ \ \ \ \  \ \ \ \ \ \ \ \ \ \ \ \ \ \ \  \ \ \ \ \ \ \ \ \ \ \ \ \ \ \ \ \ \ \ \ \ \ \ \ \ \ \ \ \ \ \ \ \ \ \ \ \ \ \ \ \ \ \ \ \ \ \ \ \ \ \ \ \ \ \ \ \ \ \ \ \ \ \ \ \ \ \ \ \ \ \ \ \ \ \ \ \ \ \ \ \ \ \ \ \ \ \ \ \ \ \ \ \ \ \  \ \ \ \ \ \ \ \ \ \ \ \ \ \ \  \ \ \ \ \ \ \ \ \ \ \ \ \ \ \ \ \ \ \ \ \ \ \ \ \ \ \ \ \ \ \ \ \ \ \ \ \ \ \ \ \ \ \ \ \ \ \ \ \ \ \ \ \ \ \ \ \ \ \ \ \ \ \ \ \ \ \ \ \ \ \ \ \ \ \ \ \ \ \ \ \ 
\end{aligned}
\right.
\end{flalign*}
Therefore there is only one isometrically inequivalent coordinate chart. \\

\vspace{5pt}

\noindent {\em Restriction to }dS$_3$ \\

\noindent The metric is given by equation \eqref{ICTmetric}. Assuming $w < v < u$, we impose the signature of the metric and the reality of the coordinates to obtain the coordinate ranges. Thus, we get \\

\noindent dS-32. {\em Null ellipsoidal web III}
\begin{flalign*}
\left \{
\begin{aligned}
&ds^2 = -\frac{(u-v)(u-w)}{4u^2(u+a)(u+1)}\, du^2 + \frac{(u-v)(v-w)}{4v^2(v+a)(v+1)}\, dv^2 - \frac{(u-w)(v-w)}{4w^2(w+a)(w+1)}\, dw^2 \\
&(t+x)^2=-\frac{uvw}{a}, \ \ \ \ -t^2 + x^2= \frac{1}{a^2}(a(uv + uw + vw) + (1+a)uvw), \\
&y^2 = -\frac{(u+a)(v+a)(w+a)}{a^2(1-a)}, \ \ \ \ z^2 = \frac{(u+1)(v+1)(w+1)}{1-a} \\ 
&w < v < -1 < u < -a < 0, \ \ w \ \mathrm{timelike} \\
&-1 < w < v < u < -a < 0, \ \ v \ \mathrm{timelike} \\
&-1 < w < -a < v < u < 0, \ \ u \ \mathrm{timelike} \ \ \ \ \ \ \ \ \ \ \ \  \ \ \ \ \ \ \ \ \ \ \ \ \ \ \  \ \ \ \ \ \ \ \ \ \ \ \ \ \ \ \ \ \  \ \ \ \ \ \ \ \ \ \ \ \ \ \ \  \ \ \ \ \ \ \ \ \ \ \ \ \ \ \ \ \ \ \ \ \ \ \ \ \ \ \ \ \ \ \ \ \ \ \ \ \ \ \ \ \ \ \ \ \ \ \ \ \ \ \ \ \ \ \ \ \ \ \ \ \ \ \ \ \ \ \ \ \ \ \ \ \ \ \ \ \ \ \ \ \ \ \ \ \ \ \ \ \ \ \ \ \ \ \  \ \ \ \ \ \ \ \ \ \ \ \ \ \ \  \ \ \ \ \ \ \ \ \ \ \ \ \ \ \ \ \ \ \ \ \ \ \ \ \ \ \ \ \ \ \ \ \ \ \ \ \ \ \ \ \ \ \ \ \ \ \ \ \ \ \ \ \ \ \ \ \ \ \ \ \ \ \ \ \ \ \ \ \ \ \ \ \ \ \ \ \ \ \ \ \ 
\end{aligned}
\right.
\end{flalign*}
Therefore there are three isometrically inequivalent coordinate charts, each one corresponding to one of the four admissible coordinate ranges above. \\

\vspace{10pt}

\noindent {\bf 4.21} \ $A = J_{3}(0)^T \oplus J_1(0)$
\vspace{10pt}

\noindent Let us now consider the case where $A$ takes the above form, in coordinates $(\eta,\xi,y,z)$, up to geometric equivalence. Since $A$ has a two-dimensional degenerate eigenspace, the CT induced by $A$ is reducible, and algorithm \ref{WP} gives a warped product $\psi$ which decomposes $A + r \odot r$ in the ambient space. By equation \eqref{nullWP}, $\psi$ is given by
\begin{align*}
\psi &: N_0 \times_{\rho} \mathbb{E}^1 \rightarrow \mathbb{E}^4_1 \\
							&(\eta \partial_\eta + \tilde{\xi} \partial_\xi + y \partial_y, p) \mapsto y\partial_y + (\tilde{\xi} - \frac{1}{2} \eta\, (Pp)^2)\partial_{\xi} + \eta \partial_\eta + \eta (Pp)
\end{align*}
where $P : \mathbb{E}^4_1 \rightarrow \mathrm{span}\{\partial_z\}$ is the orthogonal projection, $N_0 = \{ \eta \partial_\eta + \tilde{\xi} \partial_\xi + y\partial_y \in \mathbb{E}^4_1 \ | \ \eta > 0 \}$ and $\rho(\eta \partial_\eta + \tilde{\xi} \partial_\xi + y\partial_y) = \eta$. By equation \eqref{nullimage}, the image of $\psi$ consists of all points $(\eta,\xi,y,z)$ such that $\eta > 0$. Note also that $P$ is an isometry from the parabolically-embedded $\mathbb{E}^1$ to $\mathrm{span}\{\partial_z\}$. \\

\noindent {\em Restriction to $\mathbb{H}^3$} \\

\noindent $N_0(-1)$ is isometric to $\mathbb{H}^2$. The restriction of $A$ is given by $J_3(0)^T$, which induces the null elliptic web of type III on $\mathbb{H}^2$. We therefore get \\

\noindent H-33. {\em Null rotational web III}
\begin{flalign*}
\left \{
\begin{aligned}
&ds^2 = (u^{-2} + v^{-2})(du^2 + dv^2) + u^{-2}v^{-2}\, dw^2 \\
&t-x = \frac{(u^2 + v^2)^2}{4uv} + \frac{w^2}{uv}, \ \ \ \ t + x = \frac{1}{uv}, \\
&y = \frac{u^2 - v^2}{2uv}, \ \ \ \ z = \frac{w}{uv}\\ 
&0 < u < \infty, \ \ \ \ 0 < v < \infty, \ \ \ \ -\infty < w < \infty \ \ \ \ \ \ \ \ \ \ \ \  \ \ \ \ \ \ \ \ \ \ \ \ \ \ \  \ \ \ \ \ \ \ \ \ \ \ \ \ \ \ \ \ \ \ \ \ \ \ \ \ \ \ \ \ \ \ \ \ \ \ \ \ \ \ \ \ \ \ \ \ \ \ \ \ \ \ \ \  \ \ \ \ \ \ \ \ \ \ \ \ \ \ \  \ \ \ \ \ \ \ \ \ \ \ \ \ \ \ \ \ \ \ \ \ \ \ \ \ \ \ \ \ \ \ \ \ \ \ \ \ \ \ \ \ \ \ \ \ \ \ \ \ \ \ \ \ \ \ \ \ \ \ \ \ \ \ \ \ \ \ \ \ \ \ \ \ \ \ \ \ \ \ \ \ 
\end{aligned}
\right.
\end{flalign*}

\vspace{10pt}

\noindent {\em Restriction to dS$_3$} \\

\noindent $N_0(1)$ is isometric to an open subset of dS$_2$. The restriction of $A$ is given by $J_3(0)^T$, which induces the null elliptic web of type III on dS$_2$. We therefore get \\

\noindent dS-33. {\em Null rotational web III}
\begin{flalign*}
\left \{
\begin{aligned}
&ds^2 = (u^{-2} - v^{-2})(-du^2 + dv^2) + u^{-2}v^{-2}\, dw^2 \\
&t-x = \frac{(u^2 - v^2)^2}{4uv} + \frac{w^2}{uv}, \ \ \ \ t + x = \frac{1}{uv}, \\
&y = \frac{u^2 + v^2}{2uv}, \ \ \ \ z = \frac{w}{uv}\\ 
&0 < u < v < \infty, \ \ \ \ -\infty < w < \infty \ \ \ \ \ \ \ \ \ \ \ \  \ \ \ \ \ \ \ \ \ \ \ \ \ \ \  \ \ \ \ \ \ \ \ \ \ \ \ \ \ \ \ \ \ \ \ \ \ \ \ \ \ \ \ \ \ \ \ \ \ \ \ \ \ \ \ \ \ \ \ \ \ \ \ \ \ \ \ \  \ \ \ \ \ \ \ \ \ \ \ \ \ \ \  \ \ \ \ \ \ \ \ \ \ \ \ \ \ \ \ \ \ \ \ \ \ \ \ \ \ \ \ \ \ \ \ \ \ \ \ \ \ \ \ \ \ \ \ \ \ \ \ \ \ \ \ \ \ \ \ \ \ \ \ \ \ \ \ \ \ \ \ \ \ \ \ \ \ \ \ \ \ \ \ \ 
\end{aligned}
\right.
\end{flalign*}

\vspace{10pt}

\noindent {\bf 4.22} \ $A = J_3(0)^T \oplus J_1(a), \ \ \ \ a > 0$
\vspace{10pt}

\noindent We now consider the case where $A$ takes the above form, in coordinates $(\eta,y,\xi,z)$, up to geometric equivalence. In this case, since $A$ has no multidimensional eigenspaces, the induced CT $L$ is irreducible. So, in the notation of equations \eqref{ICTform}-\eqref{ICTmetric}, we have $k = 3$, and
$$B(\zeta) = \zeta^3(\zeta - a), \ \ \ \ \ \ \ \ \ B_{U^{\bot}}(\zeta) = (\zeta - a), \ \ \ \ \ \ \ \ \ p(\zeta) = (\zeta - u)(\zeta - v)(\zeta - w) $$
where $u,v,w$ are the eigenfunctions of $L$. Application of equations \eqref{ICTa} and \eqref{ICTb} yield the transformation equations between $(u,v,w)$ and $(\eta,y,\xi,z)$ on $\mathbb{H}^3$ and dS$_3$, as appropriate. As usual, we write out the transformation equations below in terms of pseudo-Cartesian coordinates $(t,x,y,z)$ associated with our null Cartesian coordinates $(\eta,y,\xi,z)$. \\

\vspace{5pt}

\noindent {\em Restriction to $\mathbb{H}^3$} \\

\noindent The metric is given by equation \eqref{ICTmetric}. Letting $w < v < u$, we impose the signature of the metric and the reality of the coordinates to obtain the coordinate ranges. \\

\noindent H-34. {\em Null ellipsoidal web IV}
\begin{flalign*}
\left \{
\begin{aligned}
&ds^2 = \frac{(u-v)(u-w)}{4u^3(u-a)}\, du^2 - \frac{(u-v)(v-w)}{4v^3(v-a)}\, dv^2 + \frac{(u-w)(v-w)}{4w^3(w-a)}\, dw^2 \\
&(t+x)^2=-\frac{uvw}{a}, \ \ \ \ -t^2 + x^2 + y^2 = -\frac{1}{a}(u+v+w) + \frac{1}{a^2}(uv + uw + vw) - \frac{1}{a^3}uvw, \\
&(t+x)y = \frac{1}{2a}(uv + uw + vw) - \frac{1}{2a^2}uvw, \ \ \ \ z^2 = -\frac{(a-u)(a-v)(a-w)}{a^3} \\ 
&w < 0 < v < a < u \ \ \ \ \ \ \ \ \ \ \ \  \ \ \ \ \ \ \ \ \ \ \ \ \ \ \  \ \ \ \ \ \ \ \ \ \ \ \ \ \ \ \ \ \  \ \ \ \ \ \ \ \ \ \ \ \ \ \ \  \ \ \ \ \ \ \ \ \ \ \ \ \ \ \ \ \ \ \ \ \ \ \ \ \ \ \ \ \ \ \ \ \ \ \ \ \ \ \ \ \ \ \ \ \ \ \ \ \ \ \ \ \ \ \ \ \ \ \ \ \ \ \ \ \ \ \ \ \ \ \ \ \ \ \ \ \ \ \ \ \ \ \ \ \ \ \ \ \ \ \ \ \ \ \  \ \ \ \ \ \ \ \ \ \ \ \ \ \ \  \ \ \ \ \ \ \ \ \ \ \ \ \ \ \ \ \ \ \ \ \ \ \ \ \ \ \ \ \ \ \ \ \ \ \ \ \ \ \ \ \ \ \ \ \ \ \ \ \ \ \ \ \ \ \ \ \ \ \ \ \ \ \ \ \ \ \ \ \ \ \ \ \ \ \ \ \ \ \ \ \ 
\end{aligned}
\right.
\end{flalign*}
Therefore there is only one isometrically inequivalent coordinate chart. \\

\vspace{5pt}

\noindent {\em Restriction to }dS$_3$ \\

\noindent The metric is given by equation \eqref{ICTmetric}. Assuming $w < v < u$, we impose the signature of the metric and the reality of the coordinates to obtain the coordinate ranges. Thus, we get \\

\noindent dS-34. {\em Null ellipsoidal web IV}
\begin{flalign*}
\left \{
\begin{aligned}
&ds^2 = \frac{(u-v)(u-w)}{4u^3(a-u)}\, du^2 + \frac{(u-v)(v-w)}{4v^3(v-a)}\, dv^2 + \frac{(u-w)(v-w)}{4w^3(a-w)}\, dw^2 \\
&(t+x)^2=\frac{uvw}{a}, \ \ \ \ -t^2 + x^2 + y^2 = \frac{1}{a}(u+v+w) - \frac{1}{a^2}(uv + uw + vw) + \frac{1}{a^3}uvw, \\
&(t+x)y = -\frac{1}{2a}(uv + uw + vw) + \frac{1}{2a^2}uvw, \ \ \ \ z^2 = \frac{(a-u)(a-u)(a-u)}{a^3} \\ 
&w < v < 0 < u < a, \ \ w \ \mathrm{timelike} \\
&0 < w < v < u < a, \ \ v \ \mathrm{timelike} \\
&0 < w < a < v < u, \ \ u \ \mathrm{timelike} \ \ \ \ \ \ \ \ \ \ \ \  \ \ \ \ \ \ \ \ \ \ \ \ \ \ \  \ \ \ \ \ \ \ \ \ \ \ \ \ \ \ \ \ \  \ \ \ \ \ \ \ \ \ \ \ \ \ \ \  \ \ \ \ \ \ \ \ \ \ \ \ \ \ \ \ \ \ \ \ \ \ \ \ \ \ \ \ \ \ \ \ \ \ \ \ \ \ \ \ \ \ \ \ \ \ \ \ \ \ \ \ \ \ \ \ \ \ \ \ \ \ \ \ \ \ \ \ \ \ \ \ \ \ \ \ \ \ \ \ \ \ \ \ \ \ \ \ \ \ \ \ \ \ \  \ \ \ \ \ \ \ \ \ \ \ \ \ \ \  \ \ \ \ \ \ \ \ \ \ \ \ \ \ \ \ \ \ \ \ \ \ \ \ \ \ \ \ \ \ \ \ \ \ \ \ \ \ \ \ \ \ \ \ \ \ \ \ \ \ \ \ \ \ \ \ \ \ \ \ \ \ \ \ \ \ \ \ \ \ \ \ \ \ \ \ \ \ \ \ \ 
\end{aligned}
\right.
\end{flalign*}
Therefore there are three isometrically inequivalent coordinate charts, each one corresponding to one of the four admissible coordinate ranges above. \\

\vspace{10pt}

\section*{Acknowledgements}

The authors wish to thank Krishan Rajaratnam for his careful reading of the paper and a number of helpful suggestions and comments.  
We 
also 
wish to acknowledge financial support from the Natural Sciences and Engineering Research Council of Canada in the form of a Undergraduate Student Research Award (CV) and a Discovery Grant (RGM).

\appendix

\section{Self-Adjoint Operators in Minkowski Space}\label{appA}
In this appendix, we review the classification of self-adjoint operators in $n$-dimensional Minkowski space $\mathbb{E}^n_1$. We simply quote the main results in this section, and refer the reader to \cite{Rajaratnam2014} for details and proofs. 
We first define a \textit{$k$-dimensional Jordan block} with eigenvalue $\lambda$, $J_k(\lambda)$, and a \textit{$k$-dimensional skew-normal matrix} $S_k$, to be the following $k \times k$ matrices:
$$J_k(\lambda) :=
	    \begin{pmatrix}
		\lambda & 1 &  & 0 &  \\ 
		& \lambda & \ddots & &  \\ 
		&  & \ddots & 1 &  \\ 
		&  &  & \lambda & 1 \\ 
		& 0 &  &  & \lambda
    	\end{pmatrix} \ \ \ \ \ \ \ \ \ \ \ \ \ \ \ S_k := \begin{pmatrix}
	                                                    	0 &   &  &  & 1 \\ 
	                                                            &   &  & 1 & \\
		                                                        &   & \iddots &  &  \\
		                                                         & 1 &  &  &  \\ 
	                                                        	1 &   &  &  & 0
                                                              \end{pmatrix}$$ \\
A sequence of vectors in which the metric (restricted to their span) takes the form $\varepsilon S_k$ is called a \textit{skew-normal sequence}. Recall that a linear operator $A : \mathbb{E}^n_\nu \rightarrow \mathbb{E}^n_\nu$ is self-adjoint with respect to the scalar product if $\langle Ax, y \rangle = \langle x, Ay \rangle$ for all $x$ and $y$. This holds if and only if the contravariant or covariant tensor metrically equivalent to $A$ is symmetric. Since the metric is not positive definite in $\mathbb{E}^n_1$, our classification of self-adjoint operators will specify the forms taken by both $A$ \textit{and} $g$ in an appropriate basis. The canonical form for the pair $(A,g)$ is called the \textit{metric-canonical form} or \textit{metric-Jordan form} for $A$. 

For this purpose, we introduce a signed integer $\varepsilon k \in \mathbb{Z}$, where $\varepsilon = \pm 1$ and $k \in \mathbb{N}$, and write $A = J_{\varepsilon k}(\lambda)$ as a shorthand for the pair $A = J_k(\lambda)$ and $g = \varepsilon S_k$. For square matrices $A_1$ and $A_2$, we also define the block diagonal matrix
$$A_1 \oplus A_2 :=
	    \begin{pmatrix}
		A_1 & 0 \\
		0 & A_2
    	\end{pmatrix} $$
We write $J_{\varepsilon k}(\lambda) \oplus J_{\delta m}(\mu)$ as a shorthand the pair $J_k(\lambda) \oplus J_m(\mu)$ and $g = \varepsilon S_k \oplus \delta S_m$. We now summarize the different possible canonical forms for a self-adjoint operator $A$ in $\mathbb{E}^n_1$. They are as follows: \\

\noindent \textbf{Case 1}: $A$ is diagonalizable with real eigenvalues. In this case, there is a basis such that
$$A = J_{-1}(\lambda_1) \oplus J_1(\lambda_2) \oplus \dots \oplus J_1(\lambda_n)$$
Equivalently, $A$ is diagonalized in Cartesian coordinates. \\

\noindent \textbf{Case 2}: $A$ has a complex eigenvalue $\lambda = a + ib$ with $b \neq 0$. Since $A$ is real, $\bar{\lambda}$ must be another eigenvalue; in Minkowski space, all other eigenvalues must be real. Then,
$$A = J_1(\lambda) \oplus J_1(\bar{\lambda}) \oplus J_1(\lambda_3) \oplus \dots \oplus J_1(\lambda_n)$$
in some orthogonal basis where the first two vectors are complex. Notice that since they are complex, we may assume they have length squared $+1$. \\

\noindent \textbf{Case 3}: $A$ has real eigenvalues but is not diagonalizable. Then there are three possibilities for the metric-canonical form. The first two occur when
$$A = J_{\varepsilon2}(\lambda) \oplus J_1(\lambda_3) \oplus \dots \oplus J_1(\lambda_n)$$
with $\varepsilon = \pm 1$, in some basis where the first two vectors are null. The last case occurs when
$$A = J_3(\lambda) \oplus J_1(\lambda_4) \oplus \dots \oplus J_1(\lambda_n)$$
in some basis where the first and third vectors are null; the second is spacelike. Note that in Minkowski space, a metric-Jordan block $J_{-3}(\lambda)$ is inadmissible. These are all the possibilities for the canonical forms of self-adjoint endomorphisms in Minkowski space. \\

\section{Classification of Separable Webs in $\mathbb{H}^2$}\label{appB}
In this appendix we will simply list the nine separable webs in $\mathbb{H}^2$, along with (the parameter tensor in $\mathbb{E}^3_1$ corresponding to) the associated concircular tensor, up to geometric equivalence. While these webs may be found in the literature, see for instance \cite{Olevsky1950}, \cite{Kalnins1986b} or \cite{Bruce2001}, the computations in section \ref{sec4} require knowledge of the corresponding CTs, which we have tabulated here. These webs can also be easily obtained using the theory reviewed in sections \ref{sec2} and \ref{sec3}, with the computations proceeding analogously to those used in obtaining the separable webs for $\mathrm{dS}_2$ in \cite{Rajaratnam2016}. \\

\noindent 1. {\em Elliptic web I}, \ \ $A = J_{-1}(0) \oplus J_1(a) \oplus J_1(1), \ \ 0 < a < 1$
\begin{flalign*}
\left \{
\begin{aligned}
&ds^2 = (a^2  \textrm{cd}^2(v;a) + \textrm{cs}^2(w;b))(dv^2 + dw^2)\\
&t= \nd(v;a)\, \ns(w;b), \ \ \ \ x= \sd(v;a)\, \ds(w;b), \ \ \ \ y= \cd(v;a)\, \cs(w;b)  \\ 
&0 < v < K(a), \ \ 0 < w < K(b), \ \ a^2 + b^2 = 1 \ \ \ \ \ \ \ \ \ \ \ \  \ \ \ \ \ \ \ \ \ \ \ \ \ \ \  \ \ \ \ \ \ \ \ \ \ \ \ \ \ \ \ \ \  \ \ \ \ \ \ \ \ \ \ \ \ \ \ \  \ \ \ \ \ \ \ \ \ \ \ \ \ \ \ \ \ \ \ \ \ \ \ \ \ \ \ \ \ \ \ \ \ \ \ \ \ \ \ \ \ \ \ \ \ \ \ \ \ \ \ \ \ \ \ \ \ \ \ \ \ \ \ \ \ \ \ \ \ \ \ \ \ \ \ \ \ \ \ \ \ \ \ \ \ \ \ \ \ \ \ \ \ \ \  \ \ \ \ \ \ \ \ \ \ \ \ \ \ \  \ \ \ \ \ \ \ \ \ \ \ \ \ \ \ \ \ \ \ \ \ \ \ \ \ \ \ \ \ \ \ \ \ \ \ \ \ \ \ \ \ \ \ \ \ \ \ \ \ \ \ \ \ \ \ \ \ \ \ \ \ \ \ \ \ \ \ \ \ \ \ \ \ \ \ \ \ \ \ \ \ 
\end{aligned}
\right.
\end{flalign*}

\vspace{10pt}

\noindent 2. {\em Elliptic web II}, \ \ $A = J_{-1}(a) \oplus J_1(0) \oplus J_1(1), \ \ 0 < a < 1$
\begin{flalign*}
\left \{
\begin{aligned}
&ds^2 = (\dc^2(v;a) + a^2 \sac^2(w;b))(dv^2 + dw^2) \\
&t= \nc(v;a)\, \nc(w;b), \ \ \ \ x= \sac(v;a)\, \dc(w;b), \ \ \ \ y= \dc(v;a)\, \sac(w;b) \\ 
&0 < v < K(a), \ \ 0 < w < K(b), \ \ a^2 + b^2 = 1 \ \ \ \ \ \ \ \ \ \ \ \  \ \ \ \ \ \ \ \ \ \ \ \ \ \ \  \ \ \ \ \ \ \ \ \ \ \ \ \ \ \ \ \ \  \ \ \ \ \ \ \ \ \ \ \ \ \ \ \  \ \ \ \ \ \ \ \ \ \ \ \ \ \ \ \ \ \ \ \ \ \ \ \ \ \ \ \ \ \ \ \ \ \ \ \ \ \ \ \ \ \ \ \ \ \ \ \ \ \ \ \ \ \ \ \ \ \ \ \ \ \ \ \ \ \ \ \ \ \ \ \ \ \ \ \ \ \ \ \ \ \ \ \ \ \ \ \ \ \ \ \ \ \ \  \ \ \ \ \ \ \ \ \ \ \ \ \ \ \  \ \ \ \ \ \ \ \ \ \ \ \ \ \ \ \ \ \ \ \ \ \ \ \ \ \ \ \ \ \ \ \ \ \ \ \ \ \ \ \ \ \ \ \ \ \ \ \ \ \ \ \ \ \ \ \ \ \ \ \ \ \ \ \ \ \ \ \ \ \ \ \ \ \ \ \ \ \ \ \ \ 
\end{aligned}
\right.
\end{flalign*}

\vspace{10pt}

\noindent 3. {\em Spacelike rotational web}, \ \ $A = J_{-1}(1) \oplus J_1(0) \oplus J_1(0)$
\begin{flalign*}
\left \{
\begin{aligned}
&ds^2 = dv^2 + \sinh^2{v}\, dw^2 \\
&t= \cosh{v}, \ \ \ \ x= \sinh{v}\, \cos{w}, \ \ \ \ y= \sinh{v}\, \sin{w} \\ 
&0 < v < \infty, \ \ 0 < w < 2\pi \ \ \ \ \ \ \ \ \ \ \ \ \  \ \ \ \ \ \ \ \ \ \ \ \ \ \ \  \ \ \ \ \ \ \ \ \ \ \ \ \ \ \ \ \ \  \ \ \ \ \ \ \ \ \ \ \ \ \ \ \  \ \ \ \ \ \ \ \ \ \ \ \ \ \ \ \ \ \ \ \ \ \ \ \ \ \ \ \ \ \ \ \ \ \ \ \ \ \ \ \ \ \ \ \ \ \ \ \ \ \ \ \ \ \ \ \ \ \ \ \ \ \ \ \ \ \ \ \ \ \ \ \ \ \ \ \ \ \ \ \ \ \ \ \ \ \ \ \ \ \ \ \ \ \ \  \ \ \ \ \ \ \ \ \ \ \ \ \ \ \  \ \ \ \ \ \ \ \ \ \ \ \ \ \ \ \ \ \ \ \ \ \ \ \ \ \ \ \ \ \ \ \ \ \ \ \ \ \ \ \ \ \ \ \ \ \ \ \ \ \ \ \ \ \ \ \ \ \ \ \ \ \ \ \ \ \ \ \ \ \ \ \ \ \ \ \ \ \ \ \ \ 
\end{aligned}
\right.
\end{flalign*}

\vspace{10pt}

\noindent 4. {\em Timelike rotational web}, \ \ $A = J_{-1}(0) \oplus J_1(0) \oplus J_1(1)$
\begin{flalign*}
\left \{
\begin{aligned}
&ds^2 = dv^2 + \cosh^2{v}\, dw^2 \\
&t= \cosh{v}\, \cosh{w}, \ \ \ \ x= \cosh{v}\, \sinh{w}, \ \ \ \ y= \sinh{v} \\ 
&-\infty < v < \infty, \ \ -\infty < w < \infty \ \ \ \ \ \ \ \ \ \ \ \ \  \ \ \ \ \ \ \ \ \ \ \ \ \ \ \ \ \ \ \ \ \ \ \ \ \ \ \ \ \ \ \ \ \  \ \ \ \ \ \ \ \ \ \ \ \ \ \ \  \ \ \ \ \ \ \ \ \ \ \ \ \ \ \ \ \ \ \ \ \ \ \ \ \ \ \ \ \ \ \ \ \ \ \ \ \ \ \ \ \ \ \ \ \ \ \ \ \ \ \ \ \ \ \ \ \ \ \ \ \ \ \ \ \ \ \ \ \ \ \ \ \ \ \ \ \ \ \ \ \ \ \ \ \ \ \ \ \ \ \ \ \ \ \  \ \ \ \ \ \ \ \ \ \ \ \ \ \ \  \ \ \ \ \ \ \ \ \ \ \ \ \ \ \ \ \ \ \ \ \ \ \ \ \ \ \ \ \ \ \ \ \ \ \ \ \ \ \ \ \ \ \ \ \ \ \ \ \ \ \ \ \ \ \ \ \ \ \ \ \ \ \ \ \ \ \ \ \ \ \ \ \ \ \ \ \ \ \ \ \ 
\end{aligned}
\right.
\end{flalign*}

\vspace{10pt}

\noindent 5. {\em Complex elliptic web}, \ \ $A = J_{1}(i) \oplus J_1(-i) \oplus J_1(c), \ \ c \in \mathbb{R}$
\begin{flalign*}
\left \{
\begin{aligned}
&ds^2 = (\sn^2(v;a)\, \dc^2(v;a) + \sn^2(w;b)\, \dc^2(w;b))(dv^2 + dw^2) \\
&t^2 + x^2 = \frac{2\, \dn(2v;a)\, \dn(2w;b)}{ab(1+\cn(2v;a))(1+\cn(2w;b))}, \ \ \ \ t^2 - x^2 = \frac{2\, (1 + \cn(2v;a)\, \cn(2w;b))}{(1 + \cn(2v;a))(1 + \cn(2w;b))}, \\
&y = \sn(v;a)\, \dc(v;a)\, \sn(w;b)\, \dc(w;b) \\
&0 < v < K(a), \ \ 0 < w < K(b), \ \ a^2 + b^2 = 1 \ \ \ \ \ \ \ \ \ \ \ \ \  \ \ \ \ \ \ \ \ \ \ \ \ \ \ \ \ \ \ \ \ \ \ \ \ \ \ \ \ \ \ \ \ \  \ \ \ \ \ \ \ \ \ \ \ \ \ \ \  \ \ \ \ \ \ \ \ \ \ \ \ \ \ \ \ \ \ \ \ \ \ \ \ \ \ \ \ \ \ \ \ \ \ \ \ \ \ \ \ \ \ \ \ \ \ \ \ \ \ \ \ \ \ \ \ \ \ \ \ \ \ \ \ \ \ \ \ \ \ \ \ \ \ \ \ \ \ \ \ \ \ \ \ \ \ \ \ \ \ \ \ \ \ \  \ \ \ \ \ \ \ \ \ \ \ \ \ \ \  \ \ \ \ \ \ \ \ \ \ \ \ \ \ \ \ \ \ \ \ \ \ \ \ \ \ \ \ \ \ \ \ \ \ \ \ \ \ \ \ \ \ \ \ \ \ \ \ \ \ \ \ \ \ \ \ \ \ \ \ \ \ \ \ \ \ \ \ \ \ \ \ \ \ \ \ \ \ \ \ \ 
\end{aligned}
\right.
\end{flalign*}

\vspace{10pt}

\noindent 6. {\em Null elliptic web I}, \ \ $A = J_{2}(0)^T \oplus J_1(1)$
\begin{flalign*}
\left \{
\begin{aligned}
&ds^2 = (\sec^2{v} - \sech^2{w})(dv^2 + dw^2) \\
&t + x = \sec{v}\, \sech{w}, \ \ \ \ t - x = \cos{v}\, \cosh{w}\, (1 + \tan^2{v} \tanh^2{w}), \ \ \ \ y = \tan{v}\, \tanh{w} \\
&0 < v < \frac{\pi}{2}, \ \ 0 < w < \infty\ \ \ \ \ \ \ \ \ \ \ \ \  \ \ \ \ \ \ \ \ \ \ \ \ \ \ \ \ \ \ \ \ \ \ \ \ \ \ \ \ \ \ \ \ \  \ \ \ \ \ \ \ \ \ \ \ \ \ \ \  \ \ \ \ \ \ \ \ \ \ \ \ \ \ \ \ \ \ \ \ \ \ \ \ \ \ \ \ \ \ \ \ \ \ \ \ \ \ \ \ \ \ \ \ \ \ \ \ \ \ \ \ \ \ \ \ \ \ \ \ \ \ \ \ \ \ \ \ \ \ \ \ \ \ \ \ \ \ \ \ \ \ \ \ \ \ \ \ \ \ \ \ \ \ \  \ \ \ \ \ \ \ \ \ \ \ \ \ \ \  \ \ \ \ \ \ \ \ \ \ \ \ \ \ \ \ \ \ \ \ \ \ \ \ \ \ \ \ \ \ \ \ \ \ \ \ \ \ \ \ \ \ \ \ \ \ \ \ \ \ \ \ \ \ \ \ \ \ \ \ \ \ \ \ \ \ \ \ \ \ \ \ \ \ \ \ \ \ \ \ \ 
\end{aligned}
\right.
\end{flalign*}

\vspace{10pt}

\noindent 7. {\em Null elliptic web II}, \ \ $A = J_{2}(0)^T \oplus J_1(-1)$
\begin{flalign*}
\left \{
\begin{aligned}
&ds^2 = du^2 + \cosh^2{u}\, (\textrm{csch}^2v + \textrm{sec}^2w)(dv^2 + dw^2) \\
&t + x = \csch{v}\, \sec{w}, \ \ \ \ t - x = \sinh{v}\, \cos{w}\, (1 + \textrm{coth}^2v \ \textrm{tan}^2w), \ \ \ \ y = \coth{v}\, \tan{w} \\
&0 < v < \infty, \ \ 0 < w < \frac{\pi}{2} \ \ \ \ \ \ \ \ \ \ \ \  \ \ \ \ \ \ \ \ \ \ \ \ \ \ \ \ \ \ \ \ \ \ \ \ \ \ \ \ \ \ \ \ \  \ \ \ \ \ \ \ \ \ \ \ \ \ \ \  \ \ \ \ \ \ \ \ \ \ \ \ \ \ \ \ \ \ \ \ \ \ \ \ \ \ \ \ \ \ \ \ \ \ \ \ \ \ \ \ \ \ \ \ \ \ \ \ \ \ \ \ \ \ \ \ \ \ \ \ \ \ \ \ \ \ \ \ \ \ \ \ \ \ \ \ \ \ \ \ \ \ \ \ \ \ \ \ \ \ \ \ \ \ \  \ \ \ \ \ \ \ \ \ \ \ \ \ \ \  \ \ \ \ \ \ \ \ \ \ \ \ \ \ \ \ \ \ \ \ \ \ \ \ \ \ \ \ \ \ \ \ \ \ \ \ \ \ \ \ \ \ \ \ \ \ \ \ \ \ \ \ \ \ \ \ \ \ \ \ \ \ \ \ \ \ \ \ \ \ \ \ \ \ \ \ \ \ \ \ \ 
\end{aligned}
\right.
\end{flalign*}

\vspace{10pt}

\noindent 8. {\em Null rotational web}, \ \ $A = J_{2}(0)^T \oplus J_1(0)$
\begin{flalign*}
\left \{
\begin{aligned}
&ds^2 = dv^2 + e^{2v}dw^2 \\
&t+x = e^v, \ \ \ \ t-x = e^{-v} + w^2e^v, \ \ \ \ y = \cosh{u}\, we^v \\
&-\infty < v < \infty, \ \ -\infty < w < \infty \ \ \ \ \ \ \ \ \ \ \ \  \ \ \ \ \ \ \ \ \ \ \ \ \ \ \ \ \ \ \ \ \ \ \ \ \ \ \ \ \ \ \ \ \  \ \ \ \ \ \ \ \ \ \ \ \ \ \ \  \ \ \ \ \ \ \ \ \ \ \ \ \ \ \ \ \ \ \ \ \ \ \ \ \ \ \ \ \ \ \ \ \ \ \ \ \ \ \ \ \ \ \ \ \ \ \ \ \ \ \ \ \ \ \ \ \ \ \ \ \ \ \ \ \ \ \ \ \ \ \ \ \ \ \ \ \ \ \ \ \ \ \ \ \ \ \ \ \ \ \ \ \ \ \  \ \ \ \ \ \ \ \ \ \ \ \ \ \ \  \ \ \ \ \ \ \ \ \ \ \ \ \ \ \ \ \ \ \ \ \ \ \ \ \ \ \ \ \ \ \ \ \ \ \ \ \ \ \ \ \ \ \ \ \ \ \ \ \ \ \ \ \ \ \ \ \ \ \ \ \ \ \ \ \ \ \ \ \ \ \ \ \ \ \ \ \ \ \ \ \ 
\end{aligned}
\right.
\end{flalign*}

\vspace{10pt}

\noindent 9. {\em Null elliptic web III}, \ \ $A = J_{3}(0)^T$
\begin{flalign*}
\left \{
\begin{aligned}
&ds^2 = (v^{-2} + w^{-2})(dv^2 + dw^2) \\
&t+x = \frac{1}{vw}, \ \ \ \ t-x = \frac{(v^2 + w^2)^2}{4vw}, \ \ \ \ y = \frac{(w^2 - v^2)}{2vw} \\
&0 < v < \infty, \ \ 0 < w < \infty \ \ \ \ \ \ \ \ \ \ \ \  \ \ \ \ \ \ \ \ \ \ \ \ \ \ \ \ \ \ \ \ \ \ \ \ \ \ \ \ \ \ \ \ \  \ \ \ \ \ \ \ \ \ \ \ \ \ \ \  \ \ \ \ \ \ \ \ \ \ \ \ \ \ \ \ \ \ \ \ \ \ \ \ \ \ \ \ \ \ \ \ \ \ \ \ \ \ \ \ \ \ \ \ \ \ \ \ \ \ \ \ \ \ \ \ \ \ \ \ \ \ \ \ \ \ \ \ \ \ \ \ \ \ \ \ \ \ \ \ \ \ \ \ \ \ \ \ \ \ \ \ \ \ \  \ \ \ \ \ \ \ \ \ \ \ \ \ \ \  \ \ \ \ \ \ \ \ \ \ \ \ \ \ \ \ \ \ \ \ \ \ \ \ \ \ \ \ \ \ \ \ \ \ \ \ \ \ \ \ \ \ \ \ \ \ \ \ \ \ \ \ \ \ \ \ \ \ \ \ \ \ \ \ \ \ \ \ \ \ \ \ \ \ \ \ \ \ \ \ \ 
\end{aligned}
\right.
\end{flalign*}

\end{document}